%% file: main.tex
\documentclass[rmp,aps,twocolumn]{revtex4-2}

\usepackage{bm}
\usepackage{cancel}
\usepackage{wrapfig}
\usepackage{siunitx}
\usepackage{amsmath,amssymb}
\usepackage[table,xcdraw,dvipsnames]{xcolor}
\usepackage{xparse}
\usepackage[many]{tcolorbox} % saxs and sans color boxes
\usepackage{fancyhdr}        % override package footer with page numbers
\usepackage{textgreek}  	 % upright greek letters
\usepackage[commandnameprefix=always]{changes}
\usepackage{enumitem}
\usepackage{hyperref}
\usepackage{xr}

\setlist[itemize]{leftmargin=*}

\newcommand{\fref}[1]{\autoref{#1}}    					% Figure reference
\newcommand{\eref}[1]{\autoref{#1}}    					% Equation reference
\newcommand{\edref}[2]{Eqs. \ref{#1} and \ref{#2}} 		% Double equation reference
\newcommand{\equadref}[4]{Eqs. \ref{#1}, \ref{#2}, \ref{#3}, and \ref{#4}} % Quadruple equation reference
\newcommand{\sref}[1]{\autoref{#1}}   					% Section reference
\newcommand{\aref}[1]{\hyperref[#1]{App.~\ref*{#1}}}    % Appendix reference
\newcommand{\tref}[1]{\autoref{#1}}    					% Table reference

\DeclareTColorBox{sasnote}{O{orange}O{0cm}}{
	breakable,
	outer arc=0pt,
	arc=0pt,
	colback=white,
	rightrule=0pt,
	toprule=0pt,
	top=0pt,
	right=0pt,
	bottom=0pt,
	left=0pt,
	bottomrule=0pt,
	colframe=#1,
	width=\linewidth-#2,
}

\newenvironment{sansnote}
{%
	\begin{sasnote}[Emerald]%
		\paragraph*{\textcolor{Emerald}{SANS}}%
	}
	{%
	\end{sasnote}%
}

\newenvironment{saxsnote}
{%
	\begin{sasnote}[orange]%
		\paragraph*{\textcolor{orange}{SAXS}}%
	}
	{%
	\end{sasnote}%
}

\DeclareMathOperator{\sinc}{sinc}
\DeclareMathOperator{\tM}{\text{M}}
\DeclareMathOperator{\tm}{\text{m}}
\DeclareMathOperator{\tb}{\text{b}}
\DeclareMathOperator{\ts}{\text{s}}
\DeclareMathOperator{\tVM}{\text{M}}
\DeclareMathOperator{\tVm}{\text{m}}
\DeclareMathOperator{\tVb}{\text{B}}
\DeclareMathOperator{\tVs}{\text{S}}

\newcommand{\citeasnoun}[1]{\citet{#1}}
\newcommand{\onecolumn}{\onecolumngrid}
\newcommand{\twocolumn}{\twocolumngrid}

\newcommand{\includesvg}[2][]{%
	\includegraphics[#1]{svg-inkscape/#2_svg-raw.pdf}
}

\begin{document}
\title{Small-angle solution scattering: from fundamental theory to practical approximations}
\author{Kristian Lytje}
\affiliation{Theoretical Physics and Center for Biophysics, Universität des Saarlandes, Saarbrücken, Germany}
\email{kristian.lytje@uni-saarland.de, jochen.hub@uni-saarland.de}
\author{Jan Skov Pedersen}
\affiliation{Department of Chemistry and Interdisciplinary Nanoscience Center (iNANO), Aarhus University, Aarhus, Denmark}
\author{Jochen S. Hub}
\affiliation{Theoretical Physics and Center for Biophysics, Universität des Saarlandes, Saarbrücken, Germany}
\input{abstract.tex}
\maketitle

%\externaldocument{main_si}
\input{introduction.tex}
\input{vacuum_scattering.tex}

\input{solvent_scattering.tex}

\input{background_subtraction.tex}

\input{modelling.tex}

\input{md.tex}

\input{conclusion.tex}

\input{acknowledgments.tex}
\input{appendix.tex}

\input{supplementary_information.tex}

\input{bibliography.tex}

\end{document}

%% file: abstract.tex
\begin{abstract}
	Small-angle scattering (SAS) is widely used in structural biology, soft matter, and colloidal science to probe molecular structures in solution. 
	SAS rests on a single physical principle: wave interference from a distribution of scatterers, averaged over orientations. Yet the theoretical foundations of SAS are spread across the literature, often based on differing notation, definitions, and implicit assumptions.	
	We present the theory of SAS in solution from first
	principles as a continuous derivation, spanning the scattering of a
	single electron to the observed intensity of a molecular solution and its
	comparison with atomistic structural models. The derivation is explicit throughout---approximations, averaging procedures, and algebraic manipulations are stated rather than assumed---and is independent of the probe (X-ray or neutron) and applicable to both rigid and flexible molecules. The
	framework resolves several ambiguities in the current literature, notably
	the role of background subtraction as a theoretical rather than a purely experimental
	operation and the role of boundary cross-terms in justifying that subtraction. A central result is that analytical scattering calculations and approaches based on explicit-solvent molecular dynamics, typically treated as distinct traditions, are realizations of the common theoretical framework derived here.
	As the precision and reproducibility of SAS data continue to increase, this unified framework provides a basis for integrating theory, simulation, and experiment in future developments of SAS.
\end{abstract}

%% file: introduction.tex
\section{Introduction}

Small-angle scattering (SAS) is among the most widely used techniques for
probing the structure of matter under solution conditions. By measuring
the intensity of X-rays or neutrons scattered at small angles from a solution,
SAS reveals ensemble-averaged structural information about
systems---biomolecules, polymer assemblies, complex fluids---that resist
crystallization or are too disordered for high-resolution imaging. Across the
broad range of applications in structural biology, soft-condensed-matter physics,
and colloidal science, SAS rests on a simple physical principle: the
observed intensity arises from wave interference from a spatial distribution
of scatterers, averaged over orientation. The full formal structure of
the technique follows from this principle. \\

In practice, however, this simple foundation is difficult to grasp
because the theory of SAS is distributed across a scattered
literature: the scattering amplitude is derived in one source, the treatment of
solvent contrast in another, background subtraction in a third (typically as an
experimental prescription rather than a theoretical step), and the connection to
explicit-solvent molecular-dynamics (MD) simulations in a fourth, with notation and definitions that differ between sources and assumptions that are rarely made explicit. This fragmentation has consequences. The two subtraction schemes used in practice---direct and
volume-corrected---are routinely applied as if interchangeable, despite
embedding different physical approximations with different validity limits.
Likewise, conventional modeling approaches and those based on MD simulations are treated in
the literature as distinct traditions, despite describing the same underlying
physics.\\

The present work develops the theory of small-angle scattering from first
principles as a single, continuous derivation, spanning the scattering of a
single electron to the measured intensity of a multi-molecular solution and its
comparison with atomistic structural models. Three commitments shape this
development. First, the full range of SAS applications---analytical form
factors, solution scattering, background subtraction, and atomistic
modeling---is derived explicitly from the same underlying interference
formalism, rather than being introduced as separate constructions. Second, the
emphasis throughout is on intuition and fully worked derivations: every
approximation is motivated, every averaging procedure is made explicit, and no
step is relegated to a convention. Third, the formalism is developed
generally, applying equally to rigid and flexible molecules, to homogeneous and
heterogeneous systems, and to both X-ray and neutron scattering.\\

The paper is organized in two halves. The first half, comprising Sections \ref{sec:theory}--\ref{sec:bg_subtraction}, derives 
the observable SAS
signal, starting from scattering by electrons and building up to atoms,
molecules, and molecular ensembles, before treating the modification of these
results by uniform and structured solvents. This development culminates in a
formal derivation of the background-subtraction schemes used in practice,
together with the approximations on which they rely. The second half turns to
modeling. Section \ref{sec:modeling} discusses approaches that require no prior atomistic
structure: model-independent analysis, the pair-distance distribution function,
and analytical scattering models for idealized shapes. Section \ref{sec:struc_validation} addresses
SAS modeling for cases where an atomistic reference structure is available. This
section connects these methods to MD-based scattering calculations and
establishes their formal equivalence. The result is a single theoretical
foundation from which the apparent diversity of SAS methods---analytical,
numerical, and explicit-solvent---emerges as different routes through the same
underlying physics.

\begin{sansnote}
	The theoretical framework developed here applies equally to X-ray and neutron scattering; we adopt X-ray terminology for concreteness, with neutron-specific substitutions noted where relevant.
\end{sansnote}

%% file: vacuum_scattering.tex
\section{Scattering in vacuum}\label{sec:theory}

Small-angle scattering is governed by \textit{diffraction}: the
interference of scattered waves. A useful way to visualize this is to imagine an
ocean wave approaching a rocky shoreline. When the wave hits a cliff, it does
not simply vanish; instead, it generates circular ripples radiating outward. As
these ripples from different cliffs overlap, they interfere: where crest meets
crest, the water rises higher; where crest meets trough, they cancel out. This
\textit{interference} between waves results in a complex pattern undulating on
the ocean surface. \\

\begin{figure}
	\centering
	\caption{Illustration of X-ray scattering: (A) an incoming plane wave, (B)
		scattering from an electron, (C) creating a secondary spherical wave, (D)
		interfering with another scattered wave.}
	\label{fig:theory:electron_scattering}
	\includesvg[width=\linewidth]{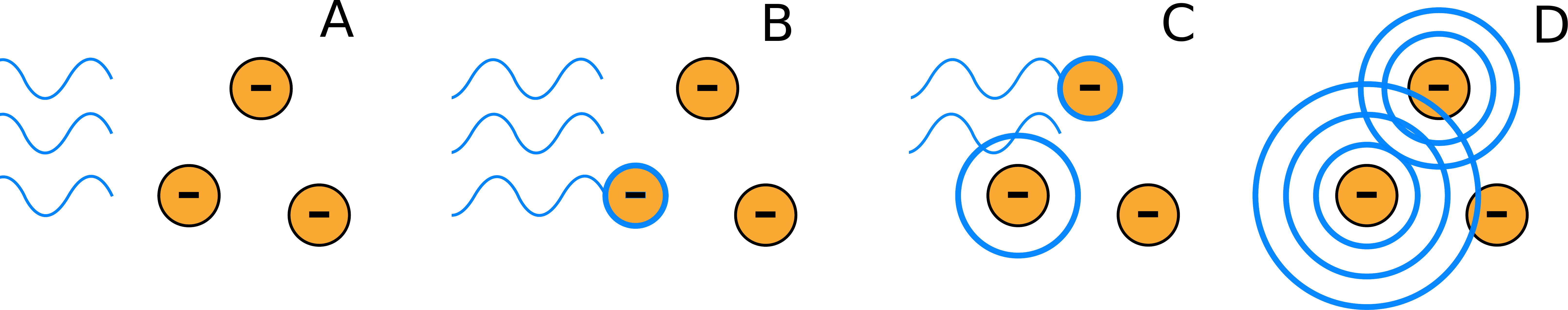}
\end{figure}

This picture is analogous to how X-rays and neutrons interact with matter. When
an incident beam of radiation encounters an electron (or a nucleus), it is
scattered in all directions as a secondary wave. Since this happens at many
scattering centers simultaneously, the outgoing waves overlap and interfere,
giving rise to the patterns measured in small-angle scattering and
crystallography. See \fref{fig:theory:electron_scattering} for an illustration
of this process. \\

To formalize the intuition, consider the scattering of an incident beam by an
electron. Because the source is far from the sample, the incident wave appears
locally planar at the scattering site. The scattered wave, in contrast,
propagates outward as a spherical wave. In the far-field with large distances
$r$, the total wave function thus takes the standard asymptotic form:
\begin{align}\label{eq:theory:wave_equation}
	\psi(\bm{r}) = e^{i\bm{k}_\text{in}\cdot\bm{r}} + A(\theta, \phi)
	\frac{e^{i k_\text{out} r }}{r}
\end{align}
where $\bm{k}_\text{in} = k_\text{in} \hat z$ is the incident wave vector, assumed parallel to the $z$ axis, and $(\theta,\phi)$ are the polar and azimuthal angles
of the outgoing wave vector $\bm{k}_\text{out}$. The
first term represents the incident plane wave, while the second term describes
the outgoing spherical wave, modulated by the angle-dependent \textit{scattering
amplitude} $A(\theta,\phi)$. This function encapsulates all information
accessible in scattering experiments and is central to this work.

\subsection{Fundamental quantities for small-angle scattering}

\subsubsection{The scattering vector}

In small-angle scattering, we are primarily concerned with \textit{elastic}
scattering, where the photon changes direction but not energy. The incident
photon may then be modeled as a particle interacting with an electron, after
which it is deflected into a new direction determined by the scattering
amplitude $A(\theta,\phi)$. \\

\begin{figure}
	\centering
	\caption{Geometry of the scattering vector $\bm{q}$. An incoming photon with
		wave vector $\bm{k}_\text{in}$ is scattered into $\bm{k}_\text{out}$, with
		$\bm{q} = \bm{k}_\text{out} - \bm{k}_\text{in}$.}
	\label{fig:theory:qvector}
	\includesvg[width=.8\linewidth]{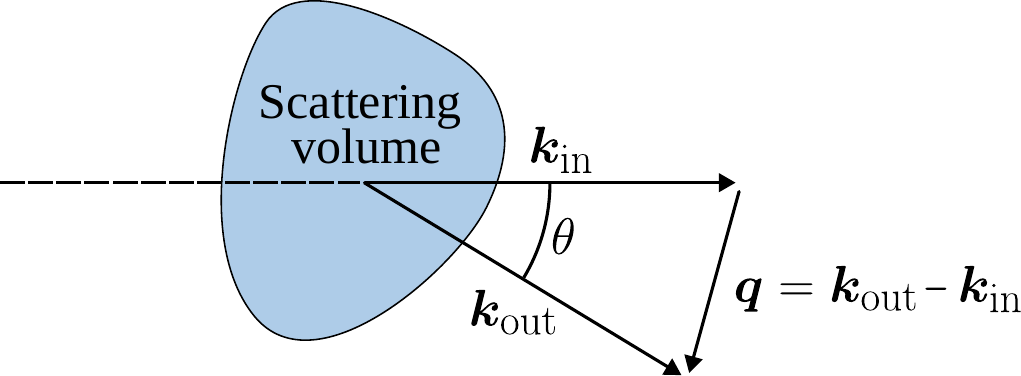}
\end{figure}

The change in momentum is described by the \textit{momentum transfer} vector,
also called the \textit{scattering vector}, denoted by $\bm{q}$. It is defined
as the difference between the outgoing and incoming wave vectors:
\begin{align*}
	|\bm{q}|^2 
	= |\bm{k}_\text{out} - \bm{k}_\text{in}|^2 
	= |\bm{k}_\text{out}|^2 + |\bm{k}_\text{in}|^2 -
	2|\bm{k}_\text{out}||\bm{k}_\text{in}|\cos\theta
\end{align*}
where $\theta$ is the scattering angle between the two wave vectors
(\fref{fig:theory:qvector}). \\

For elastic scattering, the magnitudes of the wave vectors are equal:
$|\bm{k}_\text{in}| = |\bm{k}_\text{out}| = 2\pi/\lambda$, where $\lambda$ is
the wavelength of the radiation (electromagnetic for X-rays, de Broglie for
neutrons). Substituting this gives:
\begin{align*}
	|\bm{q}| 
	= \frac{4\pi}{\lambda} \sqrt{\frac{1 - \cos\theta}{2}}
	= \frac{4\pi}{\lambda}\sin\frac{\theta}{2}
\end{align*}
This result may also be derived from the scattering geometry shown
in~\fref{fig:theory:qvector}. Expressing the measured quantity in terms of
$|\bm{q}|$ is convenient because it absorbs the experiment-specific values of
$\lambda$ and $\theta$, allowing direct comparison across different experimental
setups. Accordingly, we express scattering quantities as functions of 
$\bm{q}$ rather than $(\theta, \phi)$ throughout the remainder of this work. \\

Inelastic scattering, where both energy and momentum are exchanged (e.g.~Raman
and Compton scattering), also occurs but is typically removed experimentally for
X-rays~\cite{BookSvergun2013,Jeffries2021}. For neutrons it can be substantial
in absolute terms, but in solution measurements it originates predominantly from
the buffer~\cite{Barker2015} and therefore mostly cancels in the background
subtraction (\sref{sec:bg_subtraction}). We will therefore not consider
inelastic scattering further in this work. 

\subsubsection{The scattering intensity}

The scattering amplitude $A(\theta,\phi)$ describes the angular distribution of
scattered photons and is therefore directly related to the probability of
scattering in a given direction. Specifically, the differential cross-section
$d\sigma/d\Omega$, which quantifies the probability of scattering into a unit
solid angle for a single particle, is proportional to the squared
amplitude~\cite{Sakurai}:
\begin{align*}
	\frac{d\sigma}{d\Omega} \propto |A(\bm{q})|^2
\end{align*}
Experimentally, the detector measures the number of photons scattered into a
given direction, so the observed intensity $I(\bm{q})$ is proportional to the
differential cross-section and thus to the squared amplitude:
\begin{align}\label{eq:Iq_general}
	I(\bm{q}) = |A(\bm{q})|^2
\end{align}
Here, we have dropped the proportionality constant for brevity. Although
intensity is what we measure, interference occurs at the amplitude level.
Therefore, for modeling SAS experiments, the general strategy is:
\begin{equation*}
	\text{Compute total amplitude } A(\bm{q}) \rightarrow \text{ Square it to
		obtain } I(\bm{q})
\end{equation*}
At this stage, $I(\bm{q})$ still depends on the azimuthal angle $\phi$ through
$\bm{q}$. Later, we will remove this dependence by orientational averaging,
after which the intensity is written as $I(q)$.

\subsection{Scattering by an atom}

We now turn to the description of molecular scattering, beginning with the case of
scattering in vacuum. This serves as the foundation for the more complex
scenario of macromolecules in solution addressed further below. \\

A first-principles derivation of scattering theory would begin with the Schrödinger
equation. However, to keep this review accessible, we will omit the 
quantum-mechanical derivations and instead start from the Born series. Readers
interested in the intermediate steps are referred to standard quantum mechanics
texts such as~\citeasnoun{Sakurai} or \citeasnoun{Griffiths}. \\

For systems containing multiple electrons, a single photon could, in principle,
scatter more than once, which would greatly complicate the analysis. However, as
the scattering cross section is relatively small under typical experimental
conditions even for neutrons~\cite{Barker2015}, such consecutive scattering
events are rare and may be neglected. Assuming additionally that the incident
wave propagates through the sample essentially unperturbed---the \textit{first
Born approximation}---and neglecting polarization and magnetic effects, the
far-field scattering amplitude reduces to the Fourier transform of the local
interaction potential~\cite{Born1926}:
\begin{align}\label{eq:Aq_general}
	A(\bm{q}) = \int_V \rho(\bm{r})e^{i\bm{q}\cdot{}\bm{r}}d\bm{r}
\end{align}
where the integral runs over the entire scattering volume, and $\rho(\bm{r})$ is
the local interaction potential.\\

When evaluated for a single atom, this amplitude is commonly referred to as the
\textit{atomic form factor}, $f(\bm{q})$. These form factors have been
extensively studied and tabulated in resources such as the International Tables
for Crystallography~\cite{BookCrystTablesVolC} and~\citeasnoun{Sears1992}.

\begin{saxsnote}
	For X-rays, the interaction potential depends almost entirely on the electron
	distribution, as the contribution from protons is negligible due to their larger
	mass (\citeasnoun{BookJeu}). The scattering density is therefore well
	described by the electron density:
	\begin{align*}
		\rho(\bm{r}) \propto \rho_e(\bm{r})
	\end{align*}
	Evaluating \eref{eq:Aq_general} at $q=0$ then gives the total number of
	electrons, equal to the atomic number $Z$.\footnotemark{}\\

	Assuming the electron density of the atom is spherically symmetric, the scattering becomes independent of the azimuthal angle $\phi$ around
	the beam direction. Consequently, X-ray form factors are typically tabulated as
	a function of the magnitude $q \equiv |\bm{q}|$.\\

	Deviations from spherical symmetry can occur near atomic bonds, where
	neighboring nuclei perturb the electron distribution. These deviations provide
	valuable insight into bonding and electronic structure. Techniques such as
	\textit{anisotropic scattering} and \textit{resonant X-ray scattering} exploit
	these effects to probe chemical bonding.
\end{saxsnote}
\footnotetext{Some references absorb the classical electron (Thomson) radius
	$r_e$ into the proportionality, $\rho(\bm{r}) = r_e\rho_e(\bm{r})$, to align the
	units of the scattering amplitude with \eref{eq:theory:wave_equation} and the
	customary scattering length density convention used in the neutron scattering
	literature.}%
\begin{sansnote}
	For neutrons, the interaction potential is expressed as the scattering length
	density:
	\begin{align*}
		\rho(\bm{r}) \propto b(\bm{r})
	\end{align*}
	which varies with atomic species and isotope. \\

	Because nuclei appear effectively point-like relative to the neutron
	wavelength, they present no internal structure to the probe, and atomic
	scattering is therefore isotropic. Neutron form factors are nearly constant
	throughout the small-angle $q$-range and are thus usually tabulated as fixed
	values for each isotope. \\

	Magnetic scattering can be treated similarly by introducing a magnetic
	interaction potential in the Born
	approximation~\cite{Wiedenmann2001,Muhlbauer2019}, though this possibility will
	not be discussed further here.
\end{sansnote}

\subsection{Scattering by a molecule}
\label{sec:vacuum:molecular}

We now extend our discussion from individual atoms to larger assemblies, such as
composite particles or biomolecules. The collective scattering amplitude, still
denoted $A(\bm{q})$, is often called the \textit{molecular form factor}, while
the contribution from a single atom is denoted as the atomic form factor
introduced above.\\

The general expression for the scattering amplitude (\eref{eq:Aq_general})
remains valid for any scattering object; the distinction lies solely in the integration volume of the density function. Although this generality holds for
many of our later derivations, we will consistently refer to the scattering
objects as ``atoms'' and ``molecules'' for clarity.

\subsubsection{Summing over individual atoms}
\label{sec:theory:sum_over_scatterers}

\begin{figure}
	\centering
	\caption{A system of five overlapping spherical scatterers, each with density
		$\rho_i$ centered at $\bm{r}_i$. The total density is given by
		\eref{eq:rho_sum}.}
	\label{fig:theory:sum_of_densities}
	\includesvg[width=.7\linewidth]{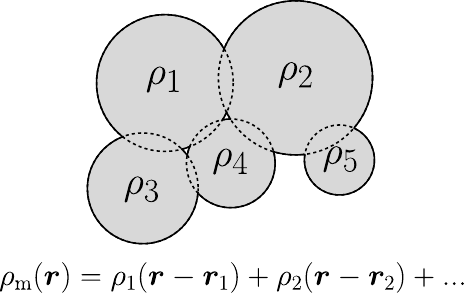}
\end{figure}

For a system composed of multiple atoms, each with density $\rho_i(\bm{r})$, the
total electron density is the sum over individual atomic densities, each
centered on its position $\bm{r}_i$:
\begin{align}\label{eq:rho_sum}
	\rho_{\tm}(\bm{r}) = \sum_i \rho_i(\bm{r} - \bm{r}_i)
\end{align}
with the subscript ``${\tm}$'' denoting the density of a single molecule. This
situation is illustrated in \fref{fig:theory:sum_of_densities}, where five
overlapping spherical scatterers contribute to the total density. \\

Inserting this into \eref{eq:Aq_general} yields:
\begin{align}\label{eq:Aq_sum}
	A_{\tm}(\bm{q}) &= \int_V \sum_i \rho_i(\bm{r} - \bm{r}_i)
	e^{i\bm{q}\cdot\bm{r}}d\bm{r} \nonumber\\
	\intertext{Because the Fourier transform is linear, the sum can be taken
		outside the integral:}
	A_{\tm}(\bm{q}) &= \sum_i \int_V \rho_i(\bm{r} - \bm{r}_i)
	e^{i\bm{q}\cdot\bm{r}}d\bm{r} \nonumber\\
	&= \sum_i e^{i\bm{q}\cdot\bm{r}_i} \int_V \rho_i(\bm{r}'_i)
	e^{i\bm{q}\cdot\bm{r}'_i}d\bm{r}'_i \nonumber\\
	&= \sum_i e^{i\bm{q}\cdot\bm{r}_i} f_i(\bm{q})
\end{align}
where we used the substitution $\bm{r}'_i = \bm{r} - \bm{r}_i$. This fundamental
result shows that the total scattering amplitude is a sum of contributions from
all atoms, each described by two components: 
\begin{itemize}
	\item The \textbf{atomic form factor} $f_i(\bm{q})$ captures the intrinsic
	scattering of the individual atom due to its shape, size, and internal
	structure, as determined by its local density distribution.
	\item The \textbf{phase term} $e^{i\bm{q}\cdot\bm{r}_i}$ accounts for how these
	contributions interfere due to the spatial arrangement of the atoms (cf. the
	ocean-and-cliffs analogy: the resulting wave pattern depends not only on the
	shape of each cliff but also on their relative positions along the shoreline). 
\end{itemize} 
This decomposition is powerful because it separates internal atomic structure
from the overall molecular geometry. The former is captured by the form factor,
while the latter is described by the phase term. The same expression applies to any collection of scatterers, not only atoms. As noted above, \eref{eq:Aq_general} places no restriction on the integration volume, so $f_i(\bm{q})$ and $\bm{r}_i$ in \eref{eq:Aq_sum} may equally be taken as the form factor and position of a molecule, a domain, or any other scattering object, yielding the scattering amplitude of an ensemble of such objects.

\subsection{Scattering by multiple molecules}

When the beam interacts with a sample containing many molecules, the measured
intensity reflects the combined scattering from all molecules in the illuminated
volume. In principle, the observed scattering depends on both the position and
orientation of every molecule; a prohibitively complex problem for most systems.
To make progress, we adopt a statistical approach: rather than modeling each
molecule individually, we approximate the ensemble by a single representative
structure. Quantities computed from this structure are indicated by an asterisk
(*).\\

This approximation works well for rigid globular systems and remains reasonable
for moderately flexible or disordered systems. When it fails, such as in
solutions containing dynamic oligomers, multiple average structures can be used,
weighted by their relative populations. For highly flexible systems, the
ensemble average can be computed over many conformations, typically obtained
from statistical models, molecular dynamics simulations, or ensemble refinement;
see \sref{sec:theory:thermal}. 

\subsubsection{Symmetries and angular dependence}

Let us consider a collection of molecules arranged in space. When the beam
interacts with this assembly, the scattered intensity depends on the relative
positions and orientations of the molecules. At this stage, there is no inherent
reason to assume that the resulting interference pattern will exhibit any
particular symmetry. \\

To make this dependence more concrete, imagine keeping the sample fixed while
rotating the detector around the beam axis. If the molecular arrangement lacks
rotational symmetry, this rotation changes the relative orientation between the
beam and the molecular geometry, altering the interference pattern. In the
general case, the scattering intensity therefore depends on both the polar angle
$\theta$ and the azimuthal angle $\phi$. \\

\begin{figure}
	\centering
	\caption{The role of rotational invariance in scattering. The top panel shows
		radially symmetric SAXS rings; the bottom panel shows discrete Bragg spots from
		a crystal. SAXS and crystallography data adapted from~\citeasnoun{Narayanan2018}
		and~\citeasnoun{Sader2009}.}
	\label{fig:theory:rotational_invariant}
	\includesvg[width=\linewidth]{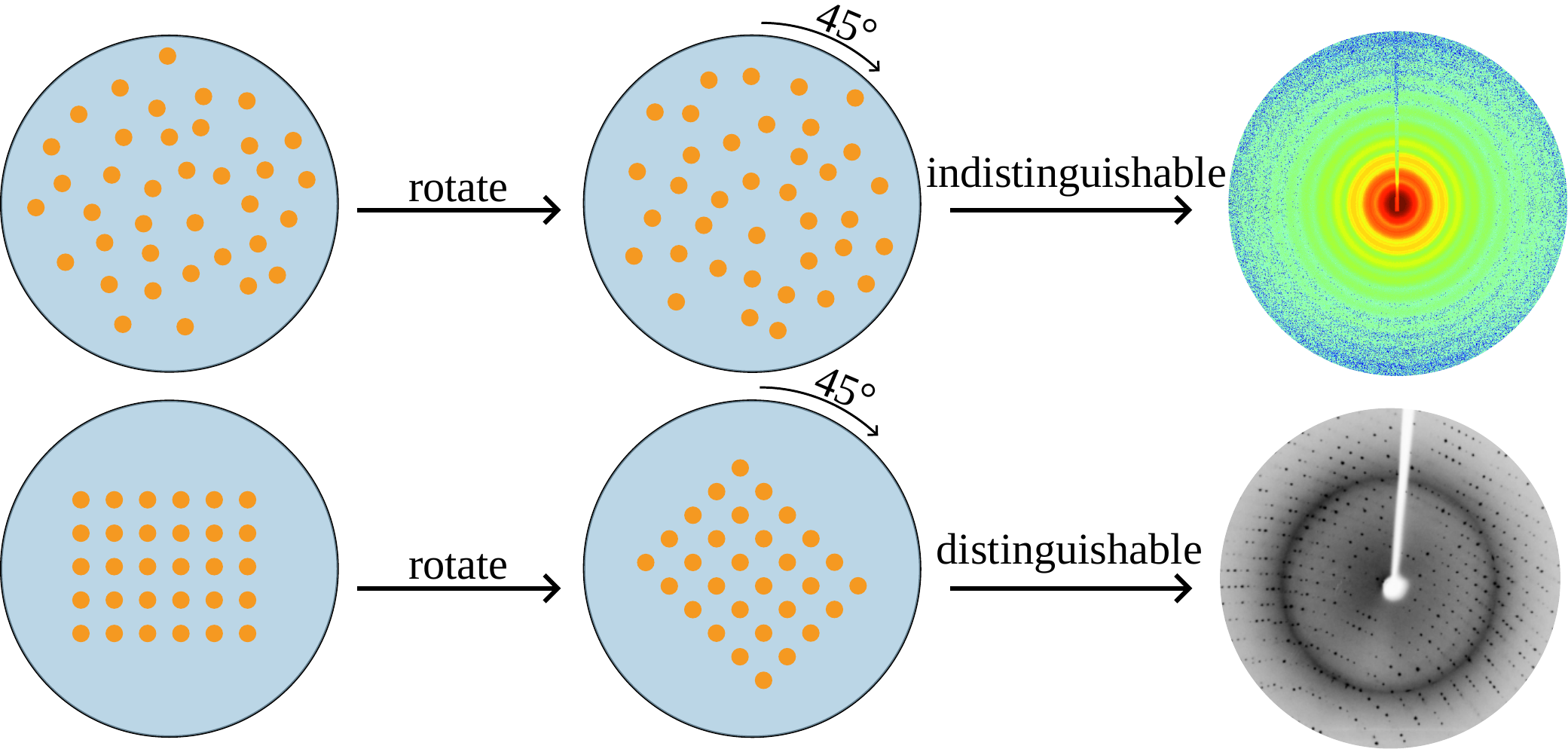}
\end{figure}

Two limiting cases illustrate this principle:
\begin{itemize}
	\item \textit{Crystals}: Their ordered lattice introduces preferred directions,
	breaking rotational symmetry about the beam axis. Rotating the detector (or
	equivalently, the sample) changes the interference conditions, and thus Bragg
	spots depend on both $\theta$ and $\phi$.
	
	\item \textit{Dilute solutions}: Randomly oriented and isotropically
	distributed molecules are invariant under rotation about the beam axis; the
	rotated system is indistinguishable from the original. This rotational symmetry
	forces the observed intensity to be independent of the azimuthal angle, leaving
	only a radial dependence on the scattering vector. The resulting interference
	pattern is radially symmetric, forming continuous rings rather than discrete
	spots (\fref{fig:theory:rotational_invariant}). When isotropy is violated---
	e.g.~due to intermolecular interactions or flow-induced alignment, particularly for
	anisotropic molecules---dependence on the azimuthal angle reappears.
\end{itemize}

\subsubsection{Scattering by \textit{uncorrelated} molecules}
\label{sec:uncorrelated_molecules}

Many SAS experiments are performed on dilute solutions, where intermolecular
interactions are negligible. We refer to such cases as \textit{uncorrelated}
systems.\\

Using \eref{eq:Aq_sum} to compute the scattering amplitude of multiple
molecules rather than multiple atoms, the amplitude of the system becomes 
$A_{\ts}(\bm{q}) = \sum_i A_{\tm}(\bm{q},\omega_i)e^{i\bm{q}\cdot\bm{r}_i}$, 
where $\omega_i$ denotes the conformation and orientation of molecule $i$ and 
the subscript ``s'' denotes the entire system. Squaring to obtain the intensity:
\begin{align*}
	I_{\ts}(q) 
	= |A_{\ts}(\bm{q})|^2
	&= \sum^N_{ij} A_{\tm}(\bm{q},\omega_i)\bar{A}_{\tm}(\bm{q},\omega_j)
	e^{i\bm{q}\cdot{}\bm{r}_{ij}}
\end{align*}
where $\bar{A}$ is the complex conjugate. Splitting the sum into self-terms
($i=j$) and cross-terms ($i \neq j$):
\begin{align*}
	I_{\ts}(q) 
	&= \sum^N_i |A_{\tm}(\bm{q},\omega_i)|^2 + \sum^N_{i\neq j}
	A_{\tm}(\bm{q},\omega_i)\bar{A}_{\tm}(\bm{q},\omega_j)
	e^{i\bm{q}\cdot{}\bm{r}_{ij}}\\
	&= \sum^N_i |A^*_{\tm}(\bm{q},\omega_i)|^2 + \sum^N_{i\neq j}
	A^*_{\tm}(\bm{q},\omega_i)\bar{A}^*_{\tm}(\bm{q},\omega_j)
	e^{i\bm{q}\cdot{}\bm{r}_{ij}}
\end{align*}
where in the second equality we have replaced each molecule by the
representative structure, so that $\omega_i$ now describes orientation alone
rather than both conformation and orientation. Under this approximation, the sum
over self-terms is self-averaging: it uniformly samples all orientations of the
same molecular structure $N$ times. Thus: 
\begin{align}\label{eq:Iq_uncorrelated_intermediate1}
	I_{\ts}(q) 
	&= N\langle |A^*_{\tm}(\bm{q},\omega_i)|^2\rangle_\Omega + \sum^N_{i\neq j}
	A^*_{\tm}(\bm{q},\omega_i)\bar{A}^*_{\tm}(\bm{q},\omega_j)
	e^{i\bm{q}\cdot{}\bm{r}_{ij}},
\end{align}
where $\langle\cdot\rangle_\Omega$ denotes the orientational average.
In a dilute solution, the orientations $\omega_i$ and $\omega_j$ of different
molecules are independent and uniformly distributed, and are independent of the
separation vector $r_{ij}$. The cross-term is therefore also self-averaging: for
each fixed orientation $\omega_i$, all orientations $\omega_j$ and distances
$r_{ij}$ are sampled with equal probability. The cross-term may therefore be
simplified to the following:
\begin{align*}
	I_{\ts}(q) 
	&= N\langle|A^*_{\tm}(\bm{q})|^2\rangle_\Omega \\ 
	&\quad+ N(N-1)\langle
	A^*_{\tm}(\bm{q})\rangle_\Omega\langle \bar{A}^*_{\tm}(\bm{q})\rangle_\Omega
	\langle e^{i\bm{q}\cdot{}\bm{r}}\rangle_\Omega \\
	&= N\langle|A^*_{\tm}(\bm{q})|^2\rangle_\Omega + |\langle
	A^*_{\tm}(\bm{q})\rangle_\Omega|^2 \sum^N_{i\neq j}\langle
	e^{i\bm{q}\cdot{}\bm{r}_{ij}}\rangle_\Omega
\end{align*}
With no preferred orientation or separation, the phase factors fluctuate
randomly, and their orientational average cancels out:\footnote{Strictly
	speaking, the phase-cancellation argument does not apply in the limit
	$q\rightarrow 0$, where $e^{i\bm{q}\cdot{}\bm{r}_{ij}} \rightarrow 1$ for all
	pairs. In this limit, scattering probes particle-number fluctuations within the
	illuminated volume rather than local spatial correlations, and the structure
	factor is fixed by the isothermal compressibility via $S(0) = n
	k_\text{B}T\kappa_T$.}
\begin{align}\label{eq:Iq_uncorrelated}
	I_{\ts}(q) &= N\langle |A^*_{\tm}(\bm{q})|^2\rangle_\Omega \equiv NI^*_{\tm}(q)
	\quad \text{(uncorrelated)}
\end{align}
This result is useful as it shows that, for dilute solutions of nearly
identical, non-interacting molecules, the total intensity equals the
number of molecules multiplied by the intensity from a single molecule. \\

In the literature, this result is often written as $I_{\ts}(q) = NP(q)$, where
$P(q)$ is also referred to as the molecular form factor. Since $P(q)$ is an
orientationally averaged intensity rather than a scattering amplitude, we will
avoid this terminology here.

\subsubsection{Scattering by \textit{correlated} molecules}
\label{sec:correlated_mols}

When the spatial positions of the molecules are correlated, for example due to
intermolecular interactions, the second term in
\eref{eq:Iq_uncorrelated_intermediate1} must be retained. However, this term
cannot be evaluated without detailed knowledge of the interaction potential. To
address this, various approximation schemes have been proposed (Pedersen,~J.S. in~\citeasnoun{BookLindner2024}); here, we use the simplest of
these: the \textit{decoupling approximation} proposed
by~\citeasnoun{Kotlarchyk1983}. This approximation assumes that molecular
orientations are independent of positions and of each other; interactions affect
where molecules sit, not how they point. \\

Applying this approximation, the orientational and positional averages decouple,
returning us to the form of \eref{eq:Iq_uncorrelated_intermediate1}:
\begin{align*}
	I_{\ts}(q) 
	&= N\langle|A^*_{\tm}(\bm{q})|^2\rangle_\Omega + |\langle
	A^*_{\tm}(\bm{q})\rangle_\Omega|^2 \sum^N_{i\neq j} \left\langle
	e^{i\bm{q}\cdot{}\bm{r}_{ij}}\right\rangle_\Omega \\
	&= N\langle|A^*_{\tm}(\bm{q})|^2\rangle_\Omega \left\{1 + \frac{|\langle
		A^*_{\tm}(\bm{q})\rangle_\Omega|^2}{\langle|A^*_{\tm}(\bm{q})|^2\rangle_\Omega}
	\frac{1}{N}\sum^N_{i\neq j} \left\langle
	e^{i\bm{q}\cdot{}\bm{r}_{ij}}\right\rangle_\Omega\right\}
\end{align*}
This motivates the definition of the \textit{structure factor}:
\begin{align*}
	S^*(q) \equiv 1 + \frac{\beta^*(q)}{N}\sum^N_{i\neq j} \left\langle
	e^{i\bm{q}\cdot{}\bm{r}_{ij}}\right\rangle_\Omega \qquad \beta^*(q) =
	\frac{|\langle
		A^*_{\tm}(\bm{q})\rangle_\Omega|^2}{\langle|A^*_{\tm}(\bm{q})|^2\rangle_\Omega}
\end{align*}
where we have introduced $\beta^*(q)$ for convenience. The intensity then takes
the form:
\begin{align}\label{eq:Iq_correlated}
	I_{\ts}(q) = N\langle |A^*_{\tm}(\bm{q})|^2\rangle_\Omega S^*(q) =
	NI^*_{\tm}(q)S^*(q)\ \ \ \text{(correlated)}
\end{align}
Under this approximation, the observed intensity is that of the dilute case but
modulated by the structure factor, which captures the effect of molecular
correlations. The structure factor is commonly evaluated using correlation
functions, which we explore in \aref{sec:appendix:correlation_funcs}.

\subsubsection{When to account for the structure factor}

The structure factor must be considered whenever interparticle interactions
produce non-random spatial correlations, such as in concentrated solutions,
aggregated systems, or in samples exhibiting strong long-range repulsion. In
these cases, $S^*(q)$ must be modeled or computed using additional information
about the system (L. Cipelletti, R. Klein, and E. Zaccarelli
in~\citeasnoun{BookLindner2024}). For dilute solutions, these interactions are
negligible, and $S^*(q)\approx1$ can be assumed, reducing the analysis to the
uncorrelated case described earlier. \\

Since the structure factor always appears as a multiplicative term with the
intensity $I^*_{\tm}(q)$, we will omit it for brevity.

\subsubsection{Evaluating $I^*_{\tm}(q)$: the Debye equation}
\label{sec:vacuum:debye}

In both the correlated and uncorrelated cases described above, the total
scattering depends on the orientationally averaged single-molecule intensity
$I^*_{\tm}(q)$. When an atomistic structure is available, it can be computed
analytically by averaging the phase factors in \eref{eq:Aq_sum} over all
molecular orientations to obtain a central equation in SAS: the
Debye equation. \\

To derive it, we start by computing the orientational average of the
exponential term in \eref{eq:Aq_sum}. Since the integrand depends only on
the relative angle between $\bm{q}$ and $\bm{r}$, the average can be
carried out either with $\bm{q}$ fixed and $\bm{r}$ averaged over orientations,
denoted $\langle\cdot\rangle_{\Omega_r}$, or vice versa, denoted
$\langle\cdot\rangle_{\Omega_q}$; see \sref{sec:modeling:orientational_avg}. 
The two are equivalent, and we write $\langle\cdot\rangle_\Omega$ throughout
when the distinction is immaterial. 
Aligning the polar axis with the fixed vector gives $\bm{q}\cdot{}\bm{r} =
qr\cos\vartheta$, and the average reduces to an integral in polar
coordinates $(\vartheta,\varphi)$:
\begin{align}\label{eq:exp_orientational_avg}
	\left\langle e^{i\bm{q}\cdot{}\bm{r}}\right\rangle_\Omega 
	&=\frac{1}{4\pi}\int^{2\pi}_0 d\varphi \int^\pi_0 \sin\vartheta\, e^{iqr\cos\vartheta}
	d\vartheta \nonumber\\
	&=\frac{1}{2}\int^{1}_{-1}e^{iqrx}dx = \frac{\sin qr}{qr} \equiv \sinc qr
\end{align}
where we have used the substitution $x = \cos\vartheta$. This term represents the
orientationally averaged phase factor between two scatterers separated by
distance $r$. It equals the zeroth-order spherical Bessel function $j_0(qr)$, a
notation often used in the modeling literature. A commonly used alternative to
this analytical average is to expand the phase into a spherical harmonics basis,
as described in \aref{sec:appendix:spherical_harmonics_expansion}. \\

Next, from the squared and
orientationally averaged amplitude of \eref{eq:Aq_sum}:
\begin{align*}
	I^*_{\tm}(q) =\langle|A^*_{\tm}(\bm{q})|^2\rangle_\Omega
	&=\sum^N_{ij}\left\langle f_i(\bm{q}) f_j(\bm{q})
	e^{i\bm{q}\cdot{}\bm{r}_{ij}}\right\rangle_\Omega
\end{align*}
For structures composed of individual atoms, the form factors depend only on the
magnitude of $\bm{q}$, so $f(\bm{q}) = f(q)$.\footnote{While we here derive it
	for atoms, the Debye equation also holds for larger constituents provided they
	are spherically symmetric and $f(\bm{q}) = f(q)$.} Using this assumption, the
average can be evaluated analytically to yield the Debye
equation~\cite{Debye1915}:
\begin{align}\label{eq:Debye}
	I^*_{\tm}(q)
	&=\sum^N_{ij}f_i(q) f_j(q)\left\langle
	e^{i\bm{q}\cdot{}\bm{r}_{ij}}\right\rangle_\Omega \nonumber \\
	&=\sum^N_{ij} f_i(q) f_j(q) \sinc qr_{ij}
\end{align}
where $r_{ij} \equiv |\bm{r}_{ij}|$ is the distance between atoms $i$ and $j$.
This equation provides an exact expression for the orientationally averaged
scattering resulting from any atomistic structure, regardless of its symmetry or
internal complexity.

%% file: solvent_scattering.tex
\section{Scattering by solutions}\label{sec:accounting_for_solvent}

\begin{figure}
	\centering
	\caption{A simplified view of a typical experimental setup. A monochromatic
		X-ray beam is collimated by apertures before scattering off the electrons in the
		sample at an angle $2\theta$, producing an azimuthally symmetric diffraction
		pattern. Figure adapted with permission
		from~\citeasnoun{BookModernXrayPhysics}.}
	\label{fig:theory:setup}
	\includesvg[width=\linewidth]{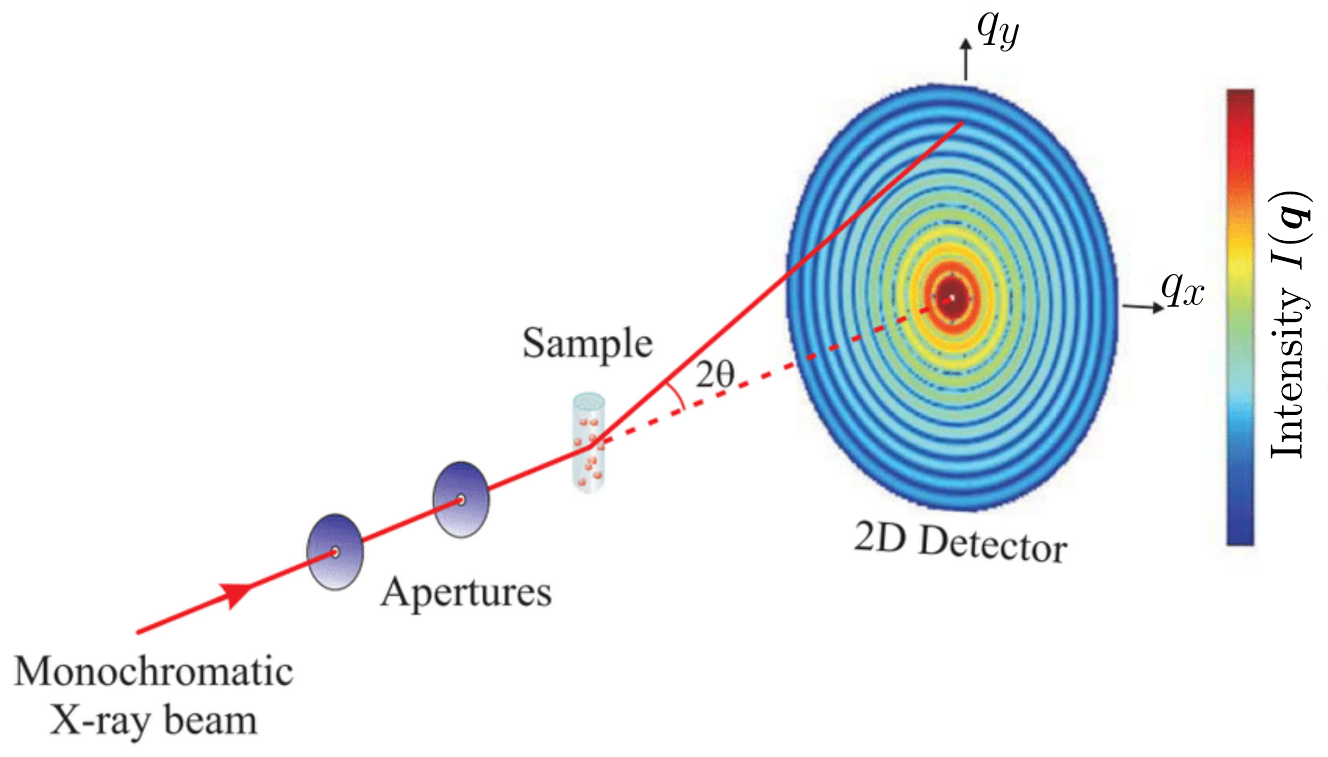}
\end{figure}

Up to this point, we assumed that the molecules are suspended in vacuum.
However, in nearly all SAS experiments (see \fref{fig:theory:setup}), there is a
solvent present which dominates the observed scattering signal. Understanding
and accounting for the solvent contribution is the focus of this section. \\

To formalize the discussion, we adopt the notation summarized in
\tref{table:notation}. The term \textit{buffer} denotes the surrounding
solution in which the molecules are embedded, while \textit{system} refers to
the complete ensemble comprising both molecules and buffer. \\

\begin{table}
	\caption{Notation used in this paper. Subscript ``${\ts}$'' denotes the entire 
		system (molecules + buffer), ``${\tb}$'' the buffer region, ``${\tm}$'' a single 
		molecule, and ``${\tM}$'' all molecules; they are capitalized in volumes for 
		clarity in nested subscripts. See \fref{fig:theory:vol_defs} for a visual overview. }
	\label{table:notation}
	\begin{tabular}{l|llll}
		Subscript & s & b & M & m \\ \hline
		Meaning & System & Buffer & All molecules & Single molecule \\
		Volume & $V_{\tVs}$ & $V_{\tVb}$ & $V_{\tVM} = V_{\tVs} - V_{\tVb}$ &
		$V_{\tVm} = V_{\tVM}/N$
	\end{tabular}
\end{table}
\begin{figure}
	\centering
	\caption{Visual overview of the notation used in this paper. 
		The system volume is denoted by $V_{\tVs}$, volume of all molecules by $V_{\tVM}$, and volume per molecule by $V_{\tVm}$. The buffer region
		is denoted as either $V_{\tVb}$ or the complement $V^\complement_{\tM}$ of $V_{\tM}$ depending on context. The exact boundary of $V_{\tVm}$ will be clarified later.}
	\label{fig:theory:vol_defs}
	\includesvg[width=\columnwidth]{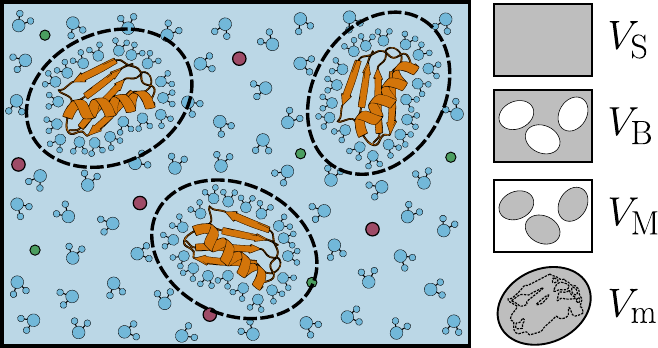}
\end{figure}

We begin with the simplest case: a solvent of uniform density, that is, 
without any internal structure. Examining this idealized system will help to understand the baseline effect of solvent on the observed scattering. In
\sref{sec:theory:inhomogeneous_solvent}, we will extend the analysis to
inhomogeneous solvents, accounting for both local density variations and time-dependent fluctuations.

\subsection{Scattering by uniform solvents}\label{sec:homogeneous_solvent}

\begin{figure*}
	\centering
	\caption{Solution scattering arises from density fluctuations. (A) A perfectly
		uniform solution produces no net scattering. (B,C) Local density variations
		(e.g.~molecules or voids) break uniformity, generating observable scattering.
		} 
	\label{fig:theory:fluctuation_scattering}
	\includesvg[width=\linewidth]{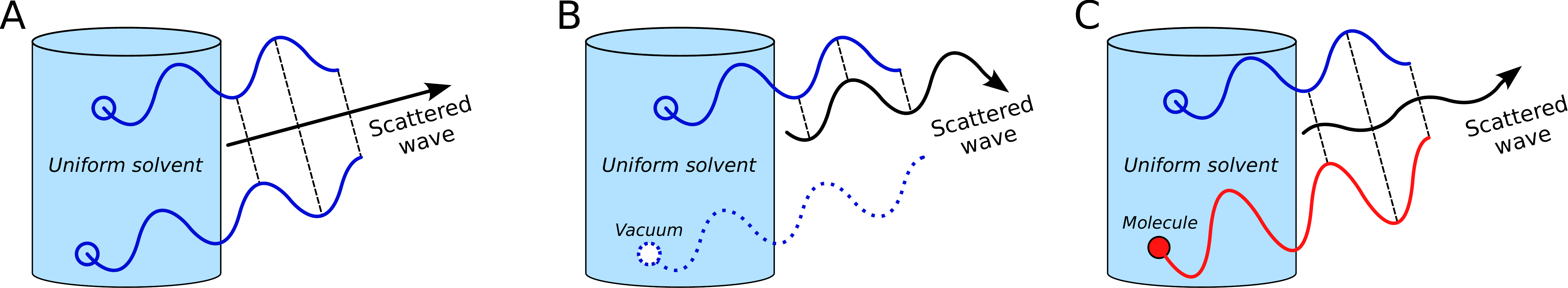}
\end{figure*}

A fundamental principle of small-angle scattering is that only spatial
variations in electron density contribute to the measured signal. To see why,
consider a system of perfectly uniform density. The scattered waves from every
point would have identical amplitudes but different phases, leading to complete
destructive interference, except at $\bm{q} = 0$
(\fref{fig:theory:fluctuation_scattering}A). Only scattering from the boundary
region near the container walls would be observed. However, when a local density
variation is introduced (\fref{fig:theory:fluctuation_scattering}B and C), the
uniformity is broken, and the scattered waves no longer cancel. Such variations
are the source of the SAS signal. \\

%\onecolumn
%\twocolumn

This intuition can be formalized by computing the scattering amplitude of a
uniform density $\rho_{\tb}$. Substituting $\rho(\bm{r}) = \rho_{\tb}$ into
\eref{eq:Aq_general} and integrating over the system volume gives:
\begin{align}\label{eq:shape_factor}
	A_{\ts}(\bm{q}) 
	= \rho_{\tb}\int_{V_{\tVs}} e^{i\bm{q}\cdot{}\bm{r}}d\bm{r}
	= \rho_{\tb}\mathcal{S}_{V_{\tVs}}(\bm{q})
\end{align}
where $\mathcal{S}_{V_{\tVs}}(\bm{q})$ is the shape integral, describing the
contribution due to the overall geometry of the scattering volume. For an
infinitely large domain, this term is proportional to the Dirac delta function: 
\begin{align}\label{eq:A_constant_is_delta}
	A_{\ts}(\bm{q}) 
	\sim \rho_{\tb}\delta(\bm{q})
\end{align}

Although this is only valid for an infinite domain, real systems are practically
large enough that $\mathcal{S}_{V_{\tVs}}(\bm{q})$ contributes only within a
narrow region near $\bm{q}=0$, which is hidden behind the beamstop; a necessary
component to protect the detector from direct beam exposure. Uniform densities
are therefore effectively invisible in SAS experiments.

\subsubsection{Scattering by a vacuum bubble: Babinet's principle}

Because constant offsets in density change only forward scattering at
$\bm{q}=0$, we may freely add or subtract uniform densities without altering the
observable scattering. This leads to an intuitive explanation for why a vacuum
bubble scatters (\fref{fig:theory:fluctuation_scattering}B). Consider a
system of uniform density $\rho_{\tb}$ except for a vacuum bubble with zero
density. By subtracting the uniform $\rho_{\tb}$, the density becomes zero in
the solvent region and $-\rho_{\tb}$ in the vacuum bubble. From this density-shifted
perspective, only the bubble contributes to the scattering.\\

To formally derive this intuition, let $V_\text{V}$ be the volume of a vacuum
bubble and $V^\complement_\text{V}$ its complement (the rest of the system). If the bubble were instead filled with solvent, the resulting amplitude would vanish at all relevant $\bm{q}$:
\begin{align*}
	\rho_{\tb}\int_{V_\text{V}}e^{i\bm{q}\cdot{}\bm{r}}d\bm{r}+\rho_{\tb}\int_{V_\text{V}^\complement}e^{i\bm{q}\cdot{}\bm{r}}d\bm{r} &=\rho_{\tb}\int_{V_{\tVs}}e^{i\bm{q}\cdot{}\bm{r}}d\bm{r}\\
	&=\cancelto{0}{\rho_{\tb}\mathcal{S}_{V_{\tVs}}(\bm{q})}
\end{align*}
since $\mathcal{S}_{V_{\tVs}}(\bm{q})$ can be neglected, as previously argued. Rearranging, we obtain the relation:
\begin{align}\label{eq:A_complementary}
	\rho_{\tb}\int_{V_\text{V}^\complement}e^{i\bm{q}\cdot{}\bm{r}}d\bm{r} 
	&= -\rho_{\tb}\int_{V_\text{V}}e^{i\bm{q}\cdot{}\bm{r}}d\bm{r}
\end{align}
This confirms our intuition when applied to the vacuum-bubble system: the scattering from the uniform solvent region is equivalent to that of the bubble alone with effective density $-\rho_{\tb}$. This
equivalence, first articulated by Jacques Babinet in the 19th century, is known
as \textit{Babinet’s principle}, which underpins the common practice of
expressing scattering in terms of excess density relative to the solvent.

\subsubsection{Deriving the excess density}
\label{sec:solv:derive_excess}

Next, we introduce molecules to the uniform solvent. When we earlier considered
the bare molecules in vacuum, a discrete density description based on individual
contributions was sufficient. Now, with the solvent occupying the space between
molecules, a continuous density description is more natural. In this framework,
the regions occupied by molecules (volume $V_{\tVM}$) exhibit a spatially
varying density $\rho_{\tM}(\bm{r})$, while the surrounding buffer is treated as
a uniform background with density $\rho_{\tb}$. The system density is thus given
as:
\begin{align}\label{eq:excess_density}
	\rho_{\ts}(\bm{r}) = 
	\begin{cases}
		\rho_{\tM}({\bm{r}}) & \text{if } \bm{r} \in V_{\tVM} \\
		\rho_{\tb} & \text{if } \bm{r} \in V^\complement_{\tVM}
	\end{cases} 
\end{align}
The volume $V_{\tVM}$ is excluded from the buffer, as this region is already
occupied by molecules. The associated scattering amplitude becomes:
\begin{align}\label{eq:A_excess_simple}
	A_{\ts}(\bm{q}) 
	&= \int_{V_{\tVs}} \rho_{\ts}(\bm{r}) e^{i\bm{q}\cdot{}\bm{r}}d\bm{r}
	\nonumber \\
	&= \int_{V_{\tVM}} \rho_{\tM}(\bm{r})e^{i\bm{q}\cdot{}\bm{r}}d\bm{r} +
	\rho_{\tb}\int_{V_{\tVM}^\complement}e^{i\bm{q}\cdot{}\bm{r}}d\bm{r} \nonumber\\
	&= \int_{V_{\tVM}} \left(\rho_{\tM}(\bm{r}) - \rho_{\tb}\right)
	e^{i\bm{q}\cdot{}\bm{r}}d\bm{r} \nonumber
	\intertext{where we have used Babinet's principle (\eref{eq:A_complementary})
		in the last equality. Defining the \textit{excess density} as
		$\Delta\rho_{\tM}(\bm{r}) \equiv \rho_{\tM}(\bm{r}) - \rho_{\tb}$, the amplitude
		simplifies to:}
	A_{\ts}(\bm{q}) &= \int_{V_{\tVM}} \Delta\rho_{\tM}(\bm{r})
	e^{i\bm{q}\cdot{}\bm{r}}d\bm{r}
\end{align}
Using \eref{eq:Iq_correlated}, the intensity becomes:
\begin{align*}
	I_{\ts}(q) = \langle|A_{\ts}(\bm{q})|^2\rangle_\Omega = N\langle|\Delta
	A^*_{\tm}(\bm{q})|^2\rangle_\Omega
\end{align*}
where $\Delta A^*_{\tm}(\bm{q})$ is the amplitude corresponding to the excess
density of the representative molecule, $\Delta\rho^*_{\tm}(\bm{r}) \equiv
\rho^*_{\tm}(\bm{r}) - \rho_{\tb}$. Here, the asterisk $*$ again indicates the
amplitude or density owing to a representative structure.
Thus, for molecules in uniform solvent, the
scattering depends only on the excess density $\Delta\rho^*_{\tm}(\bm{r})$
within the molecular volume. Note that the subtraction $\rho^*_{\tm}(\bm{r}) - \rho_{\tb}$ is a mathematical step, not a physical one: no second measurement is involved, and $A_{\ts}(\bm{q})$ still represents the total scattering from molecules and buffer combined. \\

Because of the simplicity of \eref{eq:A_excess_simple}, excess density alone may
seem to fully describe the observed scattering. This interpretation, however, holds only for a \textit{uniform} solvent: in real-world systems
with non-uniform solvents, additional buffer contributions appear (\sref{sec:theory:inhomogeneous_solvent}).

\subsubsection{Contrast variation experiments}

The excess density is historically called contrast,\footnote{The literature
generally uses \textit{contrast} with two related meanings: (i) as the average
	excess system density, determining the overall strength of the signal, and (ii)
	as a context-dependent descriptor of the relative visibility of specific
	molecular components under given solvent conditions.} and represents the portion
of the molecular density that actually contributes to the observed scattering.
As made explicit in \eref{eq:A_excess_simple}, only deviations from the buffer
density produce observable scattering, because any part of the molecule matching
the buffer blends into the background. \\

\begin{figure}
	\centering
	\caption{Illustration of contrast variation. By adjusting the solvent density,
		scattering from different molecular regions can be selectively enhanced or
		suppressed.}
	\label{fig:theory:contrast_variation}
	\includesvg[width=\columnwidth]{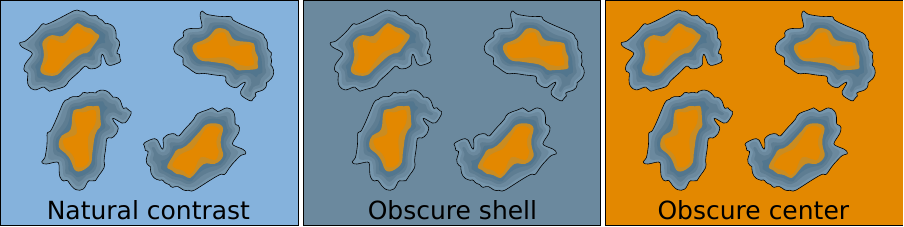}
\end{figure}

The concept of contrast underlies \textit{contrast variation} experiments, where
the buffer density is tuned such that selected molecular regions are obscured
(i.e. contrast‑matched and thus invisible to scattering); see
\fref{fig:theory:contrast_variation}. This approach enables targeted structural
analysis of complex systems, as illustrated in the following practical
implementations. \\

\begin{sansnote}
	Contrast variation is widely used with neutron scattering, where the different
	scattering lengths of hydrogen and deuterium allow control over contrast via
	isotopic substitution.\footnotemark{} The contrast can be tuned using two
	complementary approaches: adjusting the H$_2$O/D$_2$O ratio in the buffer to
	vary its density, or by varying deuteration of specific molecular components to
	enhance or suppress their relative contrasts. Targeted structural analyses of
	individual molecular subunits are routinely performed in this manner; see~\citeasnoun{Krueger2022} for a recent review.
\end{sansnote}
\footnotetext{Isotopic substitution requires caution, as it may alter
	intermolecular interactions and introduce inconsistencies (``isotope-effects'')
	between datasets.}

\begin{saxsnote}
	Contrast variation is more challenging with X-rays because the scattering
	depends on the electron density, which is harder to tune without altering
	molecular interactions. Traditionally, this has limited contrast variation to
	adding salts or sugars, though recent work employing medical contrast agents
	suggests potential for broader applications~\cite{Frank2022}.
\end{saxsnote}

\subsection{What we mean by ``molecular''}\label{sec:theory:nomenclature}

\begin{figure}
	\centering
	\caption{Illustration of the difference in nomenclature. Other literature uses
		the label ``molecular'' to denote only components from the physical molecular
		volume $V'_{\tVm}$. We instead use it to refer to an expanded volume $V_{\tVm}$,
		whose boundary is not associated with a physical interface.}
	\label{fig:theory:modeling_volume}
	\includesvg[width=.8\linewidth]{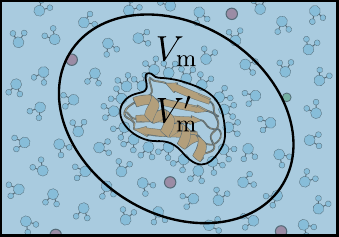}
\end{figure}

Before continuing, we clarify our use of the term \textit{molecular}.
Throughout this work, the term \textit{molecular} refers to an extended region 
$V_{\tVm}$ enclosing the physical molecule and its atoms (see \fref{fig:theory:modeling_volume}). We do not specify the precise extent of 
$V_{\tVm}$ at this stage; it is sufficient to note that $V_{\tVm}$ extends well 
beyond the immediate molecular structure, so that any structured
solvent around the molecule (hydration shell) contributes to the molecular
scattering $I_{\tm}$. This choice differs from the common convention which identifies 
the molecular volume with $V'_{\tVm}$ and therefore requires an explicit, 
molecule-dependent treatment of the hydration shell and its coupling to the bulk. \\

Our choice instead leads to a general framework that is agnostic not only to the
internal structure and chemical composition of the molecule, but also to its
shape, surface, and interaction with the solvent: all such detail is opaque to
the framework and is deferred to the modeling of $I_{\tm}$ in
\sref{sec:struc_validation}. \\

This distinction was irrelevant for the uniform solvents of
\sref{sec:homogeneous_solvent}, but becomes consequential as we extend our
attention to inhomogeneous solvent in the following.

\subsection{Scattering by non-uniform
	solvents}\label{sec:theory:inhomogeneous_solvent}

\begin{figure}
	\centering
	\caption{Decomposition of the buffer density into mean and fluctuating
		contributions.}
	\label{fig:theory:excess_density}
	\includesvg[width=.8\linewidth]{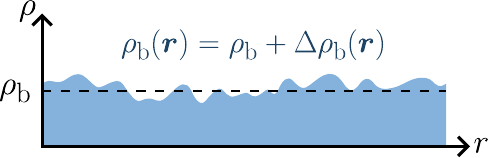}
\end{figure}

Reality is more complicated than our earlier idealization. The buffer is not a
uniform medium; it consists of discrete molecules and, optionally, additives such as salt, making the buffer density
 inhomogeneous and subject to time-dependent density fluctuations.
To account for these effects, we express the buffer density as:
\begin{align*}
	\rho_{\tb}(\bm{r},t) = \rho_{\tb} + \Delta\rho_{\tb}(\bm{r},t)
\end{align*}
where $\rho_{\tb}$ is the mean buffer density, and $\Delta\rho_{\tb}(\bm{r},t)$
is the excess buffer density, capturing spatial and temporal fluctuations,
including those from additives; see \fref{fig:theory:excess_density}. The system
density is then:
\begin{align}\label{eq:inhomogeneous_rho_system}
	\rho_{\ts}(\bm{r},t) = 
	\begin{cases}
		\rho_{\tM}({\bm{r},t}) & \text{if } \bm{r} \in V_{\tVM}(t) \\
		\rho_{\tb}({\bm{r},t}) & \text{if } \bm{r} \in V_{\tVb}(t)
	\end{cases}
\end{align}
Both densities and their associated volumes are considered time-dependent to
account for possible molecular motion. \\

\subsubsection{Evaluating the system intensity}
\label{sec:solvent:total_sample}

We start by evaluating the intensity corresponding to the given densities. The amplitude is given by:
\begin{align*}
	A_{\ts}(\bm{q},t)
	&=\int_{V_{\tVs}}\rho_{\ts}(\bm{r},t)e^{i\bm{q}\cdot{}\bm{r}}d\bm{r}\\
	&=\int_{V_{\tVM}(t)}\rho_{\tM}(\bm{r},t)e^{i\bm{q}\cdot{}\bm{r}}d\bm{r} +
	\int_{V_{\tVM}^\complement(t)}\rho_{\tb}(\bm{r},t)e^{i\bm{q}\cdot{}\bm{r}}d\bm{r}
\end{align*}
where $V_{\tVM}(t)$ and $V^\complement_{\tM}(t)$ are the (time-dependent)
volumes occupied by the molecules and buffer; see \fref{fig:theory:vol_defs}. \\

Separating the buffer contribution into mean and fluctuating parts yields:
\begin{align*}
	A_{\ts}(\bm{q},t)
	&=\int_{V_{\tVM}(t)}\rho_{\tM}(\bm{r},t)e^{i\bm{q}\cdot{}\bm{r}}d\bm{r}
	\nonumber\\
	&\quad+
	\rho_{\tb}\int_{V_{\tVM}^\complement(t)}e^{i\bm{q}\cdot{}\bm{r}}d\bm{r} +
	\int_{V_{\tVM}^\complement(t)}\Delta\rho_{\tb}(\bm{r},t)e^{i\bm{q}\cdot{}\bm{r}}d\bm{r}
\end{align*}

Applying Babinet's principle to invert the constant term and neglecting the
forward-scattering $\mathcal{S}_{V_{\tVs}}(\bm{q})$ contribution:
\begin{align}\label{eq:intermediate_inhomogeneous_amplitude}
	A_{\ts}(\bm{q},t) 
	&=\int_{V_{\tVM}(t)}\rho_{\tM}(\bm{r},t)e^{i\bm{q}\cdot{}\bm{r}}d\bm{r}
	\nonumber\\
	&\quad- \rho_{\tb}\int_{V_{\tVM}(t)}e^{i\bm{q}\cdot{}\bm{r}}d\bm{r} +
	\int_{V_{\tVM}^\complement(t)}\Delta\rho_{\tb}(\bm{r},t)e^{i\bm{q}\cdot{}\bm{r}}d\bm{r}
	\nonumber\\
	&=\int_{V_{\tVM}(t)}(\rho_{\tM}(\bm{r},t) -
	\rho_{\tb})e^{i\bm{q}\cdot{}\bm{r}}d\bm{r} \nonumber\\
	&\quad+\int_{V_{\tVb}(t)}\Delta\rho_{\tb}(\bm{r},t)e^{i\bm{q}\cdot{}\bm{r}}d\bm{r}
	\nonumber\\
	&=\Delta A_{\tM}(\bm{q},t) + \Delta A^{(V_{\tVb})}_{\tb}(\bm{q},t)
\end{align}
where we have used $V_{\tVb}(t) \equiv V^\complement_{\tM}(t)$. Here, $\Delta
A_{\tM}$ and $\Delta A^{(V_{\tVb})}_{\tb}$ are the excess molecular and buffer
amplitudes, respectively (see \fref{fig:theory:excess_density}), with the
originating volume of the latter annotated for clarity. As expected, this result
again shows that only variations in density contribute to the observable
scattering. \\

Squaring the amplitude gives the instantaneous system intensity:
\begin{align}\label{eq:solv:intermediate_Is}
	I_{\ts}(\bm{q}, t)
	&= |A_{\ts}(\bm{q},t)|^2 \nonumber\\
	&= |\Delta A_{\tM}(\bm{q},t) + \Delta
	A_{\tb}(\bm{q},t)|^2 \nonumber\\
	&= |\Delta A_{\tM}(\bm{q},t)|^2 + |\Delta A_{\tb}(\bm{q},t)|^2 + \Delta
	C^{V_{\tVM},V_{\tVb}}_{\tM,\tb}(\bm{q},t) \nonumber\\
	&= \Delta I_{\tM}(\bm{q},t) + \Delta I^{(V_{\tVb})}_{\tb}(\bm{q},t) + \Delta
	C^{V_{\tVM},V_{\tVb}}_{\tM,\tb}(\bm{q},t)
\end{align}
where the cross-term, owing to density correlations between the volumes $V_{\tVM}$ and $V_{\tVb}$, is: 
\begin{align}\label{eq:inhomogeneous_solvent_cross_term}
	C^{V_{\tVM}, V_{\tVb}}_{\tM,\tb}(\bm{q},t) &= A^{(V_{\tVM})}_{\tM}(\bm{q},t)\overline{A^{(V_{\tVb})}_{\tb}(\bm{q},t)} \nonumber\\
	&\quad+
	A^{(V_{\tVb})}_{\tb}(\bm{q},t)\overline{A^{(V_{\tVM})}_{\tM}(\bm{q},t)}
\end{align}
\eref{eq:solv:intermediate_Is} expresses the instantaneous system intensity as
the sum of three contributions: the excess molecular scattering $\Delta
I_{\tM}(\bm{q},t)$, the excess buffer scattering $\Delta
I^{(V_{\tVb})}_{\tb}(\bm{q},t)$, and the mixed term $C^{V_{\tVM},
	V_{\tVb}}_{\tM,\tb}(\bm{q},t)$. \\

Although the explicit $t$-dependence could be removed by averaging over time,
doing so would obscure the underlying physics by suggesting that long
acquisition times are required. The vector-$\bm{q}$ dependence likewise cannot
be eliminated through the usual orientational average, as the experiment probes
the system in a single fixed orientation. To make progress, we therefore examine
the three contributions in \eref{eq:solv:intermediate_Is} individually to see
how they may be simplified.

\subsubsection{Simplifying the molecular term}
\label{sec:solv:simplify_molecule}

Previously, we assumed all molecules shared a single representative
conformation, differing only in orientation. With an explicit time dependence,
this idealization no longer holds: the molecules undergo continuous thermal
motion, and the instantaneous ensemble therefore spans both orientations and
conformations. Evaluating $\Delta I_{\tM}(\bm{q},t)$ requires averaging over
this ensemble, which we denote by $\langle\cdot\rangle_N$.\footnote{In practice,
	the conformational average may be realized either by sampling many conformations
	sequentially in time, or, as is typical in SAS, by illuminating a macroscopic
	sample volume containing all conformations simultaneously.} \\

Due to the sheer number of molecules present in solution, every conformation is
present in every orientation at any given instant. The ensemble average
$\langle\cdot\rangle_N$ therefore implicitly contains an orientational average,
and the result depends only on the magnitude $q$. Furthermore, at equilibrium
the ensemble is stationary: thermal motion merely reshuffles populations among
the existing states without altering their statistics. The ensemble average
therefore also removes the explicit time dependence:
\begin{align}\label{eq:time_dependent_mol_intensity}
	\Delta I_{\tM}(\bm{q},t)
	= N\langle \Delta I^*_{\tm}(\bm{q},t) \rangle_N
	= N\Delta I^*_{\tm}(q)
\end{align}

\subsubsection{Simplifying the buffer term}
\label{sec:solv:simplify_buffer}

For the buffer, the conformational average used for the molecular term cannot be
applied directly. Unlike the molecules, the solvent does not possess a single
reference configuration: short‑range solvent–solvent correlations prevent
treating it as a set of independent scatterers. \\

Nevertheless, the buffer is an isotropic fluid at thermal equilibrium: its
statistical properties do not drift with time, and they are invariant under
rotations. The mean buffer signal is therefore independent of the exposure time
and depends only on $|\bm{q}|$. We therefore identify the observable buffer
contribution with its equilibrium mean:
\begin{align}\label{eq:inhomogeneous_solvent_intensity_average_t}
	\Delta I^{(V_{\tVb})}_{\tb}(\bm{q},t)
	\longrightarrow\Delta I^{(V_{\tVb})}_{\tb}(q).
\end{align}
A formal justification, which also clarifies what is meant by the usual
``ensemble average'' argument for a correlated bulk solvent, is given in
\aref{sec:appendix:buffer_average_t_is_orientational}.

\subsubsection{Simplifying the cross-term}
\label{sec:solv:boundary_term}

The final contribution in \eref{eq:solv:intermediate_Is} is the cross-term
$\Delta C^{V_{\tVM},V_{\tVb}}_{\tM,\tb}(\bm{q},t)$, representing scattering
arising from density correlations across the boundary of the molecular and
buffer volumes. These correlations do not vanish: the probability of finding a
solvent molecule just outside the boundary is influenced by the positions
of solvent molecules just inside the boundary. The cross-term is therefore generally nonzero and cannot be neglected, as
shown explicitly in \aref{sec:appendix:mdgamma}.\\

The cross-term contains no self-contributions: being a product of amplitudes
over disjoint volumes, all its contributions come from \textit{correlations}
between densities on the two sides of the boundary. From
\sref{sec:uncorrelated_molecules}, we know that uncorrelated phase factors
average to zero; the cross-term therefore cannot receive contributions from
regions where the densities on either side are independent. In a real solvent,
density correlations are short-ranged, so only a thin shell near the boundary
contributes: the interior of $V_{\tVm}$ plays no role. \\

\begin{figure}
	\centering
	\caption{Illustration of the volume partition. The system is divided into two
		regions separated by the solid black boundary. By making $V_{\tVm}$ large
		enough, cross-correlations across its boundary become bulk-like.}
	\label{fig:solv:cross_term_single}
	\includesvg[width=.9\columnwidth]{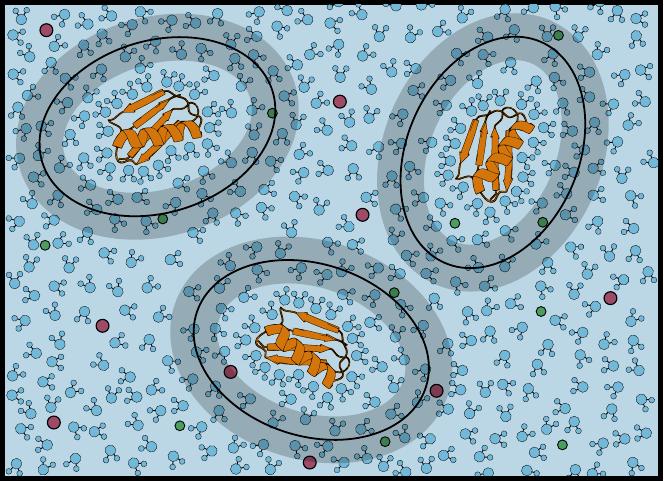}
\end{figure}

This observation implies a practical choice for $V_{\tVm}$. Provided its boundary lies
in bulk solvent, i.e. far enough from the molecule that solvent--solvent
correlations there are indistinguishable from those in pure buffer, the
cross-term reduces to a purely bulk quantity, independent of the molecule. Under
this condition, the cross-term reduces to short-range buffer--buffer
contributions:
\begin{align}\label{eq:solv:bulk_correlation}
	\Delta C^{V_{\tVM},V_{\tVb}}_{\tM,\tb}(\bm{q},t) = \Delta
	C^{V_{\tVM},V_{\tVb}}_{\tb,\tb}(\bm{q},t)
\end{align}
This construction is illustrated in \fref{fig:solv:cross_term_single}, where the
boundary is indicated and the correlated interfacial region contributing to
$\Delta C^{V_{\tVM},V_{\tVb}}_{\tb,\tb}(\bm{q},t)$ is shaded. \\

Because each such boundary region is associated with a single molecule, the
cross-term can be simplified using the same conformational average as
in \eref{eq:time_dependent_mol_intensity}:
\begin{align}\label{eq:solv:correlation_avg_struct}
	\Delta C^{V_{\tVM},V_{\tVb}}_{\tb,\tb}(\bm{q},t) 
	= N\langle\Delta C^{V_{\tVm},V_{\tVb}}_{\tb,\tb}(\bm{q},t)\rangle_N 
	= N\Delta C^{V_{\tVm},V_{\tVb}}_{\tb,\tb}(q)
\end{align}
which again removes the dependence on orientation and time. \\

Although the choice of $V_{\tVm}$ may appear arbitrary, it
introduces no ambiguity in the computed intensity. Expanding or contracting $V_{\tVm}$ merely shifts
contributions between the molecular, buffer, and cross-terms in
\eref{eq:solv:intermediate_Is}, but leaves their \textit{sum}---the total
measured intensity---unchanged. This freedom of choosing $V_{\tVm}$ 
is used advantageously in \sref{sec:bg_subtraction}, where the bulk
contribution is shown to cancel under background subtraction.

\subsubsection{Combining everything: the observable scattering}

We now have all the components required to assemble the observable 
solution scattering intensity. Starting from \eref{eq:solv:intermediate_Is} 
and applying the results of
\equadref{eq:time_dependent_mol_intensity}{eq:inhomogeneous_solvent_intensity_average_t}{eq:solv:bulk_correlation}{eq:solv:correlation_avg_struct}:
\begin{align}\label{eq:Iq_sample}
	I_{\ts}(q)
	&= \Delta I_{\tM}(\bm{q},t) + \Delta I^{(V_{\tVb})}_{\tb}(\bm{q},t) + \Delta
	C^{V_{\tVM},V_{\tVb}}_{\tM,\tb}(\bm{q},t) \nonumber\\
	&=\underbrace{N\Delta I^*_{\tm}(q)}_\text{molecular term} + \underbrace{\Delta
		I^{(V_{\tb})}_{\tb}(q) + N\Delta
		C^{V_{\tVm},V_{\tVb}}_{\tb,\tb}(q)}_\text{buffer terms}
\end{align}
\eref{eq:Iq_sample} is the central result of our derivations until now. It
expresses the observable intensity as a sum of three physically meaningful
components: (i) excess scattering from the molecules, including nearby
structured solvent, (ii) excess scattering from pure buffer, and (iii) excess
scattering from short-range buffer--buffer correlations across the interfacial
region. This decomposition makes explicit which parts of the signal originate
from the molecule of interest and which arise from its solvent environment. \\

With this analytical framework established, the remainder of this paper examines
how it can be applied to interpret and model experimentally measured scattering
intensities.

%% file: background_subtraction.tex
\section{Background subtraction}
\label{sec:bg_subtraction}

In a typical SAS measurement, the recorded intensity $I_{\ts}(q)$ is
overwhelmingly dominated by buffer contributions, obscuring the structural
information of interest. The standard remedy is to record a second measurement
of buffer alone and subtract it from the first. In this section, we examine how
this subtraction is formalized and what assumptions it requires.

\subsection{The necessity of background subtraction}
\label{sec:theory:bg_subtraction}

\begin{figure*}
	\centering
	\caption{Visualization of the two background subtraction schemes. The left side
		shows how the measured system intensities may be interpreted in terms of excess
		densities, obtained by rescaling the measured densities using Babinet’s
		principle (\sref{sec:solv:derive_excess}). The right side illustrates two
		common conventions for background subtraction using the intensities 
		from Systems A and B, used to isolate the scattering signal from the molecules.
		This schematic form understates the substantial theoretical
		derivations underlying the relations.}
	\label{fig:solv:bg_subtraction}
	
	\includesvg[width=.8\textwidth]{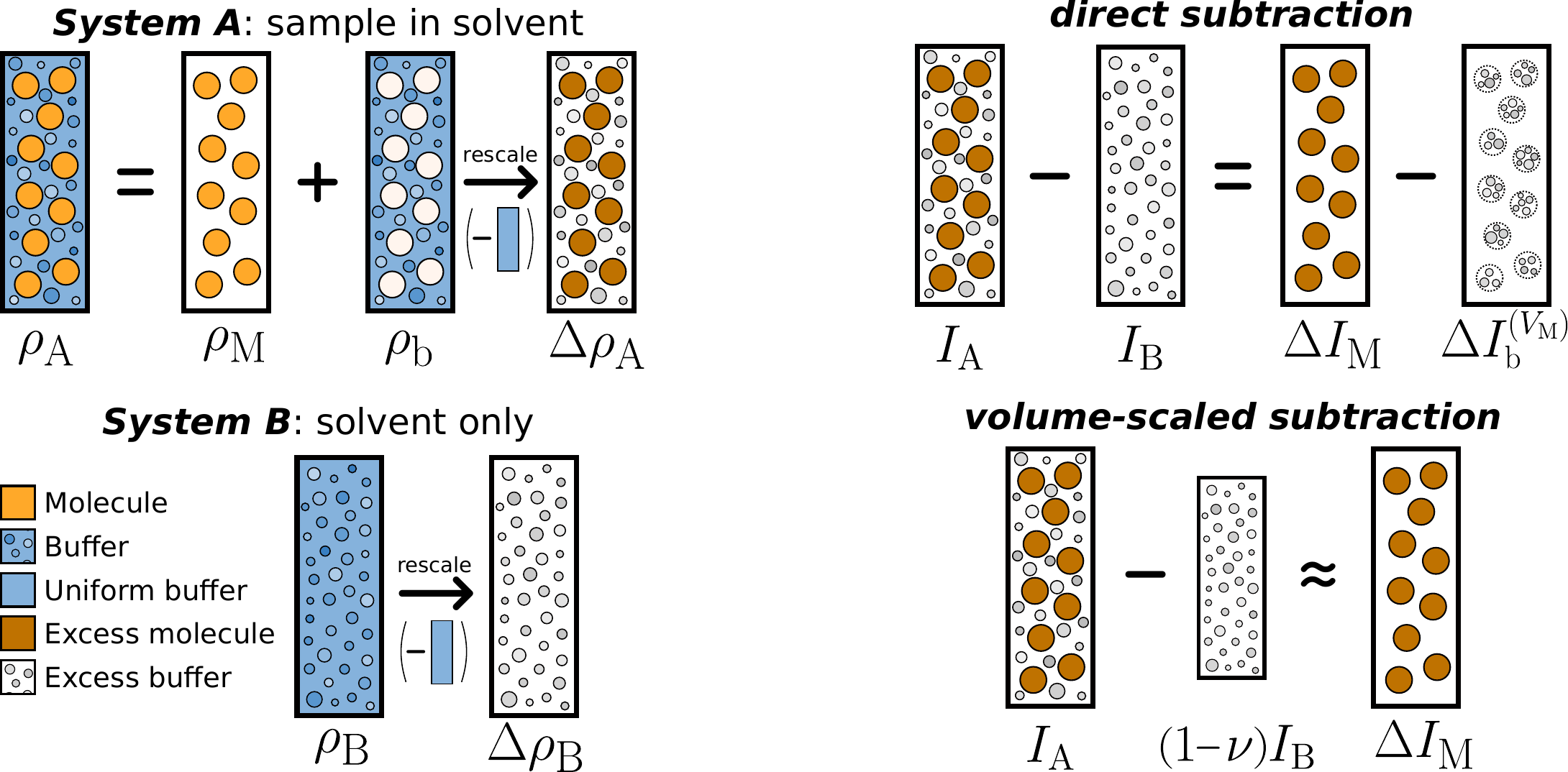}
\end{figure*}

The objective of background subtraction is to remove the excess buffer terms
from the observable intensity (\eref{eq:Iq_sample}). To achieve this, we
consider two systems measured under similar experimental conditions
(\fref{fig:solv:bg_subtraction}):
\begin{align*}
	\textit{System A}\text{: Sample in buffer.}\qquad \textit{System B}\text{:
		Buffer only.}
\end{align*}
The corresponding measured intensities are:
\begin{align}\label{eq:solv:I_exp_defs}
	I^\text{(exp)}_\text{A}(q) \equiv I_{\ts}(q) \qquad I^\text{(exp)}_\text{B}(q)
	\equiv I^{(V_{\tVs})}_{\tb}(q) = \Delta I^{(V_{\tVs})}_{\tb}(q)
\end{align}
where the uniform background density $\rho_{\tb}$ contributes only to the
(unobserved) forward scattering and is therefore omitted (see
\sref{sec:homogeneous_solvent}). Using these two measured intensities, two
subtraction schemes are commonly employed: \textit{direct subtraction} and
\textit{volume-corrected subtraction}, differing only in their treatment of the
cross-term.

%\onecolumn
%\twocolumn

\subsubsection{Scheme 1: Direct subtraction}

\begin{figure}
	\centering
	\caption{Illustration of cross-term cancellation. Partitioning system B (right)
		into the same regions as system A (left) ensures that cross-correlations across
		identical boundaries cancel upon subtraction. Because this partitioning is
		arbitrary, the final observable cannot depend on its shape or size.}
	\label{fig:solv:cross_terms}
	\includesvg[width=\columnwidth]{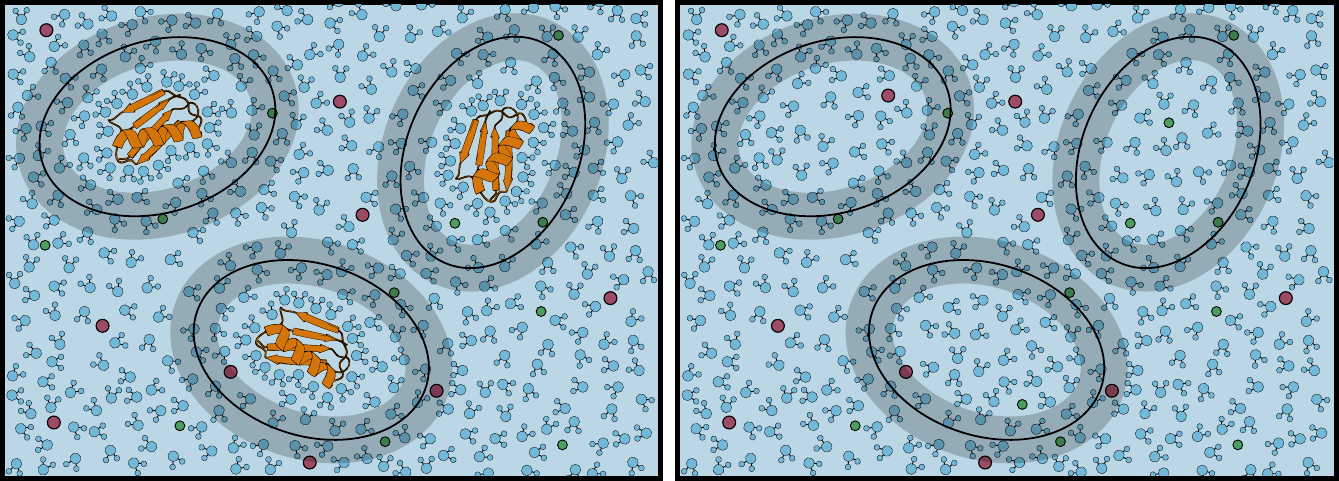}
\end{figure}

Inspired by the approach of~\citeasnoun{Park2009}, we apply to system B the same
volume partitioning used for system A. Because system B contains no molecule,
the molecular partition $V_{\tVm}$ is filled with bulk buffer, and
\eref{eq:Iq_sample} reduces to:
\begin{align}\label{eq:solv:bg_subtract_intermediate1}
	I^\text{(exp)}_\text{B}(q) = N\Delta I^{(V_{\tVm})}_{\tb}(q) + \Delta
	I^{(V_{\tVb})}_{\tb}(q) + N\Delta C^{V_{\tVm},V_{\tVb}}_{\tb,\tb}(q)
\end{align}
Because both systems use the same boundary, both their bulk and cross-terms are
identical; see Fig.~\ref{fig:solv:cross_terms}. Upon \textit{direct}
subtraction, they therefore cancel:
\begin{align}\label{eq:solv:bg_subtract_correct}
	I^\text{(exp)}_\text{A}&(q) - I^\text{(exp)}_\text{B}(q) \nonumber\\
	&= \left\{N\Delta I^*_{\tm}(q) + \Delta I^{(V_{\tVb})}_{\tb}(q) + N\Delta
	C^{V_{\tVm},V_{\tVb}}_{\tb,\tb}(q) \right\} \nonumber\\
	&\quad- \left\{N\Delta I^{(V_{\tVm})}_{\tb}(q) + \Delta I^{(V_{\tVb})}_{\tb}(q)
	+ N\Delta C^{V_{\tVm},V_{\tVb}}_{\tb,\tb}(q) \right\} \nonumber\\
	&= N\big[\Delta I^*_{\tm}(q) - \Delta I^{(V_{\tVm})}_{\tb}(q)\big]
\end{align}
Though this subtraction scheme does not fully isolate the molecular scattering,
the remaining buffer contribution has been reduced to a volume $V_{\tVm}$ of
bulk buffer. This is a significant reduction: the \textit{residual} contribution
$\Delta I^{(V_{\tVm})}_{\tb}(q)$ is purely a bulk quantity, independent of the
molecule, and can in principle be computed once and for all---for instance, from
MD simulations of pure solvent (\sref{sec:md}). Given such a computation, the
residual can be subtracted, and the direct subtraction scheme recovers the
molecular intensity exactly.

\subsubsection{Scheme 2: Volume-corrected subtraction}

To avoid explicit evaluation of the residual buffer term $\Delta
I^{(V_{\tVm})}_{\tb}(q)$ in
\eref{eq:solv:bg_subtract_correct}, a practical approximation neglects the
cross-term in \eref{eq:solv:bg_subtract_intermediate1}; this is justified for
molecules with a small surface-to-volume ratio (\aref{sec:appendix:mdgamma}).
The residual buffer contribution then reduces to:
\begin{align}\label{eq:bg:intermediate_scheme2}
	N\Delta I^{(V_{\tVm})}_{\tb}(q) 
	\approx \Delta I^{(V_{\tVs})}_{\tb}(q) - \Delta I^{(V_{\tVb})}_{\tb}(q)
\end{align}
Since $V_{\tVM} = V_{\tVs} - V_{\tVb}$, this approximation makes buffer
scattering proportional to buffer volume. Introducing the excess buffer
intensity per unit volume $\Delta\hat{I}_{\tb}(q) \equiv \Delta
I^{(V_{\tVs})}_{\tb}(q) / V_{\tVs}$, so that $\Delta I^{(V)}_{\tb}(q) =
V\Delta\hat{I}_{\tb}(q)$ for any buffer volume $V$, we can re-express the
residual buffer term in \eref{eq:solv:bg_subtract_correct} as:
\begin{align*}
	N\Delta I^{(V_{\tVm})}_{\tb}(q) = NV_{\tVm}\Delta\hat{I}_{\tb}(q) 
	= \nu V_{\tVs}\Delta\hat{I}_{\tb}(q) = \nu I^\text{(exp)}_\text{B}(q)
\end{align*}
where $\nu \equiv NV_{\tVm}/V_{\tVs}$ is the molecular volume
fraction.\footnote{Beware that this is the volume fraction of the
	\textit{modeling} volume $V_{\tVm}$, which is substantially larger than the
	\textit{physical} molecular volume $V_{\tVm}'$.} Substituting back into
\eref{eq:solv:bg_subtract_correct} and rearranging yields the
\textit{volume-corrected} subtraction:
\begin{align*}
	I^\text{(exp)}_\text{A}(q) - (1-\nu)I^\text{(exp)}_\text{B}(q) \approx N\Delta
	I^*_{\tm}(q)	
\end{align*}
This approximation is attractive because it removes the need to model bulk
solvent: after volume-scaled subtraction, only the molecular contribution
remains. However, as it relies on neglecting boundary cross-terms, the
associated error scales with interfacial area and becomes significant for
smaller molecules with their larger surface-to-volume ratio or at wider angles 
where buffer contributions dominate (\aref{sec:appendix:mdgamma}).

\subsection{Practical implications for SAS analysis}

To recap, two background subtraction schemes are used in practice:
\begin{align*}
	I^\text{(exp)}_\text{A}(q) - I^\text{(exp)}_\text{B}(q) &\qquad \text{(direct)}
	\\
	I^\text{(exp)}_\text{A}(q) - (1-\nu)I^\text{(exp)}_\text{B}(q) &\qquad
	\text{(volume-corrected)}
\end{align*}
Although the two schemes appear similar, they make fundamentally different
assumptions about how the buffer contributes, and this distinction determines
when each scheme is appropriate.\\

\textit{When the residual buffer term $\Delta I^{(V_{\tVm})}_{\tb}(q)$ can be evaluated}---e.g.~through
molecular dynamics simulations (\sref{sec:md})---the direct subtraction scheme
is exact; any remaining error depends on the residual computation, not on the subtraction itself.\\

\textit{When the residual buffer term cannot be computed}, the volume‑corrected scheme
provides a practical alternative; its approximation error arises from
neglecting boundary cross‑terms and therefore scales with interfacial area, making it
reliable for larger solutes but inaccurate for small particles and
at wider scattering angles (\aref{sec:appendix:mdgamma}).\\

These derivations demonstrate that background subtraction is not merely a step in the experimental pipeline, but an integral component of the theoretical model. Thus, SAS analysis programs must assume a convention. MD‑oriented programs insvolving explicit, all-atom solvent models typically assume direct subtraction. In contrast, more conventional programs do not model the residual buffer term $\Delta I^{(V_{\tVm})}_{\tb}(q)$ and are, therefore, only consistent with the volume-corrected scheme. Using a different scheme than the one assumed by the analysis software introduces systematic errors that are difficult to diagnose. The recommended strategy is therefore to let the analysis software perform the subtraction, ensuring consistency with its internal modeling assumptions.

\subsubsection{Accounting for instrumental scattering}

Before closing this chapter, we briefly address contributions from the system
container and instrument, which we have so far ignored. Under direct
subtraction, these contributions cancel provided both measurements share
identical geometry and normalization. In contrast, they typically do not cancel
under the volume-corrected scheme due to the $(1-\nu)$ scaling factor, requiring
explicit correction procedures~\cite{Pauw2017}. Other effects, such as beam
smearing, resolution limitations, and finite coherence lengths, are beyond the
scope of this work; for a detailed discussion, see~\citeasnoun{Pedersen1990} and~\citeasnoun{Shinohara2015}.

%% file: modelling.tex
\section{Modeling without atomistic structures}\label{sec:modeling}

With the theoretical foundations in place, we now turn to practical
applications. This section focuses on strategies for interpreting scattering
data with little to no prior structural knowledge, whereas
\sref{sec:struc_validation} addresses validation and refinement when atomistic
structures are available. Readers interested primarily in structure-based
approaches may refer directly to \sref{sec:struc_validation}. \\

In many experimental scenarios, atomistic models are either unknown or
unreliable. How, then, can we extract meaningful structural insights directly
from an observed scattering profile? To address this question, we explore a
spectrum of strategies: model-independent methods that reveal global shape
information, analytical scattering models for idealized geometries, and
numerical approaches for complex or irregular structures. For each, we examine
the underlying principles, highlight their strengths and limitations, and
outline practical considerations for application. \\

Throughout this chapter, we base our analysis on the representative molecular
density $\Delta\rho^*_{\tm}(\bm{r})$ and rely on
\edref{eq:Iq_uncorrelated}{eq:Iq_correlated} to relate these results to the
total system intensity.

\subsection{Model-independent analysis}\label{sec:modeling:model_independent}

We begin with examining what model-independent information may be extracted
directly from the scattering profile, without invoking any structural
assumptions.

\subsubsection{Guinier approximation}

André Guinier~\cite{Guinier1939,Guinier1956} demonstrated that the low-$q$ limit
of the scattering intensity contains information on overall molecular shape and
size. To obtain this result, the Taylor expansion of the $\sinc$ term in the Debye
equation (\eref{eq:Debye}) is used for small $q$:
\begin{align*}
	\sinc x &= 1 - \frac{x^2}{6} + O\left(x^4\right),
\end{align*}
where terms of order $O(x^4)$ or higher are neglected. For brevity, we
suppress the $q$-dependence of all form factors. \\

Expanding the Debye equation:
\begin{align}\label{eq:modeling:intermediate:guinier_start}
	I^*_{\tm}(q)
	&= \sum_{ij}f_if_j \sinc qr_{ij} \nonumber\\
	&\approx \sum_{ij}f_if_j \left( 1 - \frac{(qr_{ij})^2}{6} + O\left(q^4\right)
	\right)
\end{align}

The pairwise distance is:
\begin{align}\label{eq:modeling:intermediate:r_ij}
	r_{ij}^2 = |\bm{r}_i - \bm{r}_j|^2 = r_i^2 + r_j^2 - 2\bm{r}_i\cdot{}\bm{r}_j
\end{align}

We place the origin of our coordinate system at the electron-weighted center of
mass, such that:
\begin{align*}
	\sum_i f_i\, \bm{r}_i = \bm{0} 
\end{align*}

Substituting \eref{eq:modeling:intermediate:r_ij} into
\eref{eq:modeling:intermediate:guinier_start}:
\begin{align*}
	I^*_{\tm}(q)
	&=\sum_{ij}f_if_j - \frac{q^2}{6}\sum_{ij}f_if_j \left(r_i^2 + r_j^2 -
	2\bm{r}_i\cdot{}\bm{r}_j\right)
\end{align*}
The sums involving $r^2_i$ and $r^2_j$ are identical upon relabeling
the indices, and may be combined:
\begin{align*}
	I^*_{\tm}(q) &=\Big(\sum_{i}f_i\Big)^2 - \frac{q^2}{3}\sum_{ij}f_if_j r_i^2 \\
	&\quad+\frac{q^2}{3}\sum_i \Big[f_i\bm{r}_i \cdot{} \Big(\underbrace{\sum_j
		f_j\bm{r}_j}_{\bm{0}}\Big)\Big]
\end{align*}
The third term vanished by our choice of origin, leaving only:
\begin{align*}
	I^*_{\tm}(q) &=\left(\sum_{i}f_i\right)^2 - \frac{q^2}{3}\sum_if_ir_i^2 \sum_j
	f_j \\
	&=\left(\sum_{i}f_i\right)^2 \left\{ 1 -
	\frac{q^2}{3}\frac{\sum_if_ir_i^2}{\sum_{i}f_i} \right\} \\
	&\approx I^*_{\tm}(0) \exp\left( -\frac{q^2R_g^2}{3}\right)
\end{align*}
where the exponential form follows by re-exponentiating to the same order in
$q^2$, and $R^2_g = \sum_if_ir_i^2\,/\,\sum_{i}f_i$ is the electron-weighted
radius of gyration.\footnote{Note that the sum runs over \textit{all} atoms,
	including the hydration shell, in the modeling volume $V_{\tVm}$; the extracted
	$R_g$ is therefore the hydrated radius of gyration. This applies to all results
	derived in this section.} This result is consistent with our discussion in
\aref{sec:appendix:spherical_harmonics_expansion}: at small $q$, scattering is
dominated by lower-order spherical harmonics, capturing only overall shape and
size. 

\subsubsection{Porod asymptotic}

At the opposite end of the angular range, the high-$q$ regime provides insight
into fine structural details. Porod showed that, for particles with sharp
interfaces, the scattering intensity decays with a $q^{-4}$ power law reflecting
interfacial surface area. The full derivation requires asymptotic analysis of
Fourier transforms across discontinuities, which lies outside the framework
developed here; we refer the reader to~\citeasnoun{Ciccariello2002} for a
comprehensive treatment and state only the result. \\

Assuming a well-defined boundary between phases and taking the limit
$q\rightarrow\infty$, Porod obtained~\cite{Porod1951,Porod1953}:
\begin{align*}
	\lim\limits_{q\rightarrow\infty}I(q) =
	\frac{2\pi\Delta\rho_{\tm}^2}{q^4}S_\text{int}
\end{align*}
where $S_\text{int}$ is the interfacial surface area between the phases. This
result applies universally whenever a distinct interface exists, regardless of
molecular concentration. \\

Deviations from the expected $q^{-4}$ scaling can reveal additional structural
features such as surface roughness, fractal geometry, or a gradual phase
transition rather than sharp boundaries~\cite{BookSvergun2013}.

\subsubsection{Pair-distance distribution}
\label{sec:pair_distance_distribution}

The pair-distance distribution function provides some of the most detailed
model-independent real-space information obtainable from SAS. It describes the
probability-weighted distribution of distances between all pairs of points
within the molecule, and establishes a direct link between the measured
intensity and the spatial structure. \\

Starting from the definition of excess molecular intensity:
\begin{align*}
	&\Delta I^*_{\tm}(q) =\langle|\Delta A^*_{\tm}(\bm{q})|^2\rangle_\Omega \\
	&\quad=\left\langle\int_{V_{\tVm}}\Delta\rho^*_{\tm}(\bm{r}_1)e^{i\bm{q}\cdot{}\bm{r}_1}d\bm{r}_1
	\int_{V_{\tVm}}\Delta\rho^*_{\tm}(\bm{r}_2)e^{-i\bm{q}\cdot{}\bm{r}_2}d\bm{r}_2\right\rangle_\Omega \\
	&\quad=\left\langle\iint_{V_{\tVm}}\Delta\rho^*_{\tm}(\bm{r}_1)\Delta\rho^*_{\tm}(\bm{r}_2)
	e^{i\bm{q}\cdot{}\left(\bm{r}_1-\bm{r}_2\right)}d\bm{r}_1d\bm{r}_2\right\rangle_\Omega
\end{align*}
With the change of variables $\bm{r} = \bm{r}_1 - \bm{r}_2$
and $\bm{r}'= \bm{r}_2$:
\begin{align*}
	\Delta I^*_{\tm}(q) &=\left\langle\iint_{V_{\tVm}}\Delta\rho^*_{\tm}(\bm{r} +
	\bm{r}')\Delta\rho^*_{\tm}(\bm{r}')
	e^{i\bm{q}\cdot{}\bm{r}}d\bm{r}d\bm{r}'\right\rangle_\Omega \\
	&=\left\langle\int_{V_{\tVm}}e^{i\bm{q}\cdot{}\bm{r}}
	\int_{V_{\tVm}}\Delta\rho^*_{\tm}(\bm{r} +
	\bm{r}')\Delta\rho^*_{\tm}(\bm{r}')d\bm{r}'d\bm{r}\right\rangle_\Omega
\end{align*}
The inner integral is recognized as the \textit{correlation} function of the excess density:
\begin{align*}
	\Delta\gamma^*_{\tm}(\bm{r}) \equiv \int_{V_{\tVm}}\Delta\rho^*_{\tm}(\bm{r} +
	\bm{r}')\Delta\rho^*_{\tm}(\bm{r}')d\bm{r}'
\end{align*}
This correlation function, originally introduced by~\citeasnoun{Debye1949},
describes how density variations at $\bm{r}'$ correlate with those at $\bm{r}$.
Substituting this into the integral:
\begin{align}\label{eq:modeling:pair_distance_distribution}
	\Delta I^*_{\tm}(q) 
	&=\left\langle\int_{V_{\tVm}} \Delta\gamma^*_{\tm}(\bm{r})
	e^{i\bm{q}\cdot{}\bm{r}} d\bm{r}\right\rangle_\Omega \nonumber
	\intertext{There is no general analytic solution for the orientational average
		because the correlation function and phase factor are coupled. To proceed,
		recall that $\Delta\rho^*_{\tm}(\bm{r})$ is itself a representative density: it
		stands in for an ensemble of molecular orientations sampled in solution
		(\sref{sec:uncorrelated_molecules}). The correlation function
		$\Delta\gamma^*_{\tm}(\bm{r})$ inherits this average and, by isotropy of the
		solution, is rotationally symmetric, $\Delta\gamma^*_{\tm}(\bm{r}) =
		\Delta\gamma^*_{\tm}(r)$. The orientational average is:}
	\Delta I^*_{\tm}(q)
	&=\int_{V_{\tVm}}  \Delta\gamma^*_{\tm}(r) \langle
	e^{i\bm{q}\cdot{}\bm{r}}\rangle_\Omega d\bm{r} \nonumber\\
	&=\int_0^\infty 4\pi r^2 \Delta\gamma^*_{\tm}(r) \sinc qr\, dr \nonumber\\
	&=\int_0^\infty \Delta p^*_{\tm}(r) \sinc qr\, dr
\end{align}
where we have defined the pair distance distribution function $\Delta
p^*_{\tm}(r) \equiv 4\pi r^2 \Delta\gamma^*_{\tm}(r)$. \\

\eref{eq:modeling:pair_distance_distribution} shows that the intensity is given
as the Fourier sine transform of $\Delta p^*_{\tm}(r)$. In principle, we can therefore
recover $\Delta p^*_{\tm}$ from the intensity through the inverse transform:
\begin{align*}
	q\Delta I^*_{\tm}(q) &= \int_0^\infty \frac{\Delta p^*_{\tm}(r)}{r} \sin qr\,
	dr \\
	\frac{\Delta p^*_{\tm}(r)}{r} &= \frac{2}{\pi}\int_0^\infty q\Delta
	I^*_{\tm}(q) \sin qr\, dq
\end{align*}
In practice, this inversion is ill-posed due to the limited measurable $q$-range
and experimental noise. Stabilized indirect Fourier transform methods are
therefore employed practically; see~\citeasnoun{Glatter1977}
or~\citeasnoun{BookSvergun2013} for details. \\

\begin{figure}
	\centering
	\caption{The pair distance distribution function for different shapes,
		normalized along their longest dimension. Reprinted with permission
		from~\citeasnoun{Haahr2018}.}
	\label{fig:theory:pddf}
	
	\includegraphics[width=\linewidth]{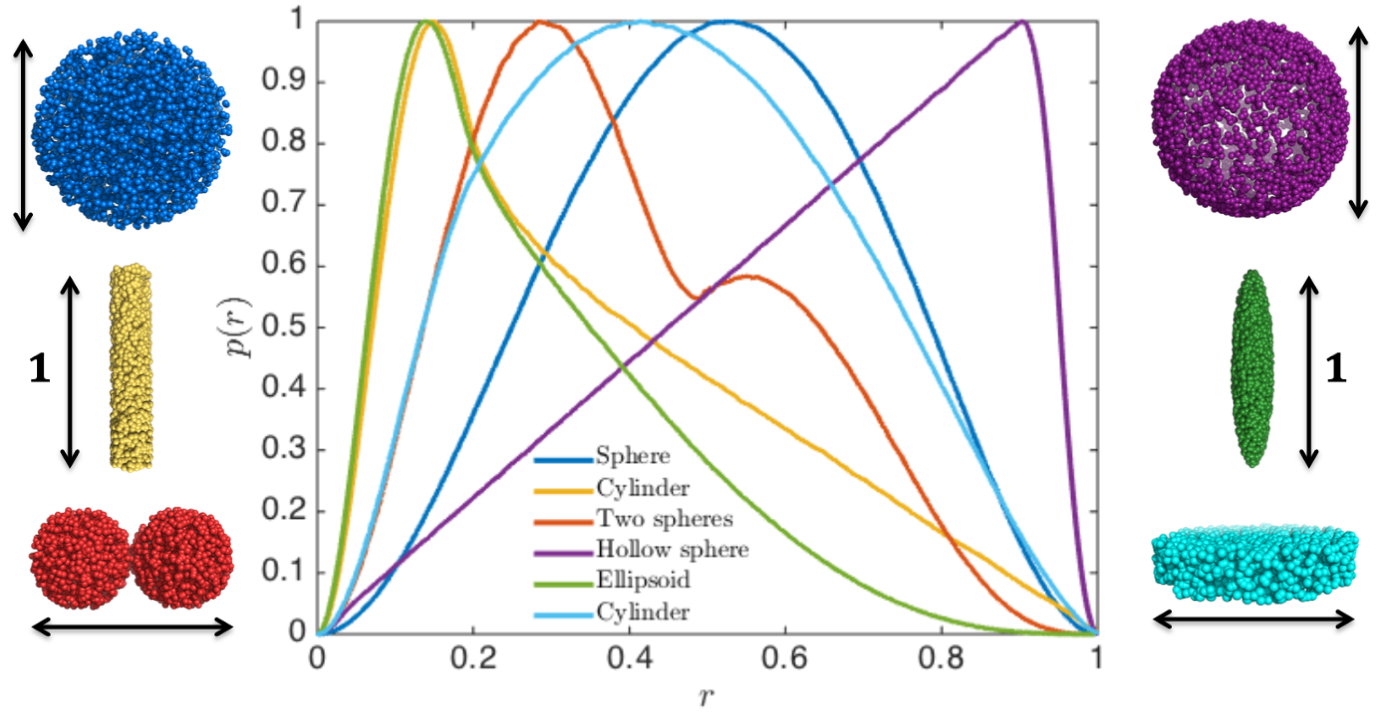}
\end{figure}

Once $\Delta p^*_{\tm}(r)$ is obtained, the maximum molecular dimension
$D_\text{max}$ is estimated as the distance where $\Delta p^*_{\tm}(r)$
vanishes. This is illustrated by $\Delta p^*_{\tm}(r)$ curves for several
shapes in \fref{fig:theory:pddf}, from which not only the
maximum extent may be estimated, but also other characteristic distances of the
shape. As an example, the thicknesses of the two elongated shapes may be
estimated as their most frequent distance (\ref{fig:theory:pddf}, yellow and green). 

\subsection{Analytical scattering models}\label{sec:modeling:model_dependent}

Many systems of interest have well-characterized overall geometry---globular
proteins, spherical micelles, rod-like assemblies---even when their internal
structure is not. In such cases, the scattering can be derived analytically by
assuming a simple geometric form for $\Delta\rho^*_{\tm}(\bm{r})$ and evaluating
the amplitude integral in closed form. The resulting expressions provide
low-resolution scattering models that can be fitted to experimental data,
allowing rapid shape characterization. We derive intensities for a few geometric cases below; for a comprehensive catalogue, see Pedersen,~J.S. in~\citeasnoun{BookLindner2024}. \\

Analytical shape models predict the scattering signal according to the volume-corrected background subtraction scheme, $I^\text{(exp)}_\text{A}(q) - (1-\nu)I^\text{(exp)}_\text{B}(q) \approx N\Delta I^*_{\tm}(q)$, which requires only the single-molecule term $\Delta I^*_{\tm}(q)$ to be modeled. The approximation holds at small $q$, where the neglected boundary cross-term and hydration-shell contributions are small; fitting at wider $q$, or to directly-subtracted data, mixes the model intensity with unmodeled buffer terms and yields biased structural parameters.

\subsubsection{Homogeneous spheres}
\label{sec:modeling:homogeneous:sphere}

The molecular density of a homogeneous sphere is defined as:
\begin{align*}
	\Delta\rho^*_{\tm}(\bm{r}) = 
	\begin{cases}
		\Delta\rho_\text{c} & \text{if } r < R \\
		0 & \text{otherwise}
	\end{cases} 
\end{align*}
The associated scattering amplitude is:
\begin{align}\label{eq:model:homogeneous_sphere}
	\Delta A^*_{\tm}(\bm{q}) &=
	\int_V\Delta\rho^*_{\tm}(\bm{r})e^{i\bm{q}\cdot\bm{r}}d\bm{r} \nonumber\\
	&=\Delta\rho_\text{c}\int^R_0 dr \int^\pi_0 d\phi \int^{2\pi}_0
	e^{iqr\cos\theta}r^2\sin\theta\, d\theta \nonumber\\
	&=\Delta\rho_\text{c} (2\pi) \int^R_0 \frac{2r\sin qr}{q}dr \nonumber\\
	&=\Delta\rho_\text{c}\frac{4\pi}{3}\left[3\frac{\sin qR - qR\cos
		qR}{q^3}\right] \nonumber\\
	&=\Delta\rho_\text{c} V_\text{sphere}\left[3\frac{j_1(qR)}{qR}\right]
\end{align}
where $j_1(x) = (\sin x - x\cos x)/x^2$ is the spherical Bessel function. The
intensity is obtained by squaring this amplitude; no orientational averaging is
needed due to spherical symmetry.

\subsubsection{Fitting analytical models to experimental data} 

Analytical shape models are compared with experimental data by fitting their
free parameters---e.g.~$\Delta\rho_\text{c}$ and $R$ for the homogeneous
sphere. In practice, tools such as SasView~\cite{SasView6.0} introduce
additional adjustable parameters to account for experimental factors not
captured by the shape model alone: the number of molecules $N$ is absorbed into
an overall scaling factor $a$, an additive constant $b$ corrects for any uniform
background, and the bulk solvent density $\rho_{\tb}$ may be used as a free parameter in addition to
$\rho_\text{c}$. The resulting model intensity for the homogeneous sphere then
takes the form:
\begin{align*}
	\Delta I^\text{(model)}_{\tm}(q) &= a\Delta I^*_{\tm}(q) + b \\
	&= a(\rho_\text{c} - \rho_{\tb})^2 \left(\frac{4\pi
		R^3}{3}\right)^2\left[3\frac{j_1(qR)}{qR}\right]^2 + b
\end{align*}
To reduce the risk of overfitting, free parameters can be constrained with independent information; this is especially important for more complex models with many such parameters.\\

A similar strategy applies to interparticle correlations: when
intermolecular interactions are significant, the structure factor $S(q)$ may be
parameterized in an analogous way and fitted simultaneously with the shape model to
account for concentration-dependent scattering; see~\citeasnoun{Larsen2020}.

\subsubsection{Core--shell structures}

The core--shell model is widely used to describe micelles, vesicles, and
nanoparticles with distinct internal and external regions of homogeneous
density. The molecular excess density is:
\begin{align*}
	\Delta\rho^*_{\tm}(\bm{r}) = 
	\begin{cases}
		\Delta\rho_\text{c} & \text{if } r < R_\text{C} \\
		\Delta\rho_\text{s} & \text{if } R_\text{C} < r < R_\text{S} \\
		0 & \text{otherwise}
	\end{cases} 
\end{align*}

Denoting by $V_\text{C}$ and $V_\text{S}$ the volumes enclosed by $R_\text{C}$ and $R_\text{S}$, the corresponding scattering amplitude is:
\begin{align}\label{eq:model:core_shell}
	&\Delta A^*_{\tm}(\bm{q}) 
	= \int_{V_\text{C}}\Delta\rho_\text{c}e^{i\bm{q}\cdot\bm{r}}d\bm{r} + \int_{V_\text{S} - V_\text{C}}\Delta\rho_\text{s}e^{i\bm{q}\cdot\bm{r}}d\bm{r} \nonumber \\
	&\quad= \int_{V_\text{C}}\Delta\rho_\text{c}e^{i\bm{q}\cdot\bm{r}}d\bm{r} + \int_{V_\text{S}}\Delta\rho_\text{s}e^{i\bm{q}\cdot\bm{r}}d\bm{r} - \int_{V_\text{C}}\Delta\rho_\text{s}e^{i\bm{q}\cdot\bm{r}}d\bm{r} \nonumber \\
	&\quad= \int_{V_\text{C}} (\Delta\rho_\text{c}-\Delta\rho_\text{s}) e^{i\bm{q}\cdot\bm{r}} d\bm{r} +\int_{V_\text{S}} \Delta\rho_\text{s} e^{i\bm{q}\cdot\bm{r}}d\bm{r} \nonumber \\
	&\quad= 3\left[\left(\Delta\rho_\text{c} -
	\Delta\rho_\text{s}\right)V_\text{C}\frac{j_1(qR_\text{C})}{qR_\text{C}} +
	\Delta\rho_\text{s}V_\text{S}\frac{j_1(qR_\text{S})}{qR_\text{S}}\right]
\end{align}
where we have used the result of \eref{sec:modeling:homogeneous:sphere}, $V_\text{C} = 4\pi R_\text{C}^3 / 3$, and $V_\text{S} = 4\pi R_\text{S}^3 /
3$. This expression reduces to the amplitude of a homogeneous sphere for
$\Delta\rho_\text{s} = 0$. \\

\begin{figure}
	\centering
	\caption{Illustration of how the total amplitude for a core–shell molecule is constructed from summing component amplitudes, leveraging the linearity of
		the Fourier transform.}
	\label{fig:model:core_shell_model}
	\includesvg[width=\linewidth]{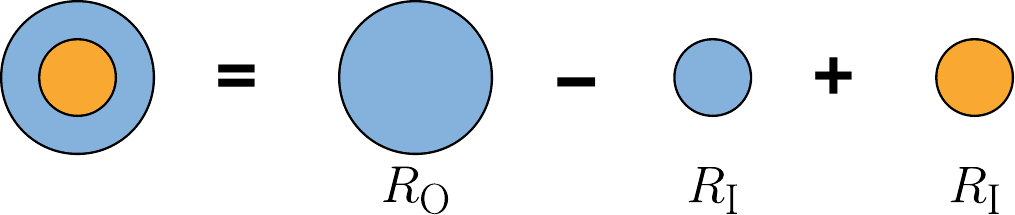}
\end{figure}

\eref{eq:model:core_shell} may also be interpreted as the sum of three spherical
components---a sphere of volume $V_\text{S}$ with density $\Delta\rho_\text{s}$,
and two spheres of volume $V_\text{C}$ with densities $\Delta\rho_\text{c}$ and
$-\Delta\rho_\text{s}$---as illustrated in \fref{fig:model:core_shell_model}.
Such decompositions are helpful and follow from
the linearity of the Fourier transform. For intensities, in contrast, such
decompositions are not valid since squaring the amplitude leads to cross-terms that generally do not vanish. \\

\subsubsection{Gaussian chains}

We now turn to a more complex model: the \textit{Gaussian chain}, a simple model
of polymers in solution. This model represents a flexible chain of $N$
scatterers connected by segments whose orientations are random and uncorrelated.
\\

Using \eref{eq:Aq_sum}, the molecular amplitude is given by $\Delta
A^*_{\tm}(\bm{q}) = \sum^N_i \Delta f_i(\bm{q})e^{i\bm{q}\cdot{}\bm{r}_i}$,
where $\Delta f_i(\bm{q})$ is an excess form factor. By restricting our analysis
to identical spherical scatterers, such as atoms, these are orientationally
independent: $\Delta f_i(\bm{q}) = \Delta f(q)$. The associated intensity for a
chain with $N$ segments is then:
\begin{align*}
	\Delta I^*_{\tm}(q) 
	&= \langle|\Delta A^*_{\tm}(\bm{q})|^2\rangle_\Omega \\
	&= \Delta f^2(q)\sum_{i,j=1}^N\big\langle e^{i\bm{q}\cdot(\bm{r}_i-\bm{r}_j)}
	\big\rangle_\Omega
\end{align*}
where $\bm{r}_i - \bm{r}_j$ is the vector connecting scatterers $i$ and $j$. Evaluating this expression requires a closer look at the chain statistics. \\

The position of the $i$th scatterer is given as the vector sum of offsets
relative to the first:
\begin{align*}
	\bm{r}_i = \bm{r}_0 + \sum_{k=1}^i \delta\bm{r}_k
\end{align*}
where the segment vectors $\delta\bm{r}_k$ are assumed independent and Gaussian-distributed with zero mean, $\langle \delta\bm{r}_k \rangle = 0$, and variance $\langle \delta\bm{r}_k^2 \rangle = l_0^2$. Assuming $i>j$, the separation between two scatterers is:
\begin{align*}
	\bm{r}_i - \bm{r}_j = \sum_{k=j+1}^i \delta\bm{r}_k
\end{align*}
and the mean squared displacement between them is:
\begin{align*}
	\langle(\bm{r}_i - \bm{r}_j)^2\rangle_\Omega
	&=\Big\langle\Big(\sum_{k=j+1}^i \delta\bm{r}_k\Big)\cdot\Big(\sum_{l=j+1}^i
	\delta\bm{r}_l\Big)\Big\rangle_\Omega \\
	&=\sum_{k=j+1}^i\langle(\delta\bm{r}_k)^2\rangle_\Omega + 2\sum_{k=j+1}^i
	\sum_{l=k+1}^i \langle\delta\bm{r}_k\cdot \delta\bm{r}_l\rangle_\Omega
\end{align*}
The second term vanishes due to independence:
$\langle\delta\bm{r}_k\cdot\delta\bm{r}_l\rangle_\Omega =
\langle\delta\bm{r}_k\rangle_\Omega\langle\delta\bm{r}_l\rangle_\Omega = 0$.
The remaining term is:
\begin{align*}
	\langle(\bm{r}_i - \bm{r}_j)^2\rangle_\Omega
	&=l_0^2 \sum_{k=j+1}^i 1 \\
	&=l_0^2|i-j|
\end{align*}
This mean squared displacement determines the orientational average in the intensity expression. A standard result for sums of independent Gaussian vectors gives~\cite{Isserlis1918}:
\begin{align*}
	\big\langle e^{i\bm{q}\cdot(\bm{r}_i-\bm{r}_j)} \big\rangle_\Omega
	= e^{-q^2 \langle(\bm{r}_i - \bm{r}_j)^2\rangle_\Omega\, / 6}
	= e^{-q^2 l_0^2 |i-j|/ 6}
\end{align*}
As the full derivation is somewhat technical, we use this result here without
proof. Substituting back into the intensity gives:
\begin{align*}
	\Delta I^*_{\tm}(q) 
	&= \Delta f^2(q)\sum_{i,j=1}^N e^{-q^2l_0^2|i-j|/6}
\end{align*}

For long chains, $|i-j|$ can be considered continuous, allowing us to replace
the sum by an integral:
\begin{align}
	\Delta I^*_{\tm}(q)
	&= \Delta f^2(q)\int_0^N\int_0^N e^{-q^2l_0^2|n-n'|/6}dn'dn \nonumber\\
	&= 2\Delta f^2(q)\int_0^N\int_0^n e^{-q^2l_0^2(n-n')/6}dn'dn \nonumber\\
	&= 2\Delta f^2(q)\int_0^N 6\frac{1 - e^{-q^2l_0^2n/6}}{q^2l_0^2}dn \nonumber\\
	&= 2\Delta f^2(q) 6\frac{6e^{-q^2l_0^2N/6} + Nq^2l_0^2 - 6}{q^4l_0^4}
	\nonumber
	\intertext{Defining $x = Nq^2l_0^2/6$, this result takes the simple form:}
	\Delta I^*_{\tm}(q) &= 2N^2\Delta f^2(q) \frac{e^{-x} + x - 1}{x^2}
\end{align}

This result, first derived by~\citeasnoun{Debye1947}, is known as the
\textit{Debye function} for Gaussian chains.\footnote{The factor $N^2$ is
	typically absorbed into an overall scaling factor in the literature.} At low
$q$, it reduces to the Guinier approximation with $R_g^2 = N l_0^2 / 6$, while
at high $q$ it exhibits a $q^{-2}$ decay characteristic of flexible polymers.\\ 

Other statistical models include star polymers~\cite{Benoit1953}, ring
polymers~\cite{Zimm1949}, and variants incorporating self-avoidance (``excluded
volume'')~\cite{Pedersen1996,Li2014}; see~\cite{Pedersen1997,Svaneborg2012,BookLindner2024} for comprehensive
reviews. Despite these advancements, complex statistical models can rarely be
solved analytically, and often require empirical parameterization of simulated
results~\cite{Pedersen1996}.

\subsection{Numerical scattering calculations}

The analytical results above cover shapes with enough symmetry to admit
closed-form solutions. For more complex geometries---branched structures,
disordered assemblies, or atomistic models---closed-form expressions are
typically not available and the intensity must be computed numerically. We outline
three widely used approaches in this section.

\subsubsection{Monte Carlo sampling of the amplitude integral}

When the structure is too complex to admit a closed-form expression, Monte Carlo
integration offers a numerical alternative for evaluating the amplitude
integral~\cite{Hansen1990,Caflisch1998,Pedersen2012,Hung2024}. Starting from the
molecular scattering amplitude:
\begin{align*}
	\Delta A^*_{\tm}(\bm{q}) 
	&= \int_{V_{\tVm}} \Delta \rho^*_{\tm}(\bm{r})e^{i\bm{q}\cdot{}\bm{r}}d\bm{r}
	\intertext{This integral can be approximated by uniformly sampling $N$ points
		from $V_{\tVm}$ and averaging:}
	\Delta A^*_{\tm}(\bm{q}) 
	&\approx \frac{V_{\tVm}}{N}\sum^N_i
	\Delta\rho^*_{\tm}(\bm{r}_i)e^{i\bm{q}\cdot{}\bm{r}_i}
\end{align*}
where the $N$ points $\{\bm{r}_i\}$ are assumed to be uniformly sampled from
$V_{\tVm}$ for simplicity. More sophisticated strategies, such as importance
sampling, can improve computational efficiency~\cite{Hung2024}. \\

Squaring and orientationally averaging to obtain the intensity:
\begin{align*}
	\Delta I^*_{\tm}(q) 
	&= \langle |\Delta A^*_{\tm}(\bm{q})|^2 \rangle_\Omega \\
	&=
	\left(\frac{V_{\tVm}}{N}\right)^2\sum^N_{ij}\Delta\rho^*_{\tm}(\bm{r}_i)\Delta\rho^*_{\tm}(\bm{r}_j)\sinc(qr_{ij})
\end{align*}
This approach has been implemented in several SAS analysis
tools~\cite{Hansen1990,Spinozzi2000,Pedersen2012,Deumer2022,Larsen2023}, where
it is used for validating analytic results or for handling complex geometries. \\

Although this expression resembles the Debye equation for point scatterers, and
some derivations indeed begin from there, the resemblance is superficial: the
points $\{\bm{r}_i\}$ are quadrature nodes, not physical scatterers, and the
factor $V_{\tVm} / N$ is an integration weight, not a structural property. The
underlying object is the continuous volume integral above, recovered as
$N\rightarrow\infty$.

\subsubsection{Representing arbitrary shapes with a grid}
\label{sec:modeling:grid_decomposition}

To evaluate the scattering of an arbitrary shape or object, instead of summing
over its atoms, it can be convenient to partition the scattering volume into a
three-dimensional grid of cells indexed by $i,j,k$. The density is then given
as:
\begin{align*}
	\rho^*_{\tm}(\bm{r}) = \sum_{ijk} \rho(\bm{r} - \bm{r}_{ijk})	
\end{align*}
where we use an absolute density as this technique is mostly useful for
evaluating uniform densities. Following the same steps used to derive the
sum-over-scatterers amplitude in \sref{sec:theory:sum_over_scatterers}, we
obtain:
\begin{align*}
	A^*_{\tm}(\bm{q}) = \sum_{ijk} e^{i\bm{q}\cdot\bm{r}_{ijk}} f_{ijk}(\bm{q})
\end{align*}
This expression closely matches the aforementioned sum-of-scatterers amplitude
(\eref{eq:Aq_sum}): in the grid-based formulation, each cell acts as an
independent scattering center. \\

\begin{figure}
	\centering
	\caption{Illustration of the grid decomposition of a homogeneous sphere. The
		scattering amplitude factors into a form factor term (cell shape) and a
		geometrical term (cell positions).}
	\label{fig:modeling:grid_sphere}
	\includesvg[width=.8\linewidth]{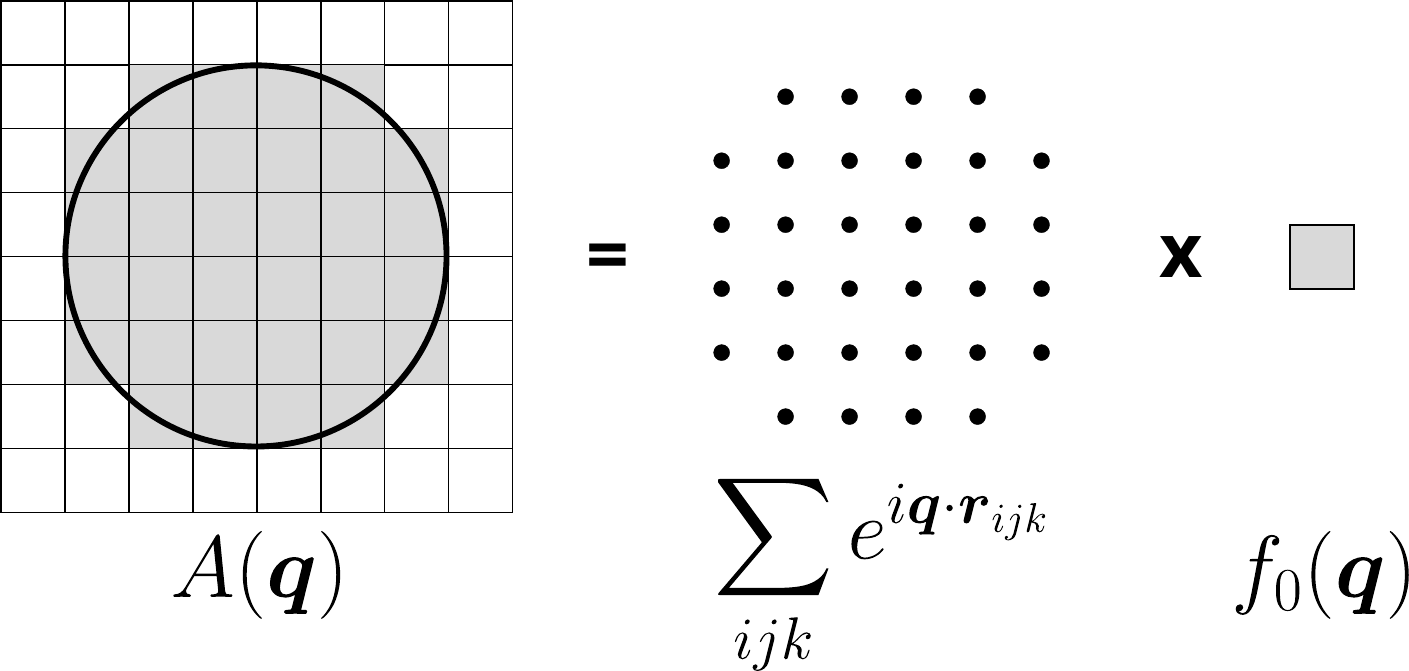}
\end{figure}

As a practical example, consider a homogeneous solid particle of density
$\rho_0$. Representing it using a regular cubic grid as in
\fref{fig:modeling:grid_sphere}, all interior cells are identical, so
$f_{ijk}(\bm{q}) = f_0(\bm{q})$. The amplitude becomes:
\begin{align}\label{eq:Aq_grid_decomposition}
	A^*_{\tm}(\bm{q}) = f_0(\bm{q})\sum_{ijk}^\text{(inside)}
	e^{i\bm{q}\cdot\bm{r}_{ijk}}
\end{align}
To compute the scattering intensity from such a grid-based structure
using the Debye equation, the cell form factors must be
orientation-independent, i.e., $f_0(\bm{q}) = f_0(q)$. Orientational
independence can be enforced by representing each cell as a Gaussian sphere, as
applied in~\citeasnoun{Lytje2025} for grid-based excluded-volume modeling; see
\sref{sec:md:exv}.

\subsubsection{Evaluating the orientational average}
\label{sec:modeling:orientational_avg}

For large or geometrically complex structures, evaluating the orientational
average analytically---for instance, via the Debye equation or a
spherical-harmonic expansion
(\aref{sec:appendix:spherical_harmonics_expansion})---can become computationally
expensive. A common alternative is to compute the average numerically by
sampling orientations directly. The orientational average takes the generic
form:
\begin{align}
	\left\langle f(\bm{q}\cdot{}\bm{r}) \right\rangle_\Omega
	&=\frac{1}{4\pi}\int_{S^2} f(\bm{q}\cdot{}\bm{r}) d\Omega
	\label{eq:modeling:int_orientation}
\end{align}
where the factor $1\,/\,4\pi$ normalizes the integral over the unit sphere. \\

As noted in \sref{sec:vacuum:debye}, this average can be carried out in two
equivalent ways: rotating the molecule with $\bm{q}$ fixed, $\langle\cdot\rangle_{\Omega_r}$, or rotating $\bm{q}$ with the molecule
fixed, $\langle\cdot\rangle_{\Omega_q}$. The latter is what enables the
numerical method below: rather than reorienting a possibly complex molecular
structure, one samples $\bm{q}$-vectors uniformly on spheres in $q$-space. \\

Using the Monte Carlo estimator introduced above, the angular integral in
\eref{eq:modeling:int_orientation} is approximated by sampling $N$ vectors
$\bm{q}_i$ uniformly on the sphere of radius $q$, evaluating the scattering for
each, and averaging. This approach is widely used in structural validation and
refinement workflows~\cite{Bardhan2009,Poitevin2011,Knight2015,Ginsburg2019}.
Uniform sampling of the sphere (not $\theta$ and $\phi$) is essential for
accuracy; see~\cite{Blech2024} for efficient strategies.

%% file: md.tex
\section{Modeling with atomistic structures}\label{sec:struc_validation}

Having explored modeling strategies for cases with limited prior information, we
now turn to situations where an atomistic structural model is
available---typically from crystallography, NMR, or cryo-EM---and ask the
central validation question of structural biology: does the model reproduce the
observed scattering? \\

Two broad traditions address this question. The conventional approach,
exemplified by tools such as \textit{CRYSOL}~\cite{Svergun1995}, decomposes the
scattering into separately modeled molecular, hydration-shell, and
excluded-volume components. The MD-based approach,
following~\citeasnoun{Park2009}, simulates molecule and solvent together and
computes the scattering directly from the trajectory. These two traditions have
largely developed in parallel and are rarely connected in the literature. In the
following, we examine each through the lens of the framework developed
previously.

\subsection{Clarifying the role of the modeling
	volume}\label{sec:modeling_volume_principle}

In \sref{sec:theory:nomenclature}, the molecular volume $V_{\tVm}$
is defined to extend beyond the physical molecule and its structured solvent
layer, with its boundary lying in bulk solvent. To compute the intensity $\Delta I^*_{\tm}(q)$ emerging from $V_{\tVm}$, intensity contributions from everything inside---the molecule, solvent, and their coupling---needs to be modeled. Hence, we refer
to it also as the \textit{modeling} volume in the following. \\

After direct background subtraction (\eref{eq:solv:bg_subtract_correct}), the
measured intensity takes the form:
\begin{align}\label{eq:struc_validation:what_to_model}
	I^\text{(exp)}_\text{A}(q) - I^\text{(exp)}_\text{B}(q) 
	= N\big[\Delta I^*_{\tm}(q) - \Delta I^{(V_{\tVm})}_{\tb}(q)\,\big]
\end{align}
Two quantities must therefore be modeled: the excess molecular scattering
$\Delta I^*_{\tm}(q)$ from the structure within $V_{\tVm}$, and the excess
buffer scattering $\Delta I^{(V_{\tVm})}_{\tb}(q)$ from an equivalent volume of
pure buffer. The remainder of this section addresses each in turn.

\subsection{Modeling the excess molecular scattering}
\label{sec:md:decomposing_molecular}

Within the modeling volume, the excess density can be decomposed into contributions from two physical components: the molecular interior and the hydration shell, relative to a uniform
excluded-volume term:
\begin{align} \label{eq:md:delta_rho_m}
	\Delta\rho^*_{\tm}(\bm{r},t) =
	\begin{cases}
		\rho'_{\tm}(\bm{r},t) & \text{if } \bm{r} \in V'_{\tm} \\
		\rho_\text{hs}({\bm{r},t}) & \text{if } \bm{r} \in V_\text{hs}
	\end{cases}\Bigg\} - \rho_\text{exv}
\end{align}
where $V'_{\tm}$ denotes the volume of the bare molecule; $V_\text{hs}$ the volume of the solvent region including the hydration shell and any additional bulk solvent. Together, $V'_{\tm}$ and $V_\text{hs}$ covers the whole modeling volume. 
The term $\rho_\text{exv} \equiv \rho_{\tb}$ is the excluded-volume
contribution, for which we use a new symbol for consistency with the literature. \\ 

This leads to the expansion:
\begin{align}\label{eq:md:intensity_components}
	\Delta I^*_{\tm}(q) 
	&= \langle|\Delta A^*_{\tm}(\bm{q},t)|^2\rangle_N \nonumber\\
	&= \langle|A'_{\tm}(\bm{q},t) - A_\text{exv}(\bm{q}) +
	A_\text{hs}(\bm{q},t)|^2\rangle_N \nonumber\\
	&= I'_{\tm}(q) + I_\text{exv}(q) + I_\text{hs}(q) + \langle
	(\text{cross-terms})\rangle_N
\end{align}
where $I'_{\tm}(q)$ is the scattering of the bare molecule in vacuum,
$I_{\text{exv}}(q)$ the contribution from the average buffer density, and
$I_{\text{hs}}(q)$ the contribution from the solvent surrounding the
molecule. These constitute the three main components that must be modeled,
together with their mutual cross-terms.

\subsubsection{Molecular scattering in vacuum}

With an atomistic reference structure available, evaluating the molecular
scattering $I'_{\tm}(q)$ is a straightforward application of the Debye equation
or spherical harmonic expansion
(\aref{sec:appendix:spherical_harmonics_expansion}). We will therefore not
discuss this contribution further here; for practical implementations, see~\cite{Lytje2025,Ginsburg2019,Grudinin2017,Gumerov2012}.

\subsubsection{Excluded volume scattering}
\label{sec:md:exv}

The excluded volume scattering amplitude is given as:
\begin{align}\label{eq:md:intermediate:exv_integral}
	A_\text{exv}(\bm{q}) = \int_{V_{\tVm}}\rho_\text{exv}\,
	e^{i\bm{q}\cdot{}\bm{r}}d\bm{r}
\end{align}
The challenge lies in determining the most efficient and accurate approach for
evaluating this integral. \\

Choosing $V_{\tVm}$ to be a simple geometric shape---a sphere or ellipsoid, for
instance---makes the integral one of the analytical shape models from
\sref{sec:modeling:model_dependent}. This is computationally convenient, but the
resulting modeling volumes are typically large, and may be prohibitively
expensive to fill with solvent in the next step.\\

To avoid this issue, one can instead discretize $V_{\tVm}$ using e.g.~the grid
decomposition described in \sref{sec:modeling:grid_decomposition}, thus
partitioning the integral into small, easily evaluated subunits. To avoid the
orientation-dependent issues arising from using cubic form factors, Gaussian
spheres may be used to approximate the uniform density~\cite{Lytje2025}. \\

A common simplification in conventional models is to split the excluded-volume
integral into two regions: the physical molecular volume $V'_{\tm}$ and the
surrounding hydration-shell region $V_\text{hs}$. The excluded-volume
contribution within $V_\text{hs}$ is then absorbed into the hydration density
via an excess form $\Delta\rho_\text{hs} \equiv \rho_\text{hs} -
\rho_\text{exv}$, leaving only $V'_{\tm}$ to be modeled explicitly. This enables the use of excluded-volume models based on the atomic positions of the molecule, including grid-decomposition
strategies either with~\cite{Lytje2025} or without~\cite{Bardhan2009} unit-cell form
factors. \\

An additional simplification approximates $V_{\tVm}'$ as a sum of atomic contributions, reducing the excluded-volume integral to a discrete sum over the molecule's atoms:
\begin{align*}
	A_\text{exv}(\bm{q}) 
	&= \int_{V_{\tVm}}\rho_\text{exv}\, e^{i\bm{q}\cdot{}\bm{r}}d\bm{r}
	\approx \sum^N_{i=1} f^\text{x}_i(\bm{q}) e^{i\bm{q}\cdot{}\bm{r}_i}
\end{align*}
where $f^\text{x}_i(\bm{q})$ are excluded-volume form factors dependent on atomic size, commonly modeled as Gaussian spheres following the Fraser model~\cite{Fraser1978,Svergun1995,Lytje2025}. Sharing atomic positions with the molecular contribution lets the two be combined into a single effective form factor:
\begin{equation}
	A_\text{m}'(\bm{q}) - A_\text{exv}(\bm{q}) 
	\approx \sum^N_{i=1} f^\text{red}_i(\bm{q}) e^{i\bm{q}\cdot{}\bm{r}_i},
\end{equation}
where $f^\text{red}_i(\bm{q}) = f_i(\bm{q}) - f^\text{x}_i(\bm{q})$ is the \textit{reduced form factor} of atom $i$. \\

This approximation introduces two systematic errors. First, replacing the continuous density $\rho_\text{exv}$ with Gaussian densities at atomic positions introduces artificial internal structure into the excluded-volume term, contributing deviations at wide angles. Second, the subtracted density depends on the chosen atomic volume table---still typically taken from~\citeasnoun{Traube1895}---and on local atomic packing, producing spurious variation across the molecule; for chemically heterogeneous systems such as protein/RNA or protein/detergent complexes, these variations affect the large-scale density distribution and can bias the model even at small angles. Given these problems and the additional issues identified in recent work~\cite{Lytje2025,Chatzimagas2022,Bardhan2009}, we recommend against the Fraser approximation in new work.

\subsubsection{Hydration shell scattering}
\label{sec:theory:hydration_shell}

\begin{figure}
	\centering
	\caption{Solvent density shell from MD simulations. (A) Molecule (green)
		surrounded by solvent, color-coded by density. (B) Radial solvent density
		profile showing two density peaks; the second shell is faintly visible in (A).
		Adapted with permission from~\citeasnoun{Linse2023}
		and~\citeasnoun{Chen2014}.}
	\label{fig:theory:hydration_shell_density}
	\includesvg[width=\linewidth]{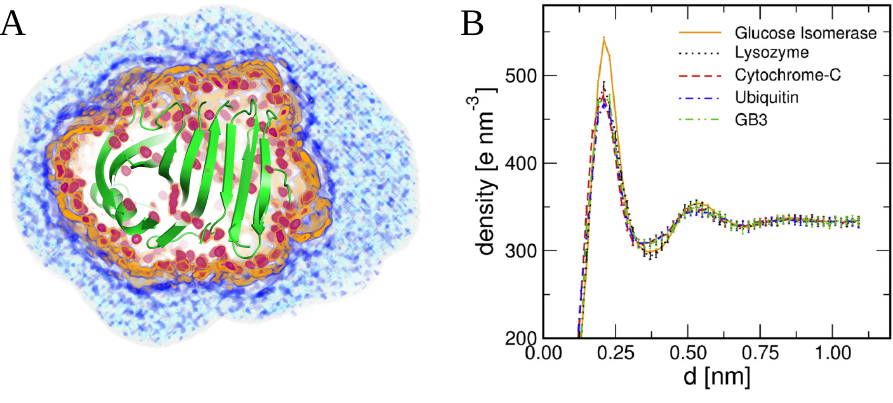}
\end{figure}

Formed by polar and hydrogen-bonding interactions with the molecular surface,
the hydration shell represents a pronounced perturbation of the local solvent
density relative to bulk water (\fref{fig:theory:hydration_shell_density}). To
achieve a bulk-like density at the boundary of $V_{\tVm}$, the modeling volume
should ideally encompass \textit{all} solvent perturbations around the molecule.
\\

In practice, most models compromise by including only the primary overdense
shell (\fref{fig:theory:hydration_shell_density}B), and neglecting secondary
hydration layers. In the language of \sref{sec:modeling_volume_principle}, this
corresponds to an effective modeling volume smaller than $V_{\tm}$, truncating
solvent dentity variations that would otherwise contribute to the scattering signal. \\

Modeling approaches fall into two broad categories: \textit{explicit} models, in
which solvent molecules are represented individually (e.g.~MD-based methods),
and \textit{implicit} models, in which the solvent distribution is described
analytically using e.g.~envelope functions~\cite{Svergun1995} or
Poisson--Boltzmann-based techniques~\cite{Poitevin2011}. These categories differ
in how they handle the truncation: explicit models can capture the full
solvent density profile within the modeling volume,
while implicit models typically restrict their attention to the primary shell. 
For comparative reviews of the implementations across software packages, see~\citeasnoun{Trewhella2024} and~\citeasnoun{Lytje2025}.\\

A separate consideration is that larger cavities inside the molecule may contain
solvent that contributes to the signal~\cite{Matthews2009} and must be modeled
in addition to the surface hydration shell.

\subsubsection{The conformational average in practice}
\label{sec:theory:thermal}

The discussion so far has treated the conformational average
$\langle\cdot\rangle_N$ as a formal operation. In practice, the full equilibrium
ensemble is rarely available, and modeling reduces to choosing how to
approximate this average from limited data. \\

Within this framework, the various strategies in the literature---rigid-body
refinement, conformer pool generation, MD trajectories, Debye--Waller
corrections, and normal-mode analysis---are all approximations of
$\langle\cdot\rangle_N$. They differ primarily in the timescales they capture
(from fast atomic fluctuations to slow collective motions) and in how they
sample conformational space (from single structures to heterogeneous all-atom ensembles). Each
adds modeling freedom and potential for overfitting; the appropriate choice
depends on data quality, the dominant motions of the system, and on the availability
of complementary data (e.g.~from NMR spectroscopy) that may help to test proposed structural models against 
overfitting.

\subsection{Modeling the excess buffer scattering}
\label{sec:md:excess_buffer}

We now turn to the second term in \eref{eq:struc_validation:what_to_model}:
$\Delta I^{(V_{\tVm})}_{\tb}(q)$, representing the excess scattering of pure
buffer within the modeling volume. As noted in \sref{sec:bg_subtraction},
volume-corrected background subtraction absorbs this term into a volume
rescaling, removing the need to evaluate it explicitly. For modeling of wide-angle SAS data,
however, the volume-corrected approximation is problematic owing to 
neglecting the cross-term $\Delta C^{V_{\tVm},V_{\tVb}}_{\tb,\tb}(q)$.
Hence, for such cases, the direct subtraction scheme is preferable, yet
requires computation of $\Delta I^{(V_{\tVm})}_{\tb}(q)$. \\

Evaluating the buffer intensity analytically is complicated by three
interlocking issues: buffer composition (water plus experiment-specific
additives), geometry (the integral must be evaluated over $V_{\tm}$, whose shape
is determined by the molecule), and the need to describe density fluctuations
relative to bulk rather than absolute densities. Taken together, a general
closed-form expression becomes impossible. \\

Progress can be made by reframing the problem in terms of local solvent
structure. As shown in \aref{sec:appendix:buffer_average_t_is_orientational},
the total buffer scattering can be decomposed into contributions from locally
correlated solvent environments; alternatively, the cross-term can be expressed
in terms of the bulk pair-correlation function (\aref{sec:appendix:mdgamma}).
Both routes reduce the problem from evaluating scattering over the full modeling
volume to tabulating short-range bulk solvent properties, which are accessible
from e.g.~MD simulations. The remainder of this section therefore takes MD
as a modeling approach, and describes how it fits into the theoretical
framework developed above.

\subsection{The theory behind molecular dynamics simulations}\label{sec:md}

In \sref{sec:md:excess_buffer} we introduced molecular dynamics (MD) as a
practical route to evaluate the pure buffer contribution. MD, however, is not
restricted to this isolated task: by explicitly simulating both solute and
solvent inside the modeling volume, it provides a direct route to estimating the
molecular term and, ultimately, the full observable scattering signal. \\

Unlike the separate excluded-volume, hydration-shell, and
conformational-ensemble treatments discussed earlier, MD captures all these
effects naturally through the explicit time evolution of the entire modeling
volume. At the same time, MD predictions depend on the forcefield, water model,
and sampling length, and remain computationally demanding~\cite{Linse2023}. MD
is therefore a complementary tool, best suited for high-resolution or wide-angle
data where detailed solvation and conformational sampling are essential.\\

Although the discussion below is framed in terms of
\texttt{WAXSiS}~\cite{Knight2015}, the underlying principles apply to any
explicit‑solvent scattering calculation based on MD trajectories. This section
focuses on how MD-based calculations connect to the theoretical framework
established above, rather than on the practical details of running MD
simulations for SAS. For the latter, see~\citeasnoun{Chatzimagas2022}. 

\subsubsection{Modeling the observed scattering with MD}

To model both intensities in \eref{eq:struc_validation:what_to_model}, two
simulation boxes are required:
\begin{enumerate}
	\itemindent=20pt
	\item[\textit{System A}:] The molecule in explicit solvent, within a large
	simulation box. The scattering calculation is later restricted to the modeling
	volume $V_{\tVm}$.
	\item[\textit{System B}:] An otherwise identical simulation box containing only
	solvent molecules.
\end{enumerate}

From System A we compute the excess molecular intensity:
\begin{align*}
	\Delta I^\text{(MD)}_\text{A}(q) \equiv \langle \Delta
	I^*_{\tm}(\bm{q},t)\rangle_N
\end{align*}
which is in principle sufficient for comparison with volume-corrected experimental profiles.
To match the more accurate direct subtraction scheme, however, we also
need the corresponding pure-buffer term:
\begin{align*}
	\Delta I^\text{(MD)}_\text{B}(q) \equiv \langle \Delta
	I^{(V_{\tVm})}_{\tb}(\bm{q},t) \rangle_N
\end{align*}
representing the scattering from a buffer-only realization of the same modeling
volume.\\ 

Subtracting the two simulated contributions yields:
\begin{align}\label{eq:md:match_measurement}
	I^\text{(exp)}_\text{A}(q) - I^\text{(exp)}_\text{B}(q) = N\big[\Delta
	I^\text{(MD)}_\text{A}(q) - \Delta I^\text{(MD)}_\text{B}(q)\big]
\end{align}
showing that MD naturally implements the exact direct subtraction scheme.

\begin{figure}
	\centering
	\caption{Two-box setup for accurate MD-based scattering calculations. System A
		(molecule + solvent) and System B (solvent only) together enable exact modeling
		of the experimental scattering signal. Adapted from~\citeasnoun{Knight2015}.}
	\label{fig:solv:park_sims}
	\includesvg[width=.8\columnwidth]{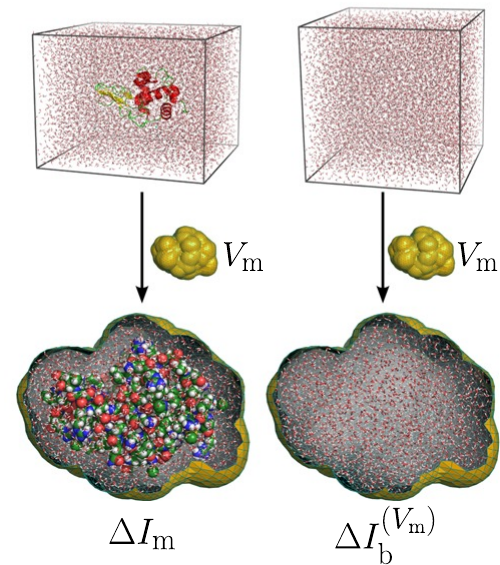}
\end{figure}

\subsubsection{Interpretation and equivalence to analytical theory}

Although \eref{eq:md:match_measurement} resembles an experimental background
subtraction, the MD calculation reconstructs the two excess terms required by
\eref{eq:struc_validation:what_to_model} and crucially does not repeat the
experiment.\\ 

As shown in \aref{sec:appendix:equivalence_to_park}, this formulation is
 equivalent to that of~\citeasnoun{Park2009}: MD-based scattering is
therefore not a separate theoretical framework but an explicit-solvent
realization of the familiar atomic + hydration-shell + excluded-volume
decomposition that underlies conventional analytical methods. The accuracy of
both is governed by the same assumptions---choice of modeling volume, treatment
of excess solvent density, boundary cross-terms---made implicit in the
simulation setup rather than explicit in the model. The conventional and
MD-based traditions, long developed in parallel, are thus two realizations of
the framework developed here.

%% file: conclusion.tex
\section{Discussion and Conclusions}

Although small-angle solution scattering is based on simple wave interference, 
the theory of SAS has often been presented as a collection
of largely separate topics, such as diffraction physics, buffer subtraction,
analytical models for idealized shapes, or SAS curve predictions based on either
implicit or explicit solvent models. In this work, we instead develop a unified
theory of small-angle scattering of generic solutes such as biomolecules or soft-matter
complexes. We emphasize that these seemingly distinct concepts are aspects of a
single framework, and that different analysis and modeling strategies emerge
from it through distinct assumptions and approximations. By deriving
the framework from first principles and by stating all assumptions and
approximations explicitly, we aim to make the theory of SAS accessible to
experimentalists, modelers, newcomers, and experienced practitioners alike. \\

Making the approximations explicit clarifies several points that are often left
implicit in introductions to SAS. First, the concept of excess density should
not be conflated with buffer subtraction. The sensitivity of SAS signals to
excess density follows from Babinet's principle and is intrinsic to the
scattering process. Buffer subtraction serves to isolate the scattering signal
from the molecule from that of the large contribution arising from the internal
structure of bulk water. As emphasized here, buffer subtraction is also an
integral part of the theoretical model. \\

Consequently, the modeling strategy must
be compatible with the applied buffer-subtraction scheme to avoid systematic errors.
Direct and volume-corrected buffer subtraction are not interchangeable,
as they lead to markedly different SAS curves at wide angles, where water
scattering dominates, and even to significant variations at intermediate
scattering angles. The two schemes furthermore differ in their treatment of 
the cross-term at the molecule–solvent boundary. In direct subtraction, the cross
term cancels exactly against the corresponding contribution from a pure-buffer
reference of equivalent volume. In contrast, in volume-corrected subtraction,
contributions from the cross-term remain but are typically neglected, rendering
this approach approximate when analyzed with standard modeling software. This
approximation is most relevant at wider angles, where water--water correlations
across the boundary contribute. The two subtraction schemes also differ in that 
only direct subtraction requires calculating the residual bulk-solvent term---a 
quantity MD simulations are well suited to provide---and avoiding this calculation 
is part of what makes the volume-corrected scheme attractive. Thus, the two schemes 
require different modeling strategies, and reliable analysis of SAS data depends 
on using compatible combinations of subtraction and modeling; otherwise, systematic 
errors may arise that are difficult to diagnose. \\

Within this framework, we derived a range of methods for analyzing SAS curves
and for validating structural models against experimental data, including
model-free analysis, MD-based approaches, and more conventional structure-based
methods. We emphasized that MD-based scattering predictions are not a
separate theoretical tradition, but rather an explicit-solvent realization of
the same framework, accounting for contributions from the molecule, hydration
shell, and excluded volume, as in conventional modeling strategies. By placing
these methods within a common framework, this work enables the future comparison and integration of these previously distinct modeling approaches.\\

Looking forward, this unified framework defines how experimental and modeling 
choices must be aligned and where their approximations matter.
This will become increasingly important as SAS advances toward higher precision,
wider scattering angles, time-resolved measurements, and integrative approaches
combining scattering with MD simulations. In this regime, distinctions that were 
negligible at low resolution become quantitatively important. By making assumptions 
explicit and placing analytical, implicit-solvent, and explicit-solvent methods 
within a single formal framework, this work provides a foundation for systematically
improving SAS methodology and for closer integration between theory, simulation,
and experiment.

%% file: acknowledgments.tex
\section{Acknowledgments}

K.L. was supported by the Deutsche Forschungsgemeinschaft (DFG, German Research Foundation) via grant HU 1971/3-2 and Danmarks Frie Forskningsfond (DFF, Independent Research Fund Denmark), grant 1026-00209B. \\

J.S.H. was supported by the DFG, grant INST 256/539-1. \\

J.S.P. was supported by the Danish Agency for Higher Education and Science (ESS Lighthouse: Colloids and Interfaces in Food and Pharma; Grant No. 726510) and DFF, grant 1026-00209B.

%% file: appendix.tex
\appendix
\renewcommand{\thefigure}{A\arabic{figure}}
\renewcommand{\theequation}{A\arabic{equation}}
\renewcommand{\thetable}{A\arabic{table}}
\renewcommand{\thesection}{A\arabic{section}}
\setcounter{figure}{0} 

\section{Computing scattering via correlation
	functions}\label{sec:appendix:correlation_funcs}

The intensity and structure factors derived in \sref{sec:correlated_mols} are
frequently evaluated in practice using real‑space correlation functions, such as
the pair correlation function appearing in
\eref{eq:modeling:pair_distance_distribution}. This appendix collects and
clarifies the connection between the derivation based on scattering amplitudes and 
the decoupling approximation (\eref{sec:correlated_mols})
and correlation‑based descriptions. We show how the 
latter are used in analytical theories and molecular simulations.

\subsection{Interpreting the structure factor in terms of correlation functions}
\label{sec:appendix:structure_factor_g}

In \sref{sec:correlated_mols}, we expressed the intensity of a correlated
molecular system under the decoupling approximation as:\footnote{The
	liquid-state literature commonly starts with the general equation $I_{\ts}(q) =
	\iint_{V_{\tVs}} \rho(\bm{r}_1)\rho(\bm{r}_2)\langle
	e^{i\bm{q}\cdot(\bm{r}_1-\bm{r}_2)}\rangle_\Omega\,d\bm{r}_1\,d\bm{r}_2$. While
	exact, this expression conflates intramolecular and intermolecular
	contributions within a single integrand: the density $\rho(\bm{r})$
	simultaneously encodes the internal structure of each molecule and the spatial
	arrangement of molecules relative to one another. Disentangling these two
	contributions leads to our starting point.}
\begin{align}\label{eq:appendix:intensity_correlation_start}
	I_{\ts}(q) = NI^*_{\tm}(q)S(q)
\end{align}
where the structure factor involves a double-sum over atoms:
\begin{align}\label{eq:appendix:Sq}
	S(q) = 1 + \frac{\beta(q)}{N}\sum^N_i\sum^N_{j\neq i}\langle
	e^{i\bm{q}\cdot\bm{r}_{ij}}\rangle_\Omega
\end{align}
While this form follows naturally from the derivation, it is not directly
useful: evaluating the double-sum over molecular positions requires detailed
knowledge of exactly how they are distributed. Hence, following
~\citeasnoun{Zernike1927}, we replace the positional sum by a statistical
description of the molecular arrangement. \\

For a fixed reference molecule $i$, define the local number density of
surrounding molecules at displacement $\bm{r}$ from $i$:
\begin{align*}
	n_i(\bm{r}) = \sum^N_{j\neq i} \delta(\bm{r}-\bm{r}_{ij})
\end{align*}
This is a microscopic quantity: it is sharply peaked at the instantaneous
positions of the other $N-1$ molecules. By the sifting property of the delta
function, the inner sum over $j$ in \eref{eq:appendix:Sq} is rewritten as an
integral over the density:
\begin{align*}
	\sum^N_{j\neq i} e^{i\bm{q}\cdot\bm{r}_{ij}}
	= \int_{V_{\tVs}} n_i(\bm{r}) e^{i\bm{q}\cdot\bm{r}} d\bm{r},
\end{align*}
where a discrete sum has been re-expressed as a 
continuous integral against a sum of delta functions. Orientationally averaging
both sides:
\begin{align*}
	\sum^N_{j\neq i} \langle e^{i\bm{q}\cdot\bm{r}_{ij}}\rangle_\Omega
	= \int_{V_{\tVs}} \langle n_i(\bm{r})
	e^{i\bm{q}\cdot\bm{r}}\rangle_\Omega\,d\bm{r}
\end{align*}

The product average on the right-hand side cannot be evaluated directly. To
continue, we adopt a statistical description and replace the microscopic density
$n_i(\bm{r})$ by its ensemble average, $\langle n_i(\bm{r})\rangle_N \equiv
n_i(r)$ and note that the latter---being spherically symmetric due to the
isotropy of the fluid---is independent of the orientation of $\bm{r}$:
\begin{align*}
	\sum^N_{j\neq i} \langle e^{i\bm{q}\cdot\bm{r}_{ij}}\rangle_\Omega
	= \int_{V_{\tVs}} n_i(r) \langle e^{i\bm{q}\cdot\bm{r}}\rangle_\Omega\,d\bm{r}
\end{align*}
Although $n_i(\bm{r})$ is highly anisotropic for any single configuration, its
ensemble average $n_i(r)$ is rotationally invariant by the isotropy of the bulk
fluid; this is what enables the reduction to radial correlation functions.\\

Because all reference molecules are statistically equivalent in the bulk,
$n_i(r)$ is identical for all molecules, allowing us to drop the index $i$.
\begin{align*}
	n(r) \equiv n  g(r)
\end{align*}
where $n = N/V_{\tVs}$ is the bulk number density and $g(r)$ is the
\textit{radial distribution function}. By construction, $n g(r) dV$ gives the
mean number of molecules found in a volume element $dV$ at distance $r$ from a
reference molecule. At large separations, where positional correlations vanish,
the local density must approach the bulk: $g(r)\rightarrow 1$ as
$r\rightarrow\infty$. The outer sum over $i$ then contributes a factor of $N$,
since the integrand is now $i$-independent:
\begin{align}\label{eq:appendix:intermediate:exponential_to_integral}
	\sum^N_i\sum^N_{j\neq i} \langle e^{i\bm{q}\cdot\bm{r}_{ij}}\rangle_\Omega
	= N \int_{V_{\tVs}} ng(r)\langle e^{i\bm{q}\cdot\bm{r}}\rangle_\Omega\, d\bm{r}
\end{align}
The structural information accessible to scattering is therefore fully encoded
in $g(r)$, with all dependencies on individual molecular positions
absorbed into this single radial function.\\

\eref{eq:appendix:intermediate:exponential_to_integral} is exact under the
assumed isotropy and homogeneity, but practically inconvenient: $g(r)$
approaches unity at large $r$, so the integral receives contributions from the
entire integrated volume. To isolate the non-trivial part, we add and subtract
unity:
\begin{align}\label{eq:appendix:intermediate:exponential_to_integral_no_diverging}
	&\sum^N_i\sum^N_{j\neq i} \langle e^{i\bm{q}\cdot\bm{r}_{ij}}\rangle_\Omega\nonumber\\
	&\quad\quad= Nn \int_{V_{\tVs}} [g(r)-1+1] \langle e^{i\bm{q}\cdot\bm{r}}\rangle_\Omega\,
	d\bm{r} \nonumber\\
	&\quad\quad= Nn \int_{V_{\tVs}} [g(r)-1] \langle e^{i\bm{q}\cdot\bm{r}}\rangle_\Omega\,
	d\bm{r} + Nn \langle\mathcal{S}_{V_{\tVs}}(\bm{q})\rangle_\Omega
\end{align}
where the shape factor:
\begin{align*}
	\langle\mathcal{S}_{V_{\tVs}}(\bm{q})\rangle_\Omega 
	= \int_{V_{\tVs}} \langle e^{i\bm{q}\cdot\bm{r}}\rangle_\Omega\,d\bm{r}
\end{align*}
is identical in form to the one introduced in \eref{eq:shape_factor}. Since this
contributes only at $q\rightarrow 0$ for macroscopic system volumes, it is
typically dropped. When the integration domain is small, however, such as when
modeling correlations within the molecular volume $V_{\tVm}$, the shape factor
may contribute significantly and must be retained. \\

Inserting \eref{eq:appendix:intermediate:exponential_to_integral_no_diverging}
into the structure factor of \eref{eq:appendix:Sq}, and dropping the shape factor for macroscopic
volumes:
\begin{align}\label{eq:appendix:Sq_exact}
	S(q) - 1 = n\,\beta(q) \int_{V_{\tVs}} [g(r)-1] \langle
	e^{i\bm{q}\cdot\bm{r}}\rangle_\Omega\,d\bm{r}
\end{align}
As the integrand decays on distances of order bulk correlation length, the
integrated volume may be truncated to a small region around the reference
molecule for practical evaluations.
Taking the orientational average with \eref{eq:exp_orientational_avg}, 
the familiar textbook identity:
\begin{align*}
	S(q) - 1 = 4\pi n \int^\infty_0 r^2 [g(r)-1]\sinc qr \,dr
\end{align*}
is recovered in the limit $\beta(q)\rightarrow 1$, which holds exactly for
spherically symmetric scatterers and is implicitly assumed in liquid-state
treatments where particles are taken as point scatterers from the outset. For
anisotropic molecules at finite $q$, $\beta(q)$ introduces a
form-factor-dependent attenuation of the apparent structure factor---a known
source of systematic error when fitting $S(q)$ to scattering data from
non-spherical particles~\cite{Kotlarchyk1983}.

\subsubsection{The excess decoupling ratio $\Delta\beta(q)$}
\label{sec:appendix:correlation_encoding}

\eref{eq:Iq_sample} shows that, for small-angle solution scattering, the relevant quantity is the \textit{excess} density distribution. The preceding section, however, followed the spirit of \sref{sec:theory} and worked throughout with \textit{absolute} densities. Since $S(q)$ depends on the density through $\beta(q)$, we here examine if introducing $\Delta\beta(q)$ is required to maintain consistency. \\

Replacing $A^*_{\tm}$ with $\Delta A^*_{\tm}$ in the definition of $\beta(q)$
yields the excess analog:
\begin{align}
	\Delta\beta(q) \equiv \frac{|\langle \Delta
		A^*_{\tm}(\bm{q})\rangle_\Omega|^2}{\langle|\Delta
		A^*_{\tm}(\bm{q})|^2\rangle_\Omega} = \frac{|\langle
		A^*_{\tm}(\bm{q})\rangle_\Omega -
		\rho_{\tb}\langle\mathcal{S}_{V_{\tVm}}(\bm{q})\rangle_\Omega|^2}{\langle|A^*_{\tm}(\bm{q})
		- \rho_{\tb}\mathcal{S}_{V_{\tVm}}(\bm{q})|^2\rangle_\Omega}
\end{align}
Although $\Delta\beta(q)$ and $\beta(q)$ are constructed from different amplitudes, they coincide exactly in two limits. For homogeneous $\rho^*_{\tm}(\bm{r})$, the two reduce to the same expression — a non-trivial cancellation, since subtracting a non-zero quantity in both numerator and denominator does not generally leave the ratio invariant.\footnote{Explicitly, $A^*_{\tm}(\bm{q}) = \rho^*_{\tm}\mathcal{S}_{V_{\tVm}}(\bm{q})$ and $\Delta A^*_{\tm}(\bm{q}) = (\rho^*_{\tm}-\rho_{\tb})\mathcal{S}_{V_{\tVm}}(\bm{q})$, so $\Delta\beta(q) = |\langle\mathcal{S}_{V_{\tVm}}(\bm{q})\rangle_\Omega|^2 / \langle|\mathcal{S}_{V_{\tVm}}(\bm{q})|^2\rangle_\Omega = \beta(q)$.} For spherically symmetric $V_{\tVm}$, the shape integral is orientation-independent, and $\beta(q) = \Delta\beta(q) = 1$ identically, regardless of the internal density. Typical biomolecules are close to those limiting cases---globular shapes are nearly spherical, and the excess density varies modestly on the typical length scales probed by SAS---so $\Delta\beta(q) \approx \beta(q)$ is a good approximation across the relevant $q$-range. We use $\Delta\beta(q)$ in subsequent derivations where it arises naturally, but note that it may be replaced by $\beta(q)$ when convenient.

\subsection{Practically using correlation functions to estimate the cross-term}
\label{sec:appendix:mdgamma}

In \sref{sec:solv:boundary_term}, the cross-term $\Delta
C^{V_{\tVm},V_{\tVb}}_{\tb,\tb}(q)$ was argued to be nonzero but to cancel upon
direct background subtraction. Here, we derive an explicit expression for it in
terms of the bulk solvent pair-correlation function, confirming that it is
indeed nonzero, and showing explicitly how it may be estimated from bulk-solvent
MD simulations. \\

We begin from the single-molecule cross-term:
\begin{align*}
	\Delta C^{V_{\tVm},V_{\tVb}}_{\tb,\tb}(q)
	= \langle \Delta A^{(V_{\tVm})}_{\tb}(\bm{q},t)\overline{\Delta
		A^{(V_{\tVb})}_{\tb}(\bm{q},t)}\rangle_N + \text{c.c.}
\end{align*}
As in \sref{sec:uncorrelated_molecules}, we express each amplitude as a discrete
sum over the individual solvent molecules in each region:
\begin{align*}
	\Delta A^{(V_\alpha)}_{\tb}(\bm{q},t) = \sum_{i \in V_\alpha} \Delta
	A_\text{bm}(\bm{q},\omega_i)e^{i\bm{q}\cdot\bm{r}_i} \qquad \alpha \in \{{\tVm},
	{\tVb}\}
\end{align*}
where $A_\text{bm}(\bm{q},\omega_i)$ is the amplitude of a single bulk molecule
(``bm'') in orientation $\omega_i$. Approximating all solvent molecules by the
representative structure:
\begin{align*}
	\Delta C^{V_{\tVm},V_{\tVb}}_{\tb,\tb}(q)
	&= \sum_{i\in V_{\tVm}}\sum_{j\in V_{\tVb}}\langle \Delta
	A^*_\text{bm}(\bm{q},\omega_i)\overline{\Delta
		A^*_\text{bm}(\bm{q},\omega_j)}e^{i\bm{q}\cdot\bm{r}_{ij}}\rangle_\Omega\\ 
		&\quad+\text{c.c.}
\end{align*}
Applying the decoupling approximation, the orientational and positional averages
factorize as in \sref{sec:correlated_mols}:
\begin{align*}
	\Delta C^{V_{\tVm},V_{\tVb}}_{\tb,\tb}(q)
	= 2|\langle \Delta A^*_\text{bm}(\bm{q})\rangle_\Omega|^2 \sum_{i\in
		V_{\tVm}}\sum_{j\in V_{\tVb}}\langle e^{i\bm{q}\cdot\bm{r}_{ij}}\rangle_\Omega
\end{align*}
where the complex conjugate term has been absorbed into the factor of $2$.
Recognizing that $|\langle \Delta A^*_\text{bm}(\bm{q})\rangle_\Omega|^2 =
\Delta I^*_\text{bm}(q)\Delta\beta(q) $:
\begin{align}\label{eq:cross_exact}
	\Delta C^{V_{\tVm},V_{\tVb}}_{\tb,\tb}(q) 
	&= 2\Delta I^*_\text{bm}(q)\Delta\beta(q)\sum_{i\in V_{\tVm}}\sum_{j\in
		V_{\tVb}}\langle e^{i\bm{q}\cdot\bm{r}_{ij}}\rangle_\Omega
\end{align}
To proceed, we rewrite the positional double sum as an integral over the
bulk-solvent radial distribution function, noting that correlations depend only
on intermolecular separation:
\begin{align}\label{eq:appendix:general_bulk_correlation}
	&\sum_{i\in V_{\tVm}}\sum_{j\in V_{\tVb}} \langle
	e^{i\bm{q}\cdot\bm{r}_{ij}}\rangle_\Omega\nonumber\\
	&\quad\quad= \sum_{i\in V_{\tVm}} n_{\tb} \int_{V_{\tVb}}
	g_{\tb}(|\bm{r}-\bm{r}_i|)\langle
	e^{i\bm{q}\cdot(\bm{r}-\bm{r}_i)}\rangle_\Omega \,d\bm{r} \nonumber\\
	&\quad\quad= n_{\tb}^2 \int_{V_{\tVm}} \int_{V_{\tVb}}
	g_{\tb}(|\bm{r}_1-\bm{r}_2|)\sinc(q|\bm{r}_1-\bm{r}_2|)\,d\bm{r}_1d\bm{r}_2
\end{align}
where $n_{\tb} = N_{\tVb}/V_{\tVb}$ is the bulk number density, $g_{\tb}(r)$ the
bulk solvent radial distribution function, and, because isotropy is broken by the boundary, the outer sum was retained as an explicit integral over $V_{\tm}$ rather than collapsed into a
factor of $N$ as in the homogeneous case. Here, the use of the
\textit{bulk}-solvent radial distribution function is justified by the fact
that the modeling volume is chosen such that its boundary lies in bulk solvent. \\

This is the most general form of the cross-correlational term. To practically
evaluate it, simplifying assumptions are required. 

\subsubsection{Practically evaluating the cross-term}

The cross-term \eref{eq:appendix:general_bulk_correlation} can be evaluated analytically if $V_{\tVm}$ is taken as a sphere with radius $R$, see the Supplementary Information~\ref{sec:si:cross_term_sphere} for the full derivation. The result is:
\begin{align}\label{eq:appendix:cross_term_sphere}
	\Delta C^{V_{\tVm},V_{\tVb}}_{\tb,\tb}(q)
		&= 8\pi^2 n_{\tb}^2 R^2\Delta\beta(q) \Delta I^*_\text{bm}(q)\nonumber\\
		&\quad\times \int_0^\infty [g_{\tb}(s)-1]\sinc(qs)s^3\,ds
\end{align}
where $\Delta I^*_\text{bm}(q)$ is the intensity of a representative bulk water molecule, and we have dropped the boundary terms. The cross-term scales with $R^2$, i.e. the surface area of the modeling volume, consistent with the surface-area argument of \sref{sec:solv:boundary_term}. This scaling reflects the fact that only solvent pairs straddling the boundary contribute. The integral is nonzero since $g_{\tb}(s) \neq 1$ for $s \lesssim \xi$, with $\xi$ being the bulk correlation length, demonstrating that the cross-term does contribute to the observable scattering. \\

To assess the practical relevance of the cross-term, we compare it to the buffer intensity of the modeling volume itself, $I^{(V_{\tVm})}_\text{b}(q) = N_\text{bm}\Delta I^*_\text{bm}(q)S_\text{interior}(q)$, where $N_\text{bm} = n_b V_{\tVm}$ is the number of solvent molecules in $V_{\tVm}$ and $S_\text{interior}(q)$ is the conventional structure factor (\eref{eq:appendix:Sq}) applied to the bulk solvent within $V_{\tVm}$. To enable a direct comparison, we recast \eref{eq:appendix:cross_term_sphere} in the same form by defining a cross-term structure factor:
\begin{align}\label{eq:appendix:cross_term_Sq_def}
	S_\text{cross}(q) \equiv \frac{8\pi^2 n_{\tb}^2 R^2 \Delta\beta(q)}{N_\text{bm}} \int_0^\infty [g_{\tb}(s)-1]\sinc(qs)s^3\,ds,
\end{align}
so that $\Delta C^{V_{\tVm},V_{\tVb}}_{\tb,\tb}(q) = N_\text{bm}\Delta I^*_\text{bm}(q)S_\text{cross}(q)$. Unlike $S_\text{interior}$, this $S_\text{cross}$ contains no self-pair contribution---the cross-region geometry forbids $i = j$ pairs---and consequently lacks the $+1$ baseline of the conventional structure factor. \\

\begin{figure}
	\centering
	\caption{
		The structure factors of the interior (solid) and cross-term (dashed) for spherical modeling volumes of different radii. The interior approaches the bulk-water structure factor as $R$ grows; the cross-term sits on a zero baseline and carries the solvation-peak structure of $g_\text{OO}$ that the small-$R$ interior has not yet developed. The two structure factors combine additively into an effective total factor, with only the total constrained to be positive, which is why the cross-term can dip below zero.
	}
	\label{fig:theory:correlation_functions}
	\includegraphics[width=\columnwidth]{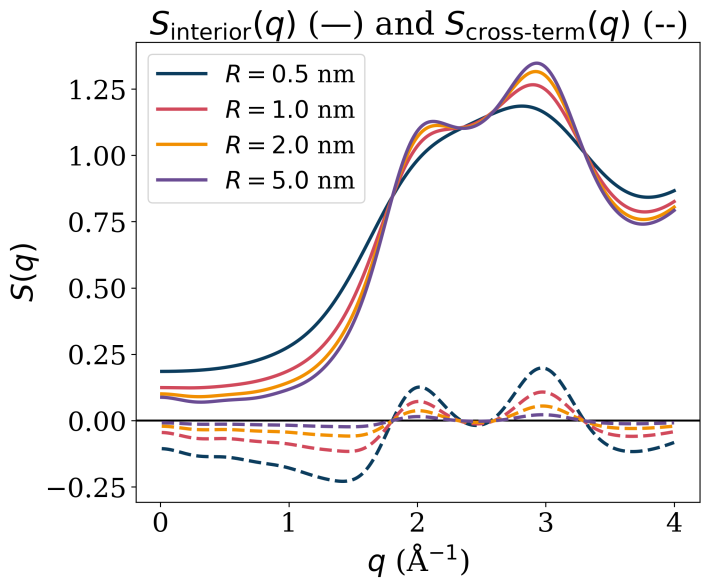}
\end{figure}

To evaluate these two contributions, we computed the oxygen-oxygen radial distribution function $g_\text{OO}$ from a $1\,$µs bulk-solvent MD simulation with the TIP4P/2005 water model~\cite{Abascal2005}. This $g_\text{OO}$ was used to evaluate both the interior structure factor (\eref{eq:si:interior_Sq}) and the cross-term (\eref{eq:appendix:cross_term_sphere}) between the interior sphere 
and a concentric shell of thickness $1.5\,$nm, beyond which $g_\text{OO}(s)$ has 
decayed to its bulk value. The result, shown in \fref{fig:theory:correlation_functions}, confirms that the cross-term is nonzero and oscillatory across the full $q$-range. The SI quantifies the cross-term's relative magnitude and its $1/R$ scaling for the bulk-only system analyzed here (\ref{sec:si:cross_term_scaling}); the molecular case requires separate consideration and is discussed there as well.

\section{Expanding the amplitude into spherical
	harmonics}\label{sec:appendix:spherical_harmonics_expansion}

\begin{figure}
	\centering
	\caption{Visualization of the first few spherical harmonic functions. Figure
		adapted from~\citeasnoun{SphericalHarmonicsVisual}.}
	\label{fig:theory:spherical_harmonics}
	\includesvg[width=\linewidth]{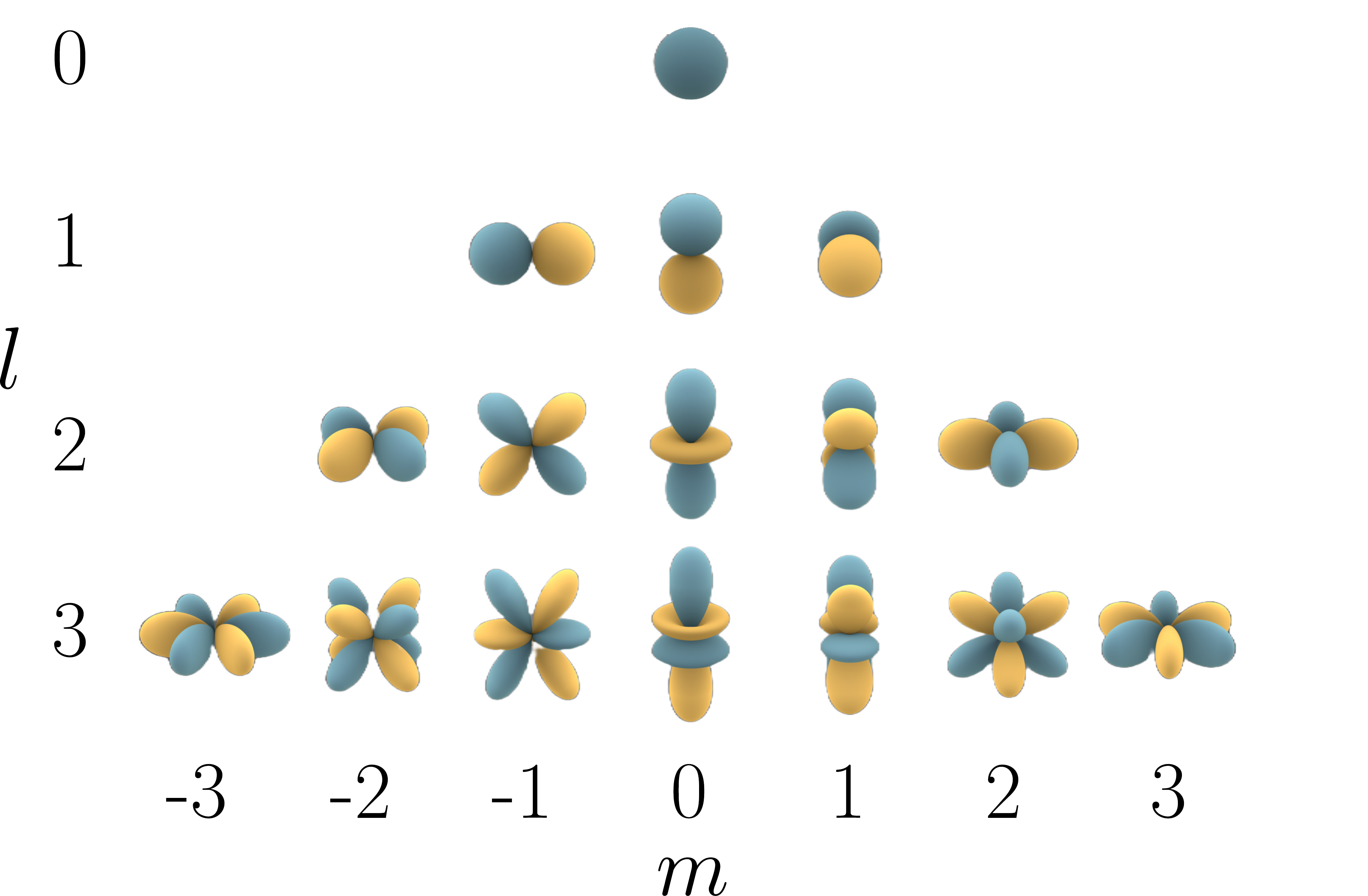}
\end{figure}

In \sref{sec:vacuum:debye}, we derived the Debye equation, which expresses the
orientationally averaged scattering intensity of a representative molecular
structure. A popular alternative is to expand the scattering amplitude in a
series of \textit{spherical harmonics}. These functions form a complete
orthogonal basis on the surface of a sphere, dependent only on the angular
coordinates $(\theta, \phi)$. The harmonic basis functions are indexed by the
degree $l$ (controlling the overall angular variation) and the order $m$
(labeling the dependence on the azimuthal angle)\footnote{As $m$ is an index, it
	is written in cursive ($m$), while the ``single molecule'' label (m) is upright.}.
\fref{fig:theory:spherical_harmonics} illustrates the first few harmonics,
showing how angular variation increases with increasing $l$. \\

Lord Rayleigh showed that a plane wave may be expanded in terms of spherical
harmonic functions as:
\begin{align*}
	e^{i\bm{q}\cdot{}\bm{r}} = 4\pi \sum^\infty_{l=0}\sum_{m=-l}^l i^l j_l(qr)
	Y^m_l(\hat{\bm{q}})\bar{Y}^m_l(\hat{\bm{r}})
\end{align*}
where $j_l(qr)$ are spherical Bessel functions and $Y^m_l(\hat{\bm{r}})$ are
spherical harmonics. \\

Substituting this into the definition of the molecular scattering amplitude
gives:
\begin{align*}
	A^*_{\tm}(\bm{q})
	&= \int_{V_{\tVm}} \rho^*_{\tm}(\bm{r}) e^{i\bm{q}\cdot{}\bm{r}}d\bm{r} \\
	&= 4\pi\sum^\infty_{l=0}\sum_{m=-l}^l i^lY^m_l(\hat{\bm{q}}) \int_{V_{\tVm}}
	\rho^*_{\tm}(\bm{r}) j_l(qr)\bar{Y}^m_l(\hat{\bm{r}})d\bm{r} \\
	&= \sum^\infty_{l=0}\sum_{m=-l}^l Y^m_l(\hat{\bm{q}}) A^m_l(q)
\end{align*}
where the coefficients are:
\begin{align*}
	A^m_l(q) \equiv 4\pi i^l \int_{V_{\tVm}} \rho^*_{\tm}(\bm{r})
	j_l(qr)\bar{Y}^m_l(\hat{\bm{r}})d\bm{r}
\end{align*}
Squaring and averaging over orientations gives:
\begin{align}\label{eq:Iq_spherical}
	I_{\tm}(q)
	= \langle \left|A^*_{\tm}(\bm{q})\right|^2\rangle_{\Omega_q}
	&= \frac{1}{4\pi}\int_\Omega \left|A^*_{\tm}(\bm{q})\right|^2 d\Omega_q \nonumber\\
	&= \sum^\infty_{l=0}\sum_{m=-l}^{l}\left|A^{m}_{l}(q)\right|^2
\end{align}
where the angular dependence vanishes due to the orthogonality of spherical
harmonics. Thus, the intensity is expressed as the squared sum of the expansion
coefficients. 

\subsection{Practically evaluating the spherical harmonic series}

Following~\citeasnoun{Stuhrmann1970}, the coefficients can be computed as a sum
over atoms:
\begin{align*}
	A^m_l(q)=4\pi i^l\sum^N_jf_j(q)j_l(qr_j)\bar{Y}^m_l(\hat{\bm{r}}_j)
\end{align*}
The function $A_l^m(q)$ describes how the atomic form factors $f_j(q)$, each
modulated by the spherical Bessel function $j_l(qr_j)$, contribute to the
overall scattering amplitude. Since $j_l(qr_j)$ is the only term dependent on
$r_j$, it accounts for the radial variation as a function of $q$. \\

As the spherical Bessel functions are defined by:
\begin{align*}
	j_l(x) \equiv (-x)^l\left(\frac{1}{x}\frac{d}{dx}\right)^l \frac{\sin x}{x}
\end{align*}
the amplitude acquires a $q^l$ dependence at small $q$, so only low-order
harmonics contribute significantly in this regime. At larger $q$, higher-order
harmonics begin to contribute, allowing finer structural details to be
resolved.\\

To use this method in practice, the series must be truncated at some maximum
$l$, which depends on the desired resolution~\cite{Gumerov2012}. The benefit of
this widely-used approximation is that \eref{eq:Iq_spherical} is in principle
linear in the number of atoms for fixed $l$, while the Debye equation is
quadratic. This comparison is deceptive, however, as to maintain accuracy, $l$
cannot be kept fixed for higher angles, leading to a similar scaling of the two
approaches~\cite{Kofinger2013,Gumerov2012}. \\

Notably, spherical harmonics are not the only possible basis; they are only used because
of their orthogonality on the sphere and natural alignment with angular
features. Other bases, such as three-dimensional Zernike polynomials, can also
be employed. The choice depends on the geometry and desired convergence
properties: spherical harmonics often work well for smoothly varying shapes,
whereas Zernike polynomials may offer advantages for structures with internal
cavities or complex surfaces~\cite{Liu2012}. 

\section{Conformational averaging of the
	buffer}\label{sec:appendix:buffer_average_t_is_orientational}

In \sref{sec:solv:simplify_buffer}, we argued that the buffer may be expressed
in terms of locally correlated volumes. Here, we prove this statement.
For clarity, the time dependence of volumes and position vectors is suppressed.
\\

The excess instantaneous system intensity is given as:
\begin{align*}
	\Delta I^{(V_{\tVb})}_{\tb}(\bm{q},t) 
	&= |\Delta A^{(V_{\tVb})}_{\tb}(\bm{q},t)|^2 \\
	&= \int_{V_{\tVb}} \Delta\rho_{\tb}(\bm{r}',t)e^{i\bm{q}\cdot\bm{r}'}d\bm{r}'
	\int_{V_{\tVb}} \Delta\rho_{\tb}(\bm{r},t) e^{-i\bm{q}\cdot\bm{r}}d\bm{r}
\end{align*}
Expressing \textit{only} the outer integral as a sum using \eref{eq:Aq_sum}:
\begin{align*}
	\Delta I^{(V_{\tVb})}_{\tb}(\bm{q},t) 
	&= \sum^N_ie^{i\bm{q}\cdot\bm{r}_i}\Delta f_i(\bm{q}) \int_{V_{\tVb}}
	\Delta\rho_{\tb}(\bm{r},t) e^{-i\bm{q}\cdot\bm{r}}d\bm{r}
\end{align*}
where $\Delta f(\bm{q})$ now denotes an excess form factor. Assuming short‑range
solvent correlations, the buffer region around solvent molecule $i$ may be
partitioned into its locally correlated environment $V_\xi(i)$ and its
uncorrelated complement $V_\xi^\complement(i)$:
\begin{align*}
	\Delta I^{(V_{\tVb})}_{\tb}(\bm{q},t)
	= \sum^N_i \Delta f_i(\bm{q}) 
	&\bigg\{ \int_{V_\text{\textxi}(i)} \Delta\rho_{\tb}(\bm{r},t)
	e^{i\bm{q}\cdot(\bm{r}_i - \bm{r})}d\bm{r}\\
	&\quad+ \int_{V^\complement_\text{\textxi}(i)}  \Delta\rho_{\tb}(\bm{r},t)
	e^{i\bm{q}\cdot(\bm{r}_i - \bm{r})}d\bm{r}\bigg\}\\
	= \sum^N_i \Delta f_i(\bm{q}) 
	&\bigg\{ \int_{V_\text{\textxi}(i)} \Delta\rho_{\tb}(\bm{r},t)
	e^{i\bm{q}\cdot(\bm{r}_i - \bm{r})}d\bm{r}\\
	&\quad+ \sum^{\{V^\complement_\text{\textxi}(i)\}}_j \Delta f_j(\bm{q})
	e^{i\bm{q}\cdot\bm{r}_{ij}} \bigg\}
\end{align*}
where the second sum iterates over all molecules in the uncorrelated region. As
in the derivation of \eref{eq:Iq_uncorrelated}, the separation vectors
$\bm{r}_{ij}$ in the uncorrelated region are randomly distributed, generating a
random walk in the complex plane. For sufficiently large $V_\xi^\complement(i)$,
the uncorrelated sum vanishes:
\begin{align*}
	\Delta I^{(V_{\tVb})}_{\tb}(\bm{q},t)
	= \sum^N_i \Delta f_i(\bm{q}) \bigg\{ \int_{V_\text{\textxi}(i)}
	\Delta\rho_{\tb}(\bm{r}) e^{i\bm{q}\cdot(\bm{r}_i - \bm{r})}d\bm{r}\bigg\} \\
	= \sum^N_i\sum^{\{V_\text{\textxi}(i)\}}_j \Delta f_i(\bm{q})\Delta f_j(\bm{q})
	e^{i\bm{q}\cdot\bm{r}_{ij}}
\end{align*}
Because the buffer is at equilibrium, all time‑evolved conformational states are
simultaneously populated, and each solvent molecule samples the same equilibrium
distribution of local environments. As in \sref{sec:uncorrelated_molecules}, the
outer sum therefore acts as an average over molecules with---on average---identical
environments:
\begin{align}
	\Delta I^{(V_{\tVb})}_{\tb}(\bm{q},t)
	&= \sum^N_i \Delta f_i(\bm{q})\sum^{\{V_\text{\textxi}(i)\}}_j \Delta
	f_j(\bm{q}) e^{i\bm{q}\cdot\bm{r}_{ij}} \nonumber\\
	&\approx N \bigg\langle \Delta f^*(\bm{q})\sum^{\{V_\text{\textxi}(i)\}}_j
	\Delta f_j(\bm{q}) e^{i\bm{q}\cdot\bm{r}_{ij}} \bigg\rangle_N \nonumber\\
	&= \Delta I^{(V_{\tVb})}_{\tb}(q) \label{eq:appendix:buffer_avg:final}
\end{align}
where both the vector‑$\bm{q}$ and time dependence vanish as argued in
\sref{sec:solv:simplify_molecule}. For multiple solvent species, the outer sum
separates naturally into species‑specific terms of the same form.\\

Intuitively, the buffer may be viewed as a collection of \textit{overlapping}
correlation regions centered on each solvent molecule. These locally correlated
volumes are the primary contributors to the scattering and are the ones over
which the conformational average is performed.

\section{Equivalence to Park \textit{et al.}}\label{sec:appendix:equivalence_to_park}

Several papers concerned with the development of MD-based SAS calculations
are based on the seminal paper by~\citeasnoun{Park2009}. We show that
their main result is equivalent to \eref{eq:md:match_measurement}. For
context, their $\tilde{A}_1(\bm{q})$ and $\tilde{B}_1(\bm{q})$ refer to the
total scattering amplitude of the simulated volume $V_{\tVm}$ in systems A and
B, respectively. \\

\begin{table}
	\caption{Overview of how our notation corresponds to that of
		~\citeasnoun{Park2009}. }
	\label{table:md:notation}
	\begin{tabular}{l|llll}
		Notation & Density & Amplitude & Intensity & Averaging \\\hline
		Our work & $\rho(\bm{r})$ & $A(\bm{q})$ & $I(q)$ & $\langle\cdot\rangle_N$
		(conformations) \\
		Park \textit{et al.} & $A(\bm{r})$ & $\tilde{A}(\bm{q})$ & $I(q)$ &
		$\langle\cdot\rangle_w$ (conformations)
	\end{tabular}
\end{table}

Using our notation (\tref{table:md:notation}), their key result is Eq.~26 in their paper:
\begin{align*}
	\tilde{I}(q) = N\langle D_{11}(\bm{q}) \rangle_\Omega
\end{align*}
where:
\onecolumn
\rule[-1ex]{.49\columnwidth}{1pt}\rule[-1ex]{1pt}{1.5ex}
\begin{align}\label{eq:md:D_ours}
	D_{11}(\bm{q}) 
	= \big|\langle A_{\tm}(\bm{q},t)\rangle_N 
	- \langle A^{(V_{\tVm})}_{\tb}(\bm{q},t)\rangle_N\big|^2 
	+ \big[\langle| A_{\tm}(\bm{q},t)|^2\rangle_N 
	- |\langle A_{\tm}(\bm{q},t)\rangle_N|^2\big] 
	-\big[\langle| A^{(V_{\tVm})}_{\tb}(\bm{q},t)|^2\rangle_N 
	- |\langle A^{(V_{\tVm})}_{\tb}(\bm{q},t)\rangle_N|^2\big]
\end{align}
\hfill\rule[1ex]{1pt}{1.5ex}\rule[2.3ex]{.49\columnwidth}{1pt}
\twocolumn

Here, $A_{\tm}(\bm{q},t)$ is the amplitude due to the \textit{absolute}
molecular density $\rho_{\tm}(\bm{r},t)$, and $A^{(V_{\tVm})}_{\tb}(\bm{q},t)$
the amplitude due to the \textit{absolute} buffer density, $\rho_{\tb}(\bm{r},t)
= \rho_{\tb} + \Delta \rho_{\tb}(\bm{r},t)$. To simplify the notation in the
following derivations, we define: 
\begin{align*}
	A^{(V_{\tVm})}_{\tb}(\bm{q},t) 
	&= \int_{V_{\tVm}} \rho_{\tb} e^{i\bm{q}\cdot\bm{r}}d\bm{r} + \int_{V_{\tVm}}
	\Delta \rho_{\tb}(\bm{r},t) e^{i\bm{q}\cdot\bm{r}}d\bm{r} \\
	&\equiv \Theta(\bm{q}) + \mathcal{F}(\bm{q},t)
\end{align*}
where the symbols $\Theta(\bm{q})$ and $\mathcal{F}(\bm{q},t)$ are used to more
clearly indicate later substitutions. Written this way, the amplitude has the
two special properties: 
\begin{align*}
	\text{i)} &\ \langle A^{(V_{\tVm})}_{\tb}(\bm{q},t)\rangle_N 
	= \Theta(q) 
	+ \cancel{\langle \mathcal{F}(\bm{q},t)\rangle_N}
	\intertext{since the fluctuations average to zero over all conformations. This
		 implies:}
	\text{ii)} &\ \langle | A^{(V_{\tVm})}_{\tb}(\bm{q},t)|^2 \rangle_N 
	= \langle|\Theta(\bm{q})|^2\rangle_N
	+ \langle\left|\mathcal{F}(\bm{q},t)\right|^2\rangle_N\\
	&\hspace*{2.35cm} 
	-
	\cancel{\langle\Theta(\bm{q})\rangle_N\langle\overline{\mathcal{F}(\bm{q},t)}\rangle}_N
	- \cancel{\langle\mathcal{F}(\bm{q},t)\rangle_N
		\langle\overline{\Theta(\bm{q})}\rangle_N}
\end{align*}
which follows since the average of the product
$\langle\mathcal{F}(\bm{q},t)\Theta(\bm{q})\rangle_N$ factorizes into the
product of averages. The first term in \eref{eq:md:D_ours} may be expanded in
terms of these as:

\onecolumn
\rule[-1ex]{.49\columnwidth}{1pt}\rule[-1ex]{1pt}{1.5ex}
\begin{align*}
	\big|\langle A_{\tm}(\bm{q},t)\rangle_N 
	- \langle A^{(V_{\tVm})}_{\tb}(\bm{q},t)\rangle_N\big|^2
	&=|\langle A_{\tm}(\bm{q},t)\rangle_N|^2
	+ |\langle A^{(V_{\tVm})}_{\tb}(\bm{q},t)\rangle_N|^2
	- \langle A_{\tm}(\bm{q},t)\rangle_N\langle
	\overline{A^{(V_{\tVm})}_{\tb}(\bm{q},t)}\rangle_N 
	- \langle A^{(V_{\tVm})}_{\tb}(\bm{q},t)\rangle_N\langle
	\overline{A_{\tm}(\bm{q},t)}\rangle_N \\
	&=|\langle A_{\tm}(\bm{q},t)\rangle_N|^2
	+ \langle|\Theta(\bm{q})|^2\rangle_N 
	- \langle A_{\tm}(\bm{q},t)\rangle_N \langle\overline{\Theta(\bm{q})}\rangle_N
	- \langle\Theta(\bm{q})\rangle_N \langle \overline{A_{\tm}(\bm{q},t)}\rangle_N
\end{align*}
where we have once again used the independence of $\Theta(\bm{q})$ to distribute
the average. Inserting this expansion back into \eref{eq:md:D_ours}:
\begin{align*}
	D_{11}(\bm{q}) 
	&=\big[\cancel{\left|\langle A_{\tm}(\bm{q},t)\rangle_N\right|^2} 
	+ \langle|\Theta(\bm{q})|^2\rangle_N 
	- \langle A_{\tm}(\bm{q},t)\rangle_N \langle\overline{\Theta(\bm{q})}\rangle_N
	- \langle\Theta(\bm{q})\rangle_N \langle
	\overline{A_{\tm}(\bm{q},t)}\rangle_N\big] 
	+ \big[\langle| A_{\tm}(\bm{q},t)|^2\big\rangle_N 
	- \cancel{|\langle A_{\tm}(\bm{q},t)\rangle_N|^2}\big] \\
	&\hspace*{11.1cm}
	- \big[\cancel{\langle|\Theta(\bm{q})|^2\rangle_N} 
	+ \langle|\mathcal{F}(\bm{q},t)|^2\rangle_N 
	- \cancel{\langle|\Theta(\bm{q})|^2\rangle_N}\big] \\
	&=\big[\big\langle| A_{\tm}(\bm{q},t)|^2\big\rangle_N 
	+ \langle|\Theta(\bm{q})|^2\rangle_N
	- \langle A_{\tm}(\bm{q},t)\rangle_N \langle\overline{\Theta(\bm{q})}\rangle_N
	- \langle\Theta(\bm{q})\rangle_N \langle
	\overline{A_{\tm}(\bm{q},t)}\rangle_N\big] 
	- \langle|\mathcal{F}(\bm{q},t)|^2\rangle_N \\
	&=\langle| A_{\tm}(\bm{q},t) 
	- \Theta(\bm{q}) |^2\rangle_N
	- \langle|\mathcal{F}(\bm{q},t)|^2\rangle_N \\
	&= \Delta I_{\tm}(q) 
	- \Delta I^{(V_{\tVm})}_{\tb}(q) \\
	&= \Delta I^\text{(MD)}_\text{A}(q) 
	- \Delta I^\text{(MD)}_\text{B}(q)
\end{align*}
\hfill\rule[1ex]{1pt}{1.5ex}\rule[2.3ex]{.49\columnwidth}{1pt}
\twocolumn
Here, we have identified the first term as the excess intensity, as the
difference of amplitudes, when integrated over the same volume $V_{\tVm}$, can
be expressed as a difference of densities. The second term was similarly
identified as the excess buffer scattering from system B. This result is
equivalent to \eref{eq:md:match_measurement} followed from our formalism.

%% file: supplementary_information.tex
\renewcommand{\thefigure}{S\arabic{figure}}
\renewcommand{\theequation}{S\arabic{equation}}
\renewcommand{\thetable}{S\arabic{table}}
\renewcommand{\thesection}{S\arabic{section}}

\setcounter{section}{0}
\renewcommand{\appendixname}{Supplementary Information}
\section{Evaluating the cross-boundary volume integral for spherical domains}
\label{sec:si:cross_term_sphere}

In \aref{sec:appendix:mdgamma}, the cross-term was reduced to the double volume
integral:
\begin{align*}
	\mathcal{I}(q)
	\equiv n_b^2 \int_{V_{\tVm}}\int_{V_{\tVb}} g_{\tb}(|\bm{r}_1 -
	\bm{r}_2|)\sinc(q|\bm{r}_1-\bm{r}_2|)d\bm{r}_2 d\bm{r}_1
\end{align*}
We evaluate this integral for a spherical modeling volume $V_{\tVm}$ of radius
$R$ with its concentric complement $V_{\tVb}$.\\

To expose the geometric structure of the integral, we introduce indicator
functions $f_{\tVm}(\bm{r}) = \mathbf{1}[\bm{r}\in V_{\tVm}]$ and
$f_{\tVb}(\bm{r}) = \mathbf{1}[\bm{r}\in V_{\tVb}]$, which are unity inside
their respective domains and zero elsewhere. Multiplying by these functions
encodes the domain restrictions as integrand factors rather than integration
limits, allowing both integrals to be extended freely to all of $V_{\tVs}$:
\begin{align*}
	\mathcal{I}(q)
	= n_b^2 &\iint_{V_{\tVs}}f_{\tVm}(\bm{r}_1) f_{\tVb}(\bm{r}_2)
	g_{\tb}(|\bm{r}_1 - \bm{r}_2|)\\
	&\phantom{n_b^2 \iint_{V_{\tVs}}f_{\tVm}}\times\sinc(q|\bm{r}_1-\bm{r}_2|)d\bm{r}_2\,d\bm{r}_1
\end{align*}
Substituting $\bm{s} = \bm{r}_1 - \bm{r}_2$:
\begin{align}\label{eq:si:radial}
	\mathcal{I}(q)
	= n_b^2 \int_{V_{\tVs}} g_{\tb}(s)\sinc(qs)\underbrace{\left[\int_{V_{\tVs}}
		f_{\tVm}(\bm{r}_1) f_{\tVb}(\bm{r}_1-\bm{s}) d\bm{r}_1\right]}_{[f_{\tVm}\star
		f_{\tVb}](s)}d\bm{s}
\end{align}
The bracketed inner integral is the cross-correlation of the two indicator
functions at separation $\bm{s}$, counting the total volume of point pairs
separated by $\bm{s}$ with one point in $V_{\tVm}$ and one in $V_{\tVb}$. By the
rotational symmetry of the concentric geometry it depends only on $s =
|\bm{s}|$, so the integral reduces to:
\begin{align*}
	\mathcal{I}(q)
	= n_b^2\int_0^\infty 4\pi s^2g_{\tb}(s)\sinc(qs)[f_{\tVm}\star f_{\tVb}](s) ds
\end{align*}
Since $V_{\tVb} = V_{\tVs} - V_{\tVm}$, the indicator functions satisfy
$f_{\tVb} = 1 - f_{\tVm}$, giving the decomposition:
\begin{align}\label{eq:si:decomposition}
	[f_{\tVm}\star f_{\tVb}](s)
	= \underbrace{[f_{\tVm}\star f_{\tVs}](s)}_{\mathcal{K}_\text{I}(s)}-
	\underbrace{[f_{\tVm}\star f_{\tVm}](s)}_{\mathcal{K}_\text{II}(s)}
\end{align}
The two terms are now evaluated separately.

\subsection{The first term}\hfill

Since $V_{\tVs}$ is taken to be all of space, $[f_{\tVm}\star f_{\tVs}](s)$
counts the volume of all points within $V_{\tVm}$ for any displacement $s$,
which is simply $V_{\tVm}$:
\begin{align*}
	\mathcal{K}_\text{I}(s) = V_{\tVm}
\end{align*}
The corresponding contribution to \eref{eq:si:radial} is:
\begin{align*}
	\mathcal{I}_\text{I}(q)
	= n_b^2 V_{\tVm}\int_0^\infty 4\pi s^2 g_{\tb}(s) \sinc(qs) ds
\end{align*}

\subsection{The second term}\hfill

The second term involves the autocorrelation of the indicator function of
$V_{\tVm}$:
\begin{align*}
	\mathcal{K}_\text{II}(s) = [f_{\tVm}\star f_{\tVm}](s)
	= \int f_{\tVm}(\bm{r}) f_{\tVm}(\bm{r}-\bm{s}) d\bm{r}
\end{align*}
This counts the volume of the intersection of $V_{\tVm}$ with a copy of itself
displaced by $\bm{s}$---the \textit{lens volume} of two identical spheres of
radius $R$ whose centres are separated by $s$.\footnote{This is commonly
	referred to as the "ghost image" in the small-angle scattering
	literature~\cite{Glatter1982,Feigin1987}.} By rotational symmetry it depends
only on $s$ and vanishes for $s > 2R$. The result is the standard geometric
formula:
\begin{align}\label{eq:si:lens}
	\Gamma(s) \equiv [f_{\tVm}\star f_{\tVm}](s)
	= V_{\tVm} \left(1 - \frac{3s}{4R} + \frac{s^3}{16R^3}\right) \qquad s \leq 2R
\end{align}
and $\Gamma(s)=0$ for $s>2R$. Substituting into \eref{eq:si:radial}:
\begin{align*}
	\mathcal{I}_\text{II}(q)
	= n_b^2\int_0^\infty 4\pi s^2 g_{\tb}(s) \sinc(qs) \Gamma(s) ds
\end{align*}

\subsection{Combining and isolating the correlation contribution}\hfill

Taking the difference from \eref{eq:si:decomposition}:
\begin{align}\label{eq:si:combined}
	\mathcal{I}(q)
	= n_b^2\int_0^\infty 4\pi s^2\,g_{\tb}(s)\,\sinc(qs)
	\bigl[V_{\tVm} - \Gamma(s)\bigr]\,ds
\end{align}
Substituting \eref{eq:si:lens} gives:
\begin{align*}
	V_{\tVm} - \Gamma(s)
	= V_{\tVm} \left(\frac{3s}{4R} - \frac{s^3}{16R^3}\right)
	= \pi R^2 s - \frac{\pi}{12}s^3
\end{align*}
To separate the structural contribution from the purely geometric one, we expand
$g_{\tb}(s) = [g_{\tb}(s) - 1] + 1$ as in \aref{sec:appendix:correlation_funcs}:
\begin{align}\label{eq:si:split}
	\mathcal{I}(q)
	&= n_b^2\int_0^\infty 4\pi s^2\,[g_{\tb}(s)-1]\,\sinc(qs)\bigl[V_{\tVm} -
	\Gamma(s)\bigr]\,ds \nonumber\\
	&\quad + n_b^2\int_0^\infty 4\pi s^2\,\sinc(qs)\bigl[V_{\tVm} -
	\Gamma(s)\bigr]\,ds
\end{align}
The second integral is a purely geometric boundary term. Since $\Gamma(s)$ is by
construction the autocorrelation of $f_{\tVm}$, and $\mathcal{S}_{V_{\tVm}}(q) =
\int f_{\tVm}(\bm{r})\,e^{i\bm{q}\cdot\bm{r}}d\bm{r}$ is its Fourier transform,
by the convolution theorem, the Fourier transform of $\Gamma(s)$ is
$|\mathcal{S}_{V_{\tVm}}(q)|^2$. The two parts of the geometric term therefore
evaluate to:
\begin{align*}
	&n_b^2\!\int_0^\infty 4\pi s^2 \sinc(qs)\,\bigl[V_{\tVm} - \Gamma(s)\bigr]\,ds\\
	&\quad= n_b^2\bigl[V_{\tVm} \mathcal{S}_{V_{\tVm}}(q) -
	|\mathcal{S}_{V_{\tVm}}(q)|^2\bigr]
\end{align*}
This term is concentrated near forward scattering for macroscopic volumes and is
dropped in what follows.\\

For the correlation integral in \eref{eq:si:split}, $g_{\tb}(s)-1$ decays to
zero beyond the bulk correlation length $\xi \ll R$, so only $s \lesssim \xi$
contributes. In this regime $V_{\tVm} - \Gamma(s) \approx \pi R^2
s$,\footnote{This amounts to neglecting curvature terms.} and the cubic
correction is negligible:
\begin{align}\label{eq:si:final}
	\mathcal{I}(q) \approx n_b^2 \pi R^2\int_0^\infty 4\pi s^3 [g_{\tb}(s)-1]
	\sinc(qs) ds
\end{align}
Collecting the prefactors from \aref{sec:appendix:mdgamma}:
\begin{align}\label{eq:si:cross_term_final}
	\Delta C^{V_{\tVm},V_{\tVb}}_{\tb,\tb}(q)
	&= 8\pi^2 n_{\tb}^2 R^2 \Delta\beta(q) \Delta I^*_\text{bm}(q)\nonumber\\
	&\quad\times\int_0^\infty s^3
	[g_{\tb}(s)-1] \sinc(qs) ds
\end{align}
The cross-term scales with $R^2$, i.e. the surface area of the modeling volume,
reflecting the fact that only solvent pairs straddling the boundary of
$V_{\tVm}$ contribute to the integral.

\subsection{Comparing with the interior scattering}

To assess the relative importance of the cross-term, we compare it with the
interior scattering $I^*_{\tm}(q) = N I^*_\text{bm}(q) S(q)$. Using
\eref{eq:appendix:Sq_exact} with $\Delta\beta(q)$ and dropping the shape-factor
term:
\begin{align}\label{eq:si:interior_Sq}
	I^*_{\tm}(q)
	&= N\Delta I^*_\text{bm}(q)\nonumber\\
	&\quad\times\left\{1 + 4\pi n_b \Delta\beta(q)\int_0^\infty
	s^2[g_{\tb}(s)-1]\sinc(qs)ds\right\}
\end{align}
Comparing \eref{eq:si:interior_Sq} with \eref{eq:si:cross_term_final}, the ratio
of the cross-term to the interior scattering is:
\begin{align}\label{eq:si:ratio}
	\frac{\Delta C^{V_{\tVm},V_{\tVb}}_{\tb,\tb}(q)}{I^*_{\tm}(q)}
		&=\frac{8\pi^2 n_{\tb}^2 R^2\Delta\beta(q)}{N1 + 4\pi
		n_b\Delta\beta(q)}\nonumber\\
		&\quad\times \frac{\displaystyle\int_0^\infty
		s^3[g_{\tb}(s)-1]\sinc(qs)ds}{\displaystyle\int_0^\infty
		s^2[g_{\tb}(s)-1]\sinc(qs)ds}
\end{align}
Using $N = 4\pi n_{\tb} R^3/3$, the prefactor simplifies to $6\pi
n_{\tb}\Delta\beta(q)/R$.\\

The denominator of \eref{eq:si:ratio} interpolates between two physically
distinct regimes. At low $q$, the structure-factor correction is negligible and
the denominator is dominated by the form-factor term: $S(q)\approx 1$. In this
regime the ratio reduces to:
\begin{align*}
	\frac{\Delta C^{V_{\tVm},V_{\tVb}}_{\tb,\tb}(q)}{I^*_{\tm}(q)}
	\,\xrightarrow{q\to 0}\,&\frac{6\pi n_{\tb}}{R}\int_0^\infty
	s^3[g_{\tb}(s)-1]ds\\
	\,\sim\, &\frac{n_{\tb}\xi^4}{R}
\end{align*}
where $\Delta\beta(q) \rightarrow 1$. \\

$\Delta I^*_\text{bm}(q)$ decays rapidly with $q$ as it is the Fourier transform
of a smooth single-molecule density. The observed buffer scattering, however,
retains structure at wider angles---the nearest-neighbor peaks of liquid water,
for example. This structure originates from $g_{\tb}(s)-1$ entering through the
structure factor, so at high $q$ the denominator of \eref{eq:si:ratio} is
dominated by the correlation integral rather than the form-factor term. Using
this approximation:
\begin{align*}
	\frac{\Delta C^{V_{\tVm},V_{\tVb}}_{\tb,\tb}(q)}{I^*_{\tm}(q)}
	\,\longrightarrow\,&\frac{\displaystyle\frac{6\pi n_{\tb}}{R}\int_0^\infty
		s^3[g_{\tb}(s)-1]\sinc(qs)ds}
	{4\pi n_b \Delta\beta(q)\displaystyle\int_0^\infty
		s^2\,[g_{\tb}(s)-1]\,\sinc(qs)\,ds} \\
	\,\sim\,&\frac{3\xi}{2R}
\end{align*}
In both limits the cross-term is suppressed relative to the interior scattering
by a factor $\xi/R$, the ratio of the bulk correlation length to the radius of
the modeling volume. For typical aqueous buffers ($\xi\sim 3$--$5\,\text{\AA}$)
and modeling volumes of radius $R\sim 30$--$100\,\text{\AA}$, this amounts to a
suppression of one to two orders of magnitude, confirming that the cross-term is
a small but nonzero correction that scales with the surface area $4\pi R^2$ of
$V_{\tVm}$.

\subsection{Numerical evaluation of the cross-term}
\label{sec:si:cross_term_scaling}

\begin{figure*}
	\centering
	\caption{
		Numerical evaluation of the cross-term structure factor for spherical $V_{\tVm}$. \textit{Left:} Reproduced from \fref{fig:theory:correlation_functions} for reference. \textit{Middle:} Ratio $S_\text{cross}(q)/S_\text{interior}(q)$, demonstrating that the cross-term is a comparable fraction of the interior structure factor across $q$. \textit{Right:} Magnitude of the same ratio versus $R$ at fixed $q$; dashed lines have slope $-1$, confirming the $\xi/R$ scaling derived above. Deviations at small $R$ correspond to the breakdown of the $\xi \ll R$ approximation.
	}
	\label{fig:si:cross_term_full}
	\includegraphics[width=\linewidth]{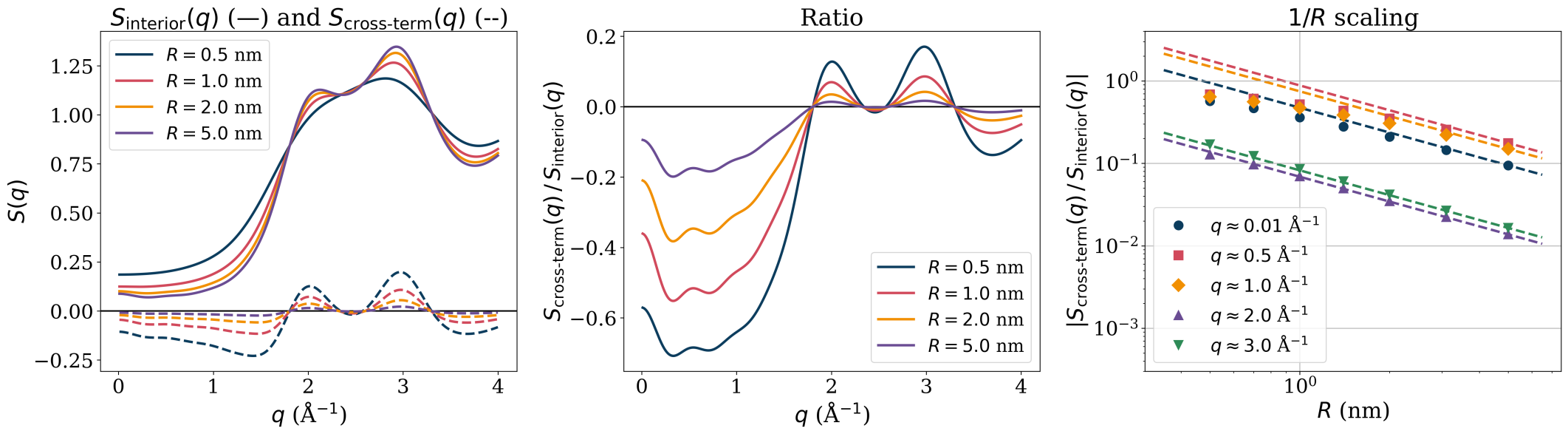}
\end{figure*}

The structure factor of the interior \eref{eq:si:interior_Sq} and the cross-term \eref{eq:si:cross_term_final} were evaluated using $g_\text{OO}$ from a $1\,\mu$s bulk-solvent simulation with TIP4P/2005~\cite{Abascal2005}, with the cross-term integrated against a concentric shell of $1.5\,$nm thickness. The results are shown in \fref{fig:si:cross_term_full}.

\subsubsection{Relative importance in the molecular system}

The $\xi/R$ scaling derived above is for System B, where $V_{\tVm}$ is filled entirely with bulk solvent. The cross-term can be read as a boundary correction: a finite interior does not scatter exactly like bulk water near its surface, and the cross-term restores the missing correlations. As established in \sref{sec:solv:boundary_term}, this correction is a purely bulk quantity, fixed by the boundary shell and independent of the molecule at the centre, so it is the same in both systems. It loses importance as $R$ grows only because the bulk-solvent interior it corrects scales as $R^3$ while the correction scales as $R^2$. In System A the molecule and its hydration shell occupy the core, leaving only a thin outer shell of bulk-like solvent inside $V_{\tVm}$. The boundary correction is unchanged, but the bulk-solvent interior it is weighed against is now far smaller, so the same correction carries much greater relative weight---qualitatively the low-$R$, or equivalently thin-shell, behaviour of \fref{fig:si:cross_term_full} (right). The all-bulk interior assumed in the figure therefore understates the ratio expected for the molecular system. A quantitative treatment would require a hollow-sphere geometry and lies outside the scope of the present work. \\

The middle panel must be read with care for the molecular system, as it shows a ratio of structure factors rather than of intensities. In the observable, the cross-term enters weighted by the single-molecule buffer intensity $\Delta I^*_\text{bm}(q)$, whereas the molecular signal carries its own form factor, which dominates at low $q$ and decays at wider angles. The cross-term is therefore negligible at low $q$ irrespective of the structure-factor ratio, and is most relevant at wider angles where solvent dominates the scattering.

%% file: bibliography.tex
%\footnotesize
%\twocolumn
\bibliographystyle{apsrmp4-2}
\bibliography{refs/export.bib}

%% file: refs/export.bib
@article{Abascal2005,
   author = {J. L.F. Abascal and C. Vega},
   doi = {10.1063/1.2121687/965459},
   issn = {00219606},
   issue = {23},
   journal = {Journal of Chemical Physics},
   month = {12},
   pmid = {16392929},
   publisher = {American Institute of Physics Inc.},
   title = {A general purpose model for the condensed phases of water: TIP4P/2005},
   volume = {123},
   url = {/aip/jcp/article/123/23/234505/965459/A-general-purpose-model-for-the-condensed-phases},
   year = {2005}
}

@article{Zernike1927,
   author = {F. Zernike and J. A. Prins},
   doi = {10.1007/BF01391926/METRICS},
   issn = {14346001},
   issue = {6-7},
   journal = {Zeitschrift für Physik},
   month = {6},
   pages = {184-194},
   publisher = {Springer-Verlag},
   title = {Die Beugung von Röntgenstrahlen in Flüssigkeiten als Effekt der Molekülanordnung},
   volume = {41},
   url = {https://link.springer.com/article/10.1007/BF01391926},
   year = {1927}
}

@article{Shinohara2015,
   author = {Yuya Shinohara and Yoshiyuki Amemiya},
   doi = {10.1107/S160057671501715X},
   issn = {1600-5767},
   issue = {6},
   journal = {urn:issn:1600-5767},
   month = {10},
   pages = {1660-1664},
   publisher = {International Union of Crystallography},
   title = {Effect of finite spatial coherence length on small-angle scattering},
   volume = {48},
   url = {//journals.iucr.org/paper?vg5030},
   year = {2015}
}

@article{Barker2015,
   author = {J. G. Barker and D. F.R. Mildner},
   doi = {10.1107/S1600576715010729},
   issn = {1600-5767},
   issue = {4},
   journal = {urn:issn:1600-5767},
   month = {6},
   pages = {1055-1071},
   publisher = {International Union of Crystallography},
   title = {Survey of background scattering from materials found in small-angle neutron scattering},
   volume = {48},
   url = {//journals.iucr.org/paper?ks5470},
   year = {2015}
}

@article{Svaneborg2012,
   author = {Carsten Svaneborg and Jan Skov Pedersen},
   doi = {10.1063/1.3682778},
   issn = {0021-9606},
   issue = {10},
   journal = {The Journal of Chemical Physics},
   month = {3},
   publisher = {AIP Publishing},
   title = {A formalism for scattering of complex composite structures. I. Applications to branched structures of asymmetric sub-units},
   volume = {136},
   url = {https://pubs.aip.org/jcp/article/136/10/104105/190144/A-formalism-for-scattering-of-complex-composite},
   year = {2012}
}

@article{Pedersen1996,
   author = {Jan Skov Pedersen and Peter Schurtenberger},
   doi = {10.1021/ma9607630},
   issn = {0024-9297},
   issue = {23},
   journal = {Macromolecules},
   month = {1},
   pages = {7602-7612},
   publisher = { American Chemical Society },
   title = {Scattering Functions of Semiflexible Polymers with and without Excluded Volume Effects},
   volume = {29},
   url = {https://pubs.acs.org/doi/10.1021/ma9607630},
   year = {1996}
}

@article{Zimm1949,
   author = {Bruno H. Zimm and Walter H. Stockmayer},
   doi = {10.1063/1.1747157},
   issn = {0021-9606},
   issue = {12},
   journal = {The Journal of Chemical Physics},
   month = {12},
   pages = {1301-1314},
   publisher = {AIP Publishing},
   title = {The Dimensions of Chain Molecules Containing Branches and Rings},
   volume = {17},
   url = {https://pubs.aip.org/jcp/article/17/12/1301/72739/The-Dimensions-of-Chain-Molecules-Containing},
   year = {1949}
}

@article{Spinozzi2000,
   author = {F. Spinozzi and F. Carsughi and P. Mariani and C.V. Teixeira and L. Q. Amaral},
   doi = {10.1107/S0021889800099908},
   issn = {0021-8898},
   issue = {3},
   journal = {Journal of Applied Crystallography},
   month = {6},
   pages = {556-559},
   publisher = {International Union of Crystallography},
   title = {SAS from inhomogeneous particles with more than one domain of scattering density and arbitrary shape},
   volume = {33},
   url = {https://journals.iucr.org/paper?S0021889800099908},
   year = {2000}
}

@article{Hansen1990,
   author = {S. Hansen},
   doi = {10.1107/S0021889890002801},
   issn = {00218898},
   issue = {4},
   journal = {Journal of Applied Crystallography},
   month = {8},
   pages = {344-346},
   publisher = {International Union of Crystallography},
   title = {Calculation of small-angle scattering profiles using Monte Carlo simulation},
   volume = {23},
   url = {https://journals.iucr.org/paper?S0021889890002801},
   year = {1990}
}

@article{Deumer2022,
   author = {Jérôme Deumer and Brian R. Pauw and Sylvie Marguet and Dieter Skroblin and Olivier Taché and Michael Krumrey and Christian Gollwitzer},
   doi = {10.1107/S160057672200499X},
   issn = {0021-8898},
   issue = {Pt 4},
   journal = {Journal of applied crystallography},
   month = {8},
   pages = {993-1001},
   pmid = {35974742},
   publisher = {J Appl Crystallogr},
   title = {Small-angle X-ray scattering: characterization of cubic Au nanoparticles using Debye's scattering formula},
   volume = {55},
   url = {https://pubmed.ncbi.nlm.nih.gov/35974742/},
   year = {2022}
}

@article{Pedersen2012,
   author = {Jan Skov Pedersen and Cristiano L.P. Oliveira and Henriette Baun Hübschmann and Lise Arleth and Soren Manniche and Nicolai Kirkby and Hanne Morck Nielsen},
   doi = {10.1016/J.BPJ.2012.03.071},
   issn = {00063495},
   issue = {10},
   journal = {Biophysical Journal},
   month = {5},
   pages = {2372},
   pmid = {22677391},
   title = {Structure of Immune Stimulating Complex Matrices and Immune Stimulating Complexes in Suspension Determined by Small-Angle X-Ray Scattering},
   volume = {102},
   url = {https://pmc.ncbi.nlm.nih.gov/articles/PMC3353015/},
   year = {2012}
}

@article{Porod1953,
   author = {G. Porod},
   doi = {10.1007/BF01513908},
   issn = {0303-402X},
   issue = {1},
   journal = {Kolloid-Zeitschrift},
   month = {10},
   pages = {51-51},
   publisher = {Springer-Verlag},
   title = {Die Röntgenkleinwinkelstreuung von dichtgepackten kolloiden Systemen},
   volume = {133},
   url = {http://link.springer.com/10.1007/BF01513908},
   year = {1953}
}

@article{Guinier1939,
   author = {André Guinier},
   journal = {Annales de la Physique},
   pages = {161-236},
   title = {La Diffraction des rayons X aux très faibles angles: applications à l'étude des phénomènes ultra-microscopiques},
   volume = {12},
   year = {1939}
}

@article{Sears1992,
   author = {Varley F. Sears},
   doi = {10.1080/10448639208218770},
   issn = {1044-8632},
   issue = {3},
   journal = {Neutron News},
   month = {1},
   pages = {26-37},
   title = {Neutron scattering lengths and cross sections},
   volume = {3},
   url = {http://www.tandfonline.com/doi/abs/10.1080/10448639208218770},
   year = {1992}
}

@book{BookLindner2024,
   author = {Peter. Lindner and Julian. Oberdisse},
   isbn = {9780443291166},
   pages = {779},
   publisher = {Elsevier},
   title = {Neutrons, x-rays, and light  : scattering methods applied to soft condensed matter},
   year = {2024}
}

@article{Pedersen1990,
   author = {J. S. Pedersen and D. Posselt and K. Mortensen},
   doi = {10.1107/S0021889890003946},
   issn = {00218898},
   issue = {4},
   journal = {Journal of Applied Crystallography},
   month = {8},
   pages = {321-333},
   publisher = {International Union of Crystallography},
   title = {Analytical treatment of the resolution function for small-angle scattering},
   volume = {23},
   url = {https://journals.iucr.org/paper?S0021889890003946},
   year = {1990}
}

@article{Lytje2025,
   author = {Kristian Lytje and Jan Skov Pedersen},
   doi = {10.1107/S160057672500562X},
   issn = {1600-5767},
   issue = {4},
   journal = {Journal of Applied Crystallography},
   month = {8},
   pages = {1332-1346},
   publisher = {International Union of Crystallography},
   title = {Small-angle X-ray scattering profile calculation for high-resolution models of biomacromolecules},
   volume = {58},
   url = {https://journals.iucr.org/paper?S160057672500562X},
   year = {2025}
}

@article{Krueger2022,
   author = {Susan Krueger},
   doi = {10.1016/J.SBI.2022.102375},
   issn = {1879033X},
   journal = {Current opinion in structural biology},
   month = {6},
   pages = {102375},
   pmid = {35490650},
   publisher = {Elsevier Ltd},
   title = {Small-angle neutron scattering contrast variation studies of biological complexes: Challenges and triumphs},
   volume = {74},
   url = {https://pmc.ncbi.nlm.nih.gov/articles/PMC10988784/},
   year = {2022}
}

@article{Frank2022,
   author = {Frank Gabel and Sylvain Engilberge and Emmanuelle Schmitt and Aurélien Thureau and Yves Mechulam and Javier Pérez and Eric Girard and M. Czjzek},
   doi = {10.1107/S2059798322007392/JC5050SUP1.PDF},
   issn = {20597983},
   issue = {Pt 9},
   journal = {Acta Crystallographica Section D: Structural Biology},
   month = {9},
   pages = {1120-1130},
   pmid = {36048152},
   publisher = {International Union of Crystallography},
   title = {Medical contrast agents as promising tools for biomacromolecular SAXS experiments},
   volume = {78},
   url = {https://journals.iucr.org/paper?jc5050 https://journals.iucr.org/d/issues/2022/09/00/jc5050/},
   year = {2022}
}

@article{Larsen2020,
   author = {Andreas Haahr Larsen and Jan Skov Pedersen and Lise Arleth},
   doi = {10.1107/S1600576720006500/TEXIMAGES/GE5072FI143.GIF},
   issn = {16005767},
   issue = {4},
   journal = {Journal of Applied Crystallography},
   month = {8},
   pages = {991-1005},
   publisher = {International Union of Crystallography},
   title = {Assessment of structure factors for analysis of small-angle scattering data from desired or undesired aggregates},
   volume = {53},
   url = {https://journals.iucr.org/paper?ge5072 https://journals.iucr.org/j/issues/2020/04/00/ge5072/},
   year = {2020}
}

@article{Wiedenmann2001,
   author = {Albrecht Wiedenmann},
   doi = {10.1016/S0921-4526(00)00872-3},
   issn = {0921-4526},
   issue = {1-4},
   journal = {Physica B: Condensed Matter},
   month = {3},
   pages = {226-233},
   publisher = {North-Holland},
   title = {Small-angle neutron scattering investigations of magnetic nanostructures and interfaces using polarized neutrons},
   volume = {297},
   year = {2001}
}

@article{Muhlbauer2019,
   author = {Sebastian Mühlbauer and Dirk Honecker and Élio A. Périgo and Frank Bergner and Sabrina Disch and André Heinemann and Sergey Erokhin and Dmitry Berkov and Chris Leighton and Morten Ring Eskildsen and Andreas Michels},
   doi = {10.1103/REVMODPHYS.91.015004/FIGURES/32/MEDIUM},
   issn = {15390756},
   issue = {1},
   journal = {Reviews of Modern Physics},
   month = {3},
   pages = {015004},
   publisher = {American Physical Society},
   title = {Magnetic small-angle neutron scattering},
   volume = {91},
   url = {https://journals.aps.org/rmp/abstract/10.1103/RevModPhys.91.015004},
   year = {2019}
}

@article{Kofinger2013,
   author = {Jürgen Köfinger and Gerhard Hummer},
   doi = {10.1103/PHYSREVE.87.052712/FIGURES/4/MEDIUM},
   issn = {15393755},
   issue = {5},
   journal = {Physical Review E - Statistical, Nonlinear, and Soft Matter Physics},
   month = {5},
   pages = {052712},
   pmid = {23767571},
   publisher = {American Physical Society},
   title = {Atomic-resolution structural information from scattering experiments on macromolecules in solution},
   volume = {87},
   url = {https://journals.aps.org/pre/abstract/10.1103/PhysRevE.87.052712},
   year = {2013}
}

@article{Pedersen1997,
   author = {Jan Skov Pedersen},
   doi = {10.1016/S0001-8686(97)00312-6},
   issn = {0001-8686},
   issue = {1-3},
   journal = {Advances in Colloid and Interface Science},
   month = {7},
   pages = {171-210},
   publisher = {Elsevier},
   title = {Analysis of small-angle scattering data from colloids and polymer solutions: modeling and least-squares fitting},
   volume = {70},
   year = {1997}
}

@article{Benoit1953,
   author = {H. Benoit},
   doi = {10.1002/POL.1953.120110512},
   issn = {1542-6238},
   issue = {5},
   journal = {Journal of Polymer Science},
   month = {11},
   pages = {507-510},
   publisher = {John Wiley \& Sons, Ltd},
   title = {On the effect of branching and polydispersity on the angular distribution of the light scattered by gaussian coils},
   volume = {11},
   url = {https://onlinelibrary.wiley.com/doi/full/10.1002/pol.1953.120110512 https://onlinelibrary.wiley.com/doi/abs/10.1002/pol.1953.120110512 https://onlinelibrary.wiley.com/doi/10.1002/pol.1953.120110512},
   year = {1953}
}

@article{Li2014,
   author = {Xin Li and Changwoo Do and Yun Liu and Luis Sánchez-Diáz and Gregory Smith and Wei Ren Chen},
   doi = {10.1107/S1600576714022249/HTTPS://JOURNALS.IUCR.ORG/SERVICES/RSS.HTML},
   issn = {16005767},
   issue = {6},
   journal = {Journal of Applied Crystallography},
   month = {12},
   pages = {1901-1905},
   publisher = {International Union of Crystallography},
   title = {A scattering function of star polymers including excluded volume effects},
   volume = {47},
   url = {//journals.iucr.org/paper?fs5081},
   year = {2014}
}

@misc{SasView6.0,
   author = {Michel Adams and Nouhaila Agouzal and Gervaise Alina and Ziggy Attala and Mathieu Doucet and Jurrian Bakker and Peter Beaucage and Jordan Berger and Wim Bouwman and Robert Bourne and Paul Butler and Jae Hie Cho and Iestyn Cadwallader-Jones and Kieran Campbell and Torin Cooper-Benun and James Crake-Merani and Giogos Drosos and Celine Durniak and Chris Farrow and Ricardo Ferraz Leal and Rachel Ford and Laura Forster and Jonathan Gaudet and Mariana Gerina and Peter Gilbert and Miguel Gonzalez and Richard Heenan and Ellis Hewins and Andrew Jackson and Grethe Jensen and Pavol Juhas and Stephen King and Julius Karliczek and Paul Kienzle and Jeff Krzywon and Jiao Lin and Yun Liu and Ruben Lopes and Dorian Lozano and Kristian Lytje and David Mannicke and Brian Maranville and Anders Markvardsen and Nicolas Martinez and Mike McKerns and Brayden Miller and Karolina Mothander and Ryan Murphy and Andrew Nelson and Torben Nielsen and Lewis O'Driscoll and Michael Oakley and Helen Park and Peter Parker and Maria Patrou and Pete Peterson and Wojciech Potrzebowski and Stuart Prescott and Maksim Rakitin and Tobias Richter and Jack Rooks and Piotr Rozyczko and Xael Shan and Tim Snow and Annika Stellhorn and Susana Teixeira and Jessica Tumarkin and Adam Washington and Katie Weigandt and Robert Whitley and Lucas Wilkins and Caitlyn Wolf and R Cortes Hernandez and Anita Zhang and Alex Zheng},
   doi = {10.5281/zenodo.11395968},
   month = {10},
   publisher = {Zenodo},
   title = {SasView version 6.0.0},
   url = {https://doi.org/10.5281/zenodo.11395968},
   year = {2024}
}

@article{Feigin1987,
   author = {L. A. Feigin and D. I. Svergun},
   doi = {10.1007/978-1-4757-6624-0},
   journal = {Structure Analysis by Small-Angle X-Ray and Neutron Scattering},
   publisher = {Springer US},
   title = {Structure Analysis by Small-Angle X-Ray and Neutron Scattering},
   year = {1987}
}

@article{Matthews2009,
   author = {Brian W. Matthews and Lijun Liu},
   doi = {10.1002/PRO.61},
   issn = {1469-896X},
   issue = {3},
   journal = {Protein Science},
   month = {3},
   pages = {494-502},
   pmid = {19241368},
   publisher = {John Wiley \& Sons, Ltd},
   title = {A review about nothing: Are apolar cavities in proteins really empty?},
   volume = {18},
   url = {https://onlinelibrary.wiley.com/doi/full/10.1002/pro.61 https://onlinelibrary.wiley.com/doi/abs/10.1002/pro.61 https://onlinelibrary.wiley.com/doi/10.1002/pro.61},
   year = {2009}
}

@misc{SphericalHarmonicsVisual,
   author = {Inigo Quilez},
   journal = {Wikimedia},
   month = {5},
   title = {Spherical Harmonics visualization. This work is licensed under the Creative Commons Attribution 3.0 International License. To view a copy of this license, visit https://creativecommons.org/licenses/by-sa/3.0/.},
   year = {2014}
}

@article{Larsen2023,
   author = {Andreas Haahr Larsen and Emre Brookes and Martin Cramer Pedersen and Jacob Judas Kain Kirkensgaard and J. Ilavsky},
   doi = {10.1107/S1600576723005848/JL5064SUP1.PDF},
   issn = {16005767},
   issue = {4},
   journal = {Journal of Applied Crystallography},
   month = {7},
   pages = {1287-1294},
   publisher = {International Union of Crystallography},
   title = {Shape2SAS: a web application to simulate small-angle scattering data and pair distance distributions from user-defined shapes},
   volume = {56},
   url = {https://journals.iucr.org/paper?jl5064 https://journals.iucr.org/j/issues/2023/04/00/jl5064/},
   year = {2023}
}

@article{Caflisch1998,
   author = {Russel E. Caflisch},
   doi = {10.1017/S0962492900002804},
   issn = {1474-0508},
   journal = {Acta Numerica},
   pages = {1-49},
   publisher = {Cambridge University Press},
   title = {Monte Carlo and quasi-Monte Carlo methods},
   volume = {7},
   url = {https://www.cambridge.org/core/journals/acta-numerica/article/abs/monte-carlo-and-quasimonte-carlo-methods/FE7C779B350CFEA45DB2A4CCB2DA9B5C},
   year = {1998}
}

@article{Hung2024,
   author = {Ying Chao Hung},
   doi = {10.1002/WICS.1637},
   issn = {1939-0068},
   issue = {1},
   journal = {Wiley Interdisciplinary Reviews: Computational Statistics},
   month = {1},
   pages = {e1637},
   publisher = {John Wiley \& Sons, Ltd},
   title = {A review of Monte Carlo and quasi-Monte Carlo sampling techniques},
   volume = {16},
   url = {https://onlinelibrary.wiley.com/doi/full/10.1002/wics.1637 https://onlinelibrary.wiley.com/doi/abs/10.1002/wics.1637 https://wires.onlinelibrary.wiley.com/doi/10.1002/wics.1637},
   year = {2024}
}

@article{Blech2024,
   author = {Alexander Blech and Raoul M. M. Ebeling and Marec Heger and Christiane P. Koch and Daniel M. Reich},
   doi = {10.1063/5.0230569/3315373},
   issn = {10897690},
   issue = {13},
   journal = {The Journal of Chemical Physics},
   month = {7},
   pmid = {39365019},
   publisher = {AIP Publishing},
   title = {Numerical evaluation of orientation averages and its application to molecular physics},
   volume = {161},
   url = {http://arxiv.org/abs/2407.17434},
   year = {2024}
}

@article{Kotlarchyk1983,
   author = {Michael Kotlarchyk and Sow-Hsin Chen},
   doi = {10.1063/1.446055},
   issn = {0021-9606},
   issue = {5},
   journal = {The Journal of Chemical Physics},
   month = {9},
   pages = {2461-2469},
   publisher = {AIP Publishing},
   title = {Analysis of small angle neutron scattering spectra from polydisperse interacting colloids},
   volume = {79},
   url = {https://pubs.aip.org/jcp/article/79/5/2461/457361/Analysis-of-small-angle-neutron-scattering-spectra},
   year = {1983}
}

@inbook{Chatzimagas2022,
   author = {Leonie Chatzimagas and Jochen S Hub},
   doi = {https://doi.org/10.1016/bs.mie.2022.08.035},
   editor = {John A Tainer},
   isbn = {0076-6879},
   booktitle = {Methods in Enzymology},
   pages = {433-456},
   publisher = {Academic Press},
   title = {Chapter Fifteen - Predicting solution scattering patterns with explicit-solvent molecular simulations},
   volume = {677},
   url = {https://www.sciencedirect.com/science/article/pii/S0076687922003603},
   year = {2022}
}

@article{Pauw2017,
   author = {B. R. Pauw and A. J. Smith and T. Snow and N. J. Terrill and A. F. Thünemann},
   doi = {10.1107/S1600576717015096},
   issn = {16005767},
   issue = {Pt 6},
   journal = {Journal of Applied Crystallography},
   month = {12},
   pages = {1800},
   pmid = {29217992},
   publisher = {International Union of Crystallography},
   title = {The modular small-angle X-ray scattering data correction sequence},
   volume = {50},
   url = {https://pmc.ncbi.nlm.nih.gov/articles/PMC5713144/},
   year = {2017}
}

@article{Griffiths,
   author = {David J. Griffiths and Darrell F. Schroeter},
   doi = {10.1017/9781316995433},
   isbn = {9781316995433},
   journal = {Introduction to Quantum Mechanics},
   month = {8},
   publisher = {Cambridge University Press},
   title = {Introduction to Quantum Mechanics},
   year = {2018}
}

@article{Ciccariello2002,
   author = {S. Ciccariello and J. M. Schneider and B. Schönfeld and G. Kostorz},
   doi = {10.1107/S0021889802003035/PE0081FIG6.HTML},
   issn = {00218898},
   issue = {3},
   journal = {Journal of Applied Crystallography},
   month = {5},
   pages = {304-313},
   publisher = {International Union of Crystallography},
   title = {Illustration of the anisotropic Porod law},
   volume = {35},
   url = {//journals.iucr.org/paper?pe0081},
   year = {2002}
}

@article{Isserlis1918,
   author = {L. Isserlis},
   doi = {10.1093/biomet/12.1-2.134},
   issn = {0006-3444},
   issue = {1-2},
   journal = {Biometrika},
   month = {11},
   pages = {134-139},
   publisher = {Oxford Academic},
   title = {ON A FORMULA FOR THE PRODUCT-MOMENT COEFFICIENT OF ANY ORDER OF A NORMAL FREQUENCY DISTRIBUTION IN ANY NUMBER OF VARIABLES},
   volume = {12},
   url = {https://academic.oup.com/biomet/article-lookup/doi/10.1093/biomet/12.1-2.134},
   year = {1918}
}

@article{Debye1947,
   author = {P. Debye},
   doi = {10.1021/J150451A002/ASSET/J150451A002.FP.PNG_V03},
   issn = {00223654},
   issue = {1},
   journal = {Journal of Physical and Colloid Chemistry},
   pages = {18-32},
   pmid = {20286386},
   publisher = {American Chemical Society},
   title = {Molecular-weight determination by light scattering},
   volume = {51},
   url = {https://pubs-acs-org.ez.statsbiblioteket.dk/doi/abs/10.1021/j150451a002},
   year = {1947}
}

@article{Ginsburg2019,
   author = {Avi Ginsburg and Tal Ben-Nun and Roi Asor and Asaf Shemesh and Lea Fink and Roee Tekoah and Yehonatan Levartovsky and Daniel Khaykelson and Raviv Dharan and Amos Fellig and Uri Raviv},
   issue = {1},
   journal = {Journal of Applied Crystallography},
   title = {D+: software for high-resolution hierarchical modeling of solution X-ray scattering from complex structures},
   volume = {52},
   url = {https://scripts.iucr.org/cgi-bin/paper?vg5099},
   year = {2019}
}

@article{Narayanan2018,
   author = {Theyencheri Narayanan and Michael Sztucki and Pierre Van Vaerenbergh and Joachim Léonardon and Jacques Gorini and Laurent Claustre and Franc Sever and John Morse and Peter Boesecke},
   doi = {10.1107/S1600576718012748},
   issue = {6},
   journal = {Journal of Applied Crystallography},
   month = {12},
   pages = {1511-1524},
   title = {A multipurpose instrument for time-resolved ultra-small-angle and coherent X-ray scattering},
   volume = {51},
   url = {https://doi.org/10.1107/S1600576718012748},
   year = {2018}
}

@article{Sader2009,
   author = {Kasim Sader and Daniel Studer and Benoît Zuber and Helmut Gnaegi and John Trinick},
   doi = {10.1016/J.ULTRAMIC.2009.09.004},
   issn = {0304-3991},
   issue = {1},
   journal = {Ultramicroscopy},
   month = {12},
   pages = {43-47},
   pmid = {19819624},
   publisher = {North-Holland},
   title = {Preservation of high resolution protein structure by cryo-electron microscopy of vitreous sections},
   volume = {110},
   year = {2009}
}

@book{BookCrystTablesVolC,
   author = {E Prince},
   isbn = {9781402019005},
   publisher = {Springer Netherlands},
   title = {International Tables for Crystallography,Volume C: Mathematical, Physical and Chemical Tables},
   url = {https://books.google.dk/books?id=60FoFEGyShIC},
   year = {2004}
}

@misc{Haahr2018,
   author = {Andreas. Haahr Larsen},
   city = {Copenhagen},
   journal = {Analytical tools for structure determination of protein complexes with small-angle scattering},
   publisher = {University of Copenhagen, Faculty of Science, Niels Bohr Institute},
   title = {Analytical tools for structure determination of protein complexes with small-angle scattering },
   year = {2018}
}

@article{Linse2023,
   author = {Johanna Barbara Linse and Jochen S. Hub},
   doi = {10.1038/s42004-023-01067-1},
   issn = {2399-3669},
   issue = {1},
   journal = {Communications Chemistry 2023 6:1},
   month = {12},
   pages = {1-10},
   publisher = {Nature Publishing Group},
   title = {Scrutinizing the protein hydration shell from molecular dynamics simulations against consensus small-angle scattering data},
   volume = {6},
   url = {https://www.nature.com/articles/s42004-023-01067-1},
   year = {2023}
}

@article{BookModernXrayPhysics,
   author = {Jens Als-Nielsen and Des McMorrow},
   doi = {10.1002/9781119998365},
   isbn = {9780470973950},
   journal = {Elements of Modern X-ray Physics: Second Edition},
   month = {4},
   publisher = {John Wiley and Sons},
   title = {Elements of Modern X-ray Physics: Second Edition},
   url = {https://onlinelibrary.wiley.com/doi/book/10.1002/9781119998365},
   year = {2011}
}

@book{BookJeu,
   author = {Wim H de Jeu},
   doi = {10.1093/acprof:oso/9780198728665.001.0001},
   isbn = {9780198728665},
   month = {4},
   publisher = {Oxford University Press},
   title = {Basic X-Ray Scattering for Soft Matter},
   url = {https://doi.org/10.1093/acprof:oso/9780198728665.001.0001},
   year = {2016}
}

@article{Glatter1977,
   author = {O Glatter},
   doi = {https://doi.org/10.1107/S0021889877013879},
   issn = {1600-5767},
   issue = {5},
   journal = {Journal of Applied Crystallography},
   month = {10},
   pages = {415-421},
   publisher = {International Union of Crystallography (IUCr)},
   title = {A new method for the evaluation of small-angle scattering data},
   volume = {10},
   url = {https://doi.org/10.1107/S0021889877013879},
   year = {1977}
}

@book{Glatter1982,
   author = {O Glatter and O Kratky},
   isbn = {9780122862809},
   publisher = {Academic Press},
   title = {Small Angle X-ray Scattering},
   url = {https://books.google.dk/books?id=J8fvAAAAMAAJ},
   year = {1982}
}

@article{Guinier1956,
   author = {André Guinier and Gérard Fournet and C B Walker and George H Vineyard},
   doi = {10.1063/1.3060069},
   issn = {0031-9228},
   issue = {8},
   journal = {Physics Today},
   month = {8},
   pages = {38-39},
   title = {Small‐Angle Scattering of X‐Rays},
   volume = {9},
   url = {https://doi.org/10.1063/1.3060069},
   year = {1956}
}

@article{Porod1951,
   author = {G Porod},
   doi = {10.1007/BF01512792},
   issn = {1435-1536},
   issue = {2},
   journal = {Kolloid-Zeitschrift},
   pages = {83-114},
   title = {Die Röntgenkleinwinkelstreuung von dichtgepackten kolloiden Systemen},
   volume = {124},
   url = {https://doi.org/10.1007/BF01512792},
   year = {1951}
}

@article{Born1926,
   author = {Max Born},
   doi = {10.1007/BF01397184},
   issn = {14346001},
   issue = {11},
   journal = {Zeitschrift für Physik 1926 38:11},
   month = {11},
   pages = {803-827},
   publisher = {Springer},
   title = {Quantenmechanik der Stoßvorgänge},
   volume = {38},
   url = {https://link.springer.com/article/10.1007/BF01397184},
   year = {1926}
}

@book{BookSvergun2013,
   author = {Dmitri I. Svergun and Michel H. J. Koch and Peter A. Timmins and Roland P. May},
   doi = {10.1093/ACPROF:OSO/9780199639533.001.0001},
   edition = {1st},
   isbn = {9780191747731},
   journal = {Small Angle X-Ray and Neutron Scattering from Solutions of Biological Macromolecules},
   month = {12},
   publisher = {Oxford University Press},
   title = {Small Angle X-Ray and Neutron Scattering from Solutions of Biological Macromolecules},
   url = {https://academic.oup.com/book/6240},
   year = {2013}
}

@book{Sakurai,
   author = {J. J. Sakurai and Jim Napolitano},
   doi = {10.1017/9781108499996},
   edition = {2nd},
   isbn = {9781108499996},
   journal = {Modern Quantum Mechanics},
   month = {9},
   publisher = {Cambridge University Press},
   title = {Modern Quantum Mechanics},
   url = {https://www.cambridge.org/core/product/identifier/9781108499996/type/book},
   year = {2017}
}

@article{Jeffries2021,
   author = {Cy M. Jeffries and Jan Ilavsky and Anne Martel and Stephan Hinrichs and Andreas Meyer and Jan Skov Pedersen and Anna V. Sokolova and Dmitri I. Svergun},
   doi = {10.1038/s43586-021-00064-9},
   issn = {2662-8449},
   issue = {1},
   journal = {Nature Reviews Methods Primers 2021 1:1},
   month = {10},
   pages = {1-39},
   publisher = {Nature Publishing Group},
   title = {Small-angle X-ray and neutron scattering},
   volume = {1},
   url = {https://www.nature.com/articles/s43586-021-00064-9},
   year = {2021}
}

@article{Trewhella2024,
   author = {Jill Trewhella and Patrice Vachette and Andreas Haahr Larsen},
   doi = {10.1107/S205225252400486X},
   issn = {2052-2525},
   issue = {5},
   journal = {IUCrJ},
   month = {9},
   pmid = {38989800},
   publisher = {IUCrJ},
   title = {Benchmarking predictive methods for small-angle X-ray scattering from atomic coordinates of proteins using maximum likelihood consensus data},
   volume = {11},
   url = {https://pubmed.ncbi.nlm.nih.gov/38989800/},
   year = {2024}
}

@article{Debye1949,
   author = {P. Debye and A. M. Bueche},
   doi = {10.1063/1.1698419},
   issn = {0021-8979},
   issue = {6},
   journal = {Journal of Applied Physics},
   month = {6},
   pages = {518-525},
   publisher = {AIP Publishing},
   title = {Scattering by an Inhomogeneous Solid},
   volume = {20},
   url = {/aip/jap/article/20/6/518/159618/Scattering-by-an-Inhomogeneous-Solid},
   year = {1949}
}

@article{Bardhan2009,
   author = {Jaydeep Bardhan and Sanghyun Park and Lee Makowski},
   doi = {10.1107/S0021889809032919},
   issn = {0021-8898},
   journal = {J. Appl. Cryst},
   pages = {932-943},
   title = {SoftWAXS: a computational tool for modeling wide-angle X-ray solution scattering from biomolecules},
   volume = {42},
   year = {2009}
}

@article{Park2009,
   author = {Sanghyun Park and Jaydeep P. Bardhan and Benoît Roux and Lee Makowski},
   doi = {10.1063/1.3099611},
   issn = {0021-9606},
   issue = {13},
   journal = {The Journal of Chemical Physics},
   month = {4},
   title = {Simulated x-ray scattering of protein solutions using explicit-solvent models},
   volume = {130},
   url = {https://pubs.aip.org/jcp/article/130/13/134114/924081/Simulated-x-ray-scattering-of-protein-solutions},
   year = {2009}
}

@article{Poitevin2011,
   author = {Frédéric Poitevin and Henri Orland and Sebastian Doniach and Patrice Koehl and Marc Delarue},
   doi = {10.1093/NAR/GKR430},
   issn = {03051048},
   issue = {Web Server issue},
   journal = {Nucleic Acids Research},
   month = {7},
   pages = {W184},
   pmid = {21665925},
   publisher = {Oxford University Press},
   title = {AquaSAXS: a web server for computation and fitting of SAXS profiles with non-uniformally hydrated atomic models},
   volume = {39},
   url = {/pmc/articles/PMC3125794/ /pmc/articles/PMC3125794/?report=abstract https://www.ncbi.nlm.nih.gov/pmc/articles/PMC3125794/},
   year = {2011}
}

@article{Liu2012,
   author = {Haiguang Liu and Richard J. Morris and Alexander Hexemer and Scott Grandison and Peter H. Zwart},
   doi = {10.1107/S010876731104788X},
   issn = {01087673},
   issue = {2},
   journal = {Acta Crystallographica Section A: Foundations of Crystallography},
   month = {3},
   pages = {278-285},
   pmid = {22338662},
   title = {Computation of small-angle scattering profiles with three-dimensional Zernike polynomials},
   volume = {68},
   year = {2012}
}

@article{Gumerov2012,
   author = {Nail A. Gumerov and Konstantin Berlin and David Fushman and Ramani Duraiswami},
   doi = {10.1002/jcc.23025},
   issn = {0192-8651},
   issue = {25},
   journal = {Journal of Computational Chemistry},
   month = {9},
   pages = {1981-1996},
   title = {A hierarchical algorithm for fast debye summation with applications to small angle scattering},
   volume = {33},
   url = {https://onlinelibrary.wiley.com/doi/10.1002/jcc.23025},
   year = {2012}
}

@article{Traube1895,
   author = {J Traube},
   issue = {3},
   journal = {Ber. Dtsch. Chem. Ges},
   pages = {2722-2728},
   title = {Ueber das Molekularvolumen},
   volume = {28},
   year = {1895}
}

@article{Fraser1978,
   author = {R. D. B. Fraser and T. P. MacRae and E. Suzuki},
   doi = {10.1107/S0021889878014296},
   issn = {0021-8898},
   issue = {6},
   journal = {Journal of Applied Crystallography},
   month = {12},
   pages = {693-694},
   publisher = {International Union of Crystallography (IUCr)},
   title = {An improved method for calculating the contribution of solvent to the X-ray diffraction pattern of biological molecules},
   volume = {11},
   year = {1978}
}

@article{Debye1915,
   author = {P. Debye},
   doi = {10.1002/andp.19153510606},
   issn = {0003-3804},
   issue = {6},
   journal = {Annalen der Physik},
   month = {1},
   pages = {809-823},
   publisher = {John Wiley \& Sons, Ltd},
   title = {Zerstreuung von Röntgenstrahlen},
   volume = {351},
   url = {https://onlinelibrary.wiley.com/doi/10.1002/andp.19153510606},
   year = {1915}
}

@article{Chen2014,
   author = {Po Chia Chen and Jochen S. Hub},
   doi = {10.1016/j.bpj.2014.06.006},
   issn = {15420086},
   issue = {2},
   journal = {Biophysical Journal},
   month = {7},
   pages = {435-447},
   pmid = {25028885},
   publisher = {Biophysical Society},
   title = {Validating solution ensembles from molecular dynamics simulation by wide-angle X-ray scattering data},
   volume = {107},
   url = {http://www.cell.com/article/S0006349514006109/fulltext http://www.cell.com/article/S0006349514006109/abstract https://www.cell.com/biophysj/abstract/S0006-3495(14)00610-9},
   year = {2014}
}

@article{Stuhrmann1970,
   author = {H. B. Stuhrmann},
   doi = {10.1107/S0567739470000748},
   issn = {0567-7394},
   issue = {3},
   journal = {Acta Crystallographica Section A},
   month = {5},
   pages = {297-306},
   publisher = {International Union of Crystallography (IUCr)},
   title = {Interpretation of small-angle scattering functions of dilute solutions and gases. A representation of the structures related to a one-particle scattering function},
   volume = {26},
   url = {https://scripts.iucr.org/cgi-bin/paper?S0567739470000748},
   year = {1970}
}

@article{Knight2015,
   author = {Christopher J. Knight and Jochen S. Hub},
   doi = {10.1093/NAR/GKV309},
   issn = {0305-1048},
   issue = {W1},
   journal = {Nucleic Acids Research},
   month = {7},
   pages = {W225-W230},
   pmid = {25855813},
   publisher = {Oxford Academic},
   title = {WAXSiS: a web server for the calculation of SAXS/WAXS curves based on explicit-solvent molecular dynamics},
   volume = {43},
   url = {https://academic.oup.com/nar/article/43/W1/W225/2467862},
   year = {2015}
}

@article{Grudinin2017,
   author = {Sergei Grudinin and Maria Garkavenko and Andrei Kazennov},
   doi = {10.1107/S2059798317005745},
   issn = {2059-7983},
   issue = {Pt 5},
   journal = {Acta crystallographica. Section D, Structural biology},
   month = {5},
   pages = {449-464},
   pmid = {28471369},
   publisher = {Acta Crystallogr D Struct Biol},
   title = {Pepsi-SAXS: an adaptive method for rapid and accurate computation of small-angle X-ray scattering profiles},
   volume = {73},
   url = {https://pubmed.ncbi.nlm.nih.gov/28471369/},
   year = {2017}
}

@article{Svergun1995,
   author = {D. Svergun and C. Barberato and M. H. J. Koch},
   doi = {10.1107/S0021889895007047},
   issn = {0021-8898},
   issue = {6},
   journal = {Journal of Applied Crystallography},
   month = {12},
   pages = {768-773},
   title = {<i>CRYSOL</i> – a Program to Evaluate X-ray Solution Scattering of Biological Macromolecules from Atomic Coordinates},
   volume = {28},
   url = {https://scripts.iucr.org/cgi-bin/paper?S0021889895007047},
   year = {1995}
}
